%% file: main.tex
\documentclass{llncs}

\usepackage{pdfsync}
\usepackage{amsmath,mathtools}
\usepackage{amssymb}
\usepackage{pdfsync}
\usepackage{graphicx}
\usepackage{xspace}
\usepackage{mathrsfs}
\usepackage[all]{xy}
\usepackage{xspace}
\newtheorem{fact}{Fact}

\input{preamble}

\title{Weak {\scshape MSO+U} with Path Quantifiers over Infinite Trees}
\author{Miko\l{}aj Boja\'{n}czyk\thanks{Supported by ERC Starting Grant ``Sosna''}}
\institute{University of Warsaw}
\begin{document}

\maketitle

\begin{abstract}
	This paper shows that over infinite trees, satisfiability is decidable  for weak monadic second-order logic extended by the unbounding quantifier $\mathsf U$ and  quantification over infinite paths. The proof is by reduction to emptiness for a certain automaton model, while emptiness for the automaton model is decided using profinite trees.
\end{abstract}

\input{intro}

\input{applications}

\input{flatautomata}

\input{proftree}

\input{automaton-chains}
\input{flat-emptiness}

\input{conclusions}

\newcommand{\noopsort}[1]{}

\newpage
\appendix

\newpage
\begin{center}
	{\Large Appendix Part I, consisting of  Sections~\ref{sec:abstract-nesting}-\ref{sec:flat}
	
	\vspace{1cm}
	Equivalence of Logic and Automata}	
		\vspace{2cm}
\end{center}

\input{appendix-overview}

\input{appendix-abstract-nesting}
\input{appendix-weighted}
\input{appendix-nested-counter}

\input{appendix-path-separation}

\input{appendix-flat}

\newpage
\begin{center}
	{\Large Appendix Part II, consisting of  Sections~\ref{sec:appendix-profinite}-\ref{sec:profinite-characterisation-appendix}
	
	\vspace{1cm}
	Emptiness of \wmsoup Automata}	
		\vspace{3cm}
\end{center}

\input{appendix-overview-emptiness}
\input{appendix-profinite}

\input{appendix-automaton-chains}

\input{appendix-lar-form}
\input{appendix-profinite-characterisation}

\end{document}

%% file: preamble.tex
\newcommand{\svalue}{\mathrm{svalue}}

\newcommand{\api}[1]{\boldclass{\Pi}{1}{#1}}
\newcommand{\adelta}[1]{\boldclass{\Delta}{1}{#1}}

\newcommand{\boldclass}[3]{\ensuremath{\mathbf{#1}^{#2}_{#3}}}

	\newcommand{\restrict}{\restriction}

\newcommand{\wmsoup}{{\sc wmso+up}\xspace}
\newcommand{\wmsou}{{\sc wmso+u}\xspace}
\newcommand{\wmso}{{\sc wmso}\xspace}
\newcommand{\msou}{{\sc mso+u}\xspace}
\newcommand{\mso}{{\sc mso}\xspace}

\renewcommand{\phi}{\varphi}
\renewcommand{\epsilon}{\varepsilon}

\newcommand{\set}[1]{\{#1\}}

\newcommand{\eqdef}{\quad\stackrel{\mathrm{def}}{=}\quad}
\newcommand{\Nat}{\mathbb N}

\newcommand{\rrr}{R}

\newcommand{\Bb}{\mathcal B}
\newcommand{\Rr}{\mathcal R}

\newcommand{\Aa}{\mathcal A}

\newcommand{\Uu}{\mathcal U}
\newcommand{\Ll}{\mathcal L}

\newcommand{\trees}[1]{\mathrm{trees}(#1)}

\newcommand{\countertrees}[1]{\mathrm{countertrees}(#1)}

\newcommand{\weightedtrees}[1]{\mathrm{weightedtrees}(#1)}
\newcommand{\weightedwords}[1]{\mathrm{weightedwords}(#1)}

\newcommand{\weightlab}[1]{#1_{\mathrm{lab}}}
\newcommand{\weightwei}[1]{#1_{\mathrm{we}}}

\newcommand{\lreg}[1]{L_{\mathrm{reg}}(#1)}

\newcommand{\treeval}[1]{[\![#1]\!]}

\newenvironment{mproof}{\begin{proof}}{\qed\end{proof}}

\newcommand{\real}[1]{{#1_{\mathrm{r}}}}

\newcommand{\help}[1]{\mathrm{help}_{#1}}

\newcommand{\sidevalue}{\mathrm{sidevalue}}

%% file: intro.tex
% \section{Introduction}

This paper presents a logic over infinite trees with decidable satisfiability. The logic is  \emph{weak monadic second-order logic with $\mathsf U$ and path quantifiers} (\wmsoup).  A formula of the logic is evaluated in an infinite binary labelled tree. The logic can quantify over: nodes, finite sets of nodes, and paths (a path is a possibly infinite set of nodes totally ordered by the descendant relation and connected with respect to the child relation). The predicates are as usual in \mso for trees: a unary predicate for every letter of the input alphabet,  binary left and right child predicates, and membership of a node in a set (which is either a path or a finite set). Finally, formulas can use  the  \emph{unbounding quantifier}, denoted by
	\begin{align*}
		\mathsf{U} X\  \varphi(X),
	\end{align*}
which 	says that $\varphi(X)$ holds for arbitrarily large finite sets $X$.
	 As usual with quantifiers, the formula $\varphi(X)$ might have other free variables except for $X$. The main contribution of the paper is  the following theorem.
	 \begin{theorem}\label{thm:main}
Satisfiability is  decidable for \wmsoup over infinite trees.
	 \end{theorem}
	
% \begin{example}To distinguish between the various quantifiers, we assume that node variables are denoted by $x,y,z,\ldots$, path variables  are $\pi,\sigma,\ldots$ and finite set variables are $X,Y,Z,\ldots$.  Consider trees over alphabet $\set{a,b}$. Define an $a$-block to be a connected set of $a$'s, which is defined by the following first-order formula, call it $\mathrm{block(X)}$:
% 	\begin{align*}
% 		\exists x \in X \  \forall y \in X \qquad  \ (a(y) \land y \ge x \land  \forall z \ (x \le z \le y \Rightarrow z \in X))
% 	\end{align*}
% 	The following formula of \wmsoup says that for some path $\pi$, one can find  $a$-blocks of unbounded size in the subtree of every node $x$ from $\pi$:
% 	\begin{align*}
% 		\exists \pi \ \forall x \in \pi \ \mathsf U X \qquad (\forall y \in X \ y \ge x) \ \land \ block(X)
% 	\end{align*}
% \end{example}

\paragraph*{Background.} This paper is part of a program researching the logic \msou, i.e.~monadic second-order logic extended with the $\mathsf U$  quantifier. 
 The logic  was  introduced in~\cite{DBLP:conf/csl/Bojanczyk04}, where it was shown  that satisfiability is decidable over infinite trees as long as the $\mathsf U$ quantifier is used once and not under the scope of set quantification. A significantly more powerful fragment of the logic, albeit for infinite words, was shown decidable in~\cite{BojanczykColcombet06} using automata with counters. These automata  where further developed into the theory of cost functions initiated by Colcombet in~\cite{Colcombet09}. Cost functions can be seen as a special case of \msou in the sense that  decision problems regarding cost functions, such as limitedness or domination, can be easily encoded into satisfiability of \msou formulas. This encoding need not be helpful, since  the unsolved problems for cost functions get encoded into unsolved problems from \msou. % For instance: the star height problem can be solved using both cost functions and decidable fragments of \msou~\cite{BojanczykColcombet06}; the star height problem for trees can be solved using 

The logic \msou  can be used to solve problems that do not have a simple solution in \mso alone. One example is the finite model problem for two-way $\mu$-calculus, which can be solved by a reduction to satisfiability of  \msou on infinite trees; the reduction and the decidability of the fragment used by the reduction are shown in~\cite{DBLP:conf/csl/Bojanczyk04}. A  more famous problem is  the star height problem, which can be solved by a reduction to the satisfiability of \msou on infinite words; the  particular fragment of \msou used in this reduction is decidable by~\cite{BojanczykColcombet06}.  In Section~\ref{sec:notation-and-applications}  we give more examples of problems which can be reduced to satisfiability for  \msou, examples which use the fragment that is	 solved in this paper.
 An example of an unsolved problem that reduces to \msou  is the decidability of the nondeterministic parity index problem, see~\cite{DBLP:conf/icalp/ColcombetL08}.

The first strong evidence that \msou can be too expressive was given  in~\cite{DBLP:journals/fuin/HummelS12}, where it was shown that \msou  can define languages of infinite words that are arbitrarily high in the projective hierarchy. 
In~\cite{msoq},  the result from~\cite{DBLP:journals/fuin/HummelS12} is  used to show  that there is no algorithm which decides satisfiability of \msou on infinite trees and has a correctness proof using the axioms of  \textsc{zfc}. A challenging open question is whether satisfiability of \msou is decidable on infinite words.

The principal reason for the undecidability result above is that \msou can define languages of high topological complexity. Such problems  go away in the weak variant, where only quantification over finite sets is allowed, because weak quantification can only define Borel languages.  Indeed, satisfiability is decidable for \wmsou over infinite words~\cite{Bojanczyk11} and  infinite trees~\cite{BojanczykTorunczyk12}. This paper continues the research on weak fragments from~\cite{Bojanczyk11,BojanczykTorunczyk12}. Note that \wmsoup can, unlike \wmsou,   define non Borel-languages, e.g.~``finitely  many $a$'s on every path'', which is complete for level $\api 1$ of the projective hierarchy.  The automaton characterization of \wmsoup in this paper  implies that \wmsoup definable languages are contained in level $\adelta 2$.

What is the added value of path quantifiers? One answer is given in the following section, where we show how \wmsoup can be used to solve games winning conditions definable in \wmsou; here the use of path quantifiers is crucial. Another answer is that  solving a logic with path quantifiers is a step in the direction of tackling one of the  most notorious difficulties when dealing with the unbounding quantifier, namely  the interaction between  
quantitative properties (e.g.~some counters have small values) with qualitative limit properties (e.g.~the parity condition). The difficulty of this interaction is one of the reasons why the boundedness problem for cost-parity automata on infinite trees remains open~\cite{DBLP:conf/icalp/ColcombetL08}.  Such interaction is also a source of difficulty in the present paper, arguably more so than in the previous paper on \wmsou for infinite trees~\cite{BojanczykTorunczyk12}. One of the main contributions of the paper is a set of tools that can be used to tackle this interaction. The tools use profinite trees.
% \paragraph*{Plan of the proof.} 
%  The  proof of Therem~\ref{thm:main} uses the automata method. We introduce an automaton model in Section~\ref{sec:automata} which has the same expressive power as existential \wmsoup. Emptiness for the automaton is then shown decidable, with profinite trees playing an important role. We develop some theory of profinite trees in Section~\ref{sec:profinite-trees}, and apply it to the emptiness problem in Section~\ref{sec:emptiness-for-biginf-smallsup-automata}.
 
 \paragraph*{Acknowledgment.} I would like to thank Szymon Toru\'nczyk and Martin Zimmermann for months of discussions about this paper; in particular Szymon Toru\'nczyk suggested the use of profinite trees. Also, I would like to thank the anonymous referees for their comments.

%% file: applications.tex
\section{Notation and some applications}
\label{sec:notation-and-applications}
Let us begin by fixing notation for trees and parity automata. Notions of root, leaf, sibling, descendant, ancestor, parent are used in the usual sense. A tree in this paper is labelled, binary, possibly infinite and not necessarily complete. In other words, a tree  is a partial function from $\set{0,1}^*$ to the input alphabet, whose domain is closed under parents and siblings.
  % We write $\trees A$ for the set of trees over input alphabet $A$. 
  The logic \wmsoup, as defined in the introduction,  is used to define languages of such trees.   To recognise properties of trees, we use the following variant of parity automata. A parity automaton is given by an input alphabet $A$, a set of states $Q$, an initial state, a total order on the states, a set of accepting states, and finite sets of transitions 
\begin{align*}
 \delta_0 \subseteq Q \times A \quad \mbox{and}\quad \delta_2  \subseteq Q \times A  \times Q^2.
\end{align*}
A run of the automaton is a labeling of the input tree by states such that for every node with $i \in \set{0,2}$ children, the set $\delta_i$  contains the  tuple consisting of the node's state, label and the sequence of states in its children.
 A run is accepting if it has the initial state in the root, and on every infinite path, the maximal state appearing infinitely often is accepting. Parity automata defined this way have the same expressive power as \mso.   

Before continuing, we underline the distinction between paths, which are connected sets of nodes totally ordered by the ancestor relation,  and chains which can be possibly disconnected.  Having  chain quantification and the $\mathsf U$ quantifier would be sufficient to  express  all properties of the leftmost path definable in \msou, and therefore its decidability would imply decidability of \msou on infinite words, which is open.

The rest of this section is devoted to  describing some consequences of Theorem~\ref{thm:main}, which says that satisfiability is decidable for  \wmsoup on infinite trees.

\paragraph*{Stronger than \mso.} When deciding satisfiability of \wmsoup in Theorem~\ref{thm:main}, we ask for the existence of a tree labelled by  the input alphabet. Since the labelling is quantified existentially in the satisfiability problem, the decidability  result immediately extends to  formulas of \emph{existential \wmsoup}, which are obtained from formulas of  \wmsoup by adding a prefix of existential quantifiers over  arbitrary, possibly infinite, sets. A result equivalent to  Theorem~\ref{thm:main} is that the existential \wmsoup theory of the  unlabeled complete binary tree  is decidable. 

Existential \wmsoup contains all of \mso, because it can express that a parity tree automaton has  an accepting run. The existential prefix is used to guess the accepting run, while the path quantifiers are used to say that it is accepting.
 One can prove a stronger result. Define \emph{\wmsoup with \mso subformulas}, to be the extension of  \wmsoup where  quantification over arbitrary sets is allowed under the following condition:
 if a subformula $\exists X\ \varphi(X)$ quantifies over an arbitrary set $X$, then $\varphi(X)$ does not use the unbounding quantifier. 
 \begin{fact}\label{lem:}
 	\wmsoup with \mso subformulas is contained in existential \wmsoup.
 \end{fact}
The idea behind the fact is to use the existential prefix to label each node with the \mso-theory of its subtree.

\begin{example}
	Consider the modal $\mu$-calculus with backward modalities, as introduced in~\cite{DBLP:conf/icalp/Vardi98}. 
	As shown in~\cite{DBLP:conf/csl/Bojanczyk04}, for every formula $\varphi$ of the modal $\mu$-calculus with backward modalities, one can compute a formula $\psi(X)$ of \mso such that $\varphi$ is true in some finite Kripke structure if and only if 
	\begin{align}
		\label{eq:csl}
		\mathsf U X \varphi(X)
	\end{align} is true in some infinite tree. The paper~\cite{DBLP:conf/csl/Bojanczyk04} gives a direct algorithm for testing satisfiability of formulas of the form as in~\eqref{eq:csl}. Since this formula is in \wmsoup with \mso subformulas, Theorem~\ref{thm:main} can be used instead. 
\end{example}
	
By inspecting the proofs of~\cite{BojanczykTorunczyk12}, one can show that also~\cite{BojanczykTorunczyk12} would be enough for the above example. This is no longer the case for the following example.

\begin{example}\label{ex:games-with-wmso-conditions}
	Consider a two-player game over an arena with a finite set of  vertices $V$, where the winning condition is a subset of  $V^\omega$ defined in \wmsou over infinite words. For instance, the winning condition could say that a node $v \in V$ is visited infinitely often, but the time between visits is unbounded. 
	A winning strategy for player 1 in such a game is a subset $\sigma \subseteq V^*$, which can be visualized as a tree of branching at most $V$. The properties required of a strategy can be formalised in \wmsoup over infinite trees, using  path quantifiers to range over strategies of the opposing player. Therefore, one can write a formula of  \wmsoup over infinite trees, which is true in some tree if and only if player 1 has a winning strategy in the game. Therefore Theorem~\ref{thm:main} implies that one can decide the winner in games over finite arenas  with \wmsou winning conditions.
\end{example}

The games described in Example~\ref{ex:games-with-wmso-conditions} generalize  cost-parity games from~\cite{DBLP:conf/fsttcs/FijalkowZ12} or energy consumption games from~\cite{DBLP:conf/cav/BrazdilCKN12}, so Theorem~\ref{thm:main} implies the decidability results from those papers (but not the optimal complexities).

\begin{example}\label{ex:colcombet-style}
%	Consider a variant of LTL, called \emph{prompt LTL}, which contains a modality $\mathsf F_{\le k} \varphi$.  
	Consider a  game as in the previous example, but where the winning condition is defined by a formula $\varphi$ of \wmsou which can also use a binary predicate ``$x$ and $y$ are close''.  For $n \in \Nat$, consider the winning condition $\varphi_n$ to be the formula $\varphi$ with ``$x$ and $y$ are close'' replaced by ``the distance between $x$ and $y$ is at most $n$''. Consider the following problem: is there some $n \in \Nat$, such that player 1 has a winning strategy according to the winning condition $\varphi_n$? This problem can also be reduced to satisfiability of \wmsoup on infinite trees. The idea is to guess a strategy $\sigma \subseteq V^*$, and a set of nodes $X \subseteq \sigma$, such that 1) there is a common upper bound on the length of finite paths that do not contain nodes from $X$; 2) every infinite path consistent with $\sigma$ satisfies the formula $\varphi$ with  ``$x$ and $y$ are close'' replaced by ``between $x$ and $y$ there is at most one node from $X$''. Using the same idea, one can solve  the realizability problem for Prompt LTL~\cite{KupfermanPitermanVardi09}.
\end{example}

%% file: flatautomata.tex
\section{Automata}
\label{sec:automata}
In this section,
 we define an automaton model with  the same expressive power as existential \wmsoup, which is called a \emph{\wmsoup automaton}. The automaton uses a  labellings of trees by counter operations called \emph{counter trees}, so we begin by describing these.
 
%  A tree walking automaton is a sequential two-way automaton model for trees. In a node of the tree, the automaton can query the label of the node, and whether or not the node is a left child, a right child or a leaf (or a combination of these, e.g.~a left child leaf) 
%  Based on the current state, information about whether the current node is the root, a le
% A \emph{weighted tree-walking automaton} is a nondeterministic tree-walking automaton plus a mapping which assigns to every transition a weight in $\set{0,1}$. The weight of a run of such an automaton is the sum of weights of transitions it uses.  The weight of a configuration (a tree node and a state) is the supremum of weights of runs that start in the configuration and which end in an accepting state.
 
 \paragraph*{Counter trees.}
 Let $C$ be a finite set of counters. A \emph{counter tree} over a set of counters $C$ is defined to be a tree where every node is labelled by a subset of 
 \begin{align}\label{eq:counterops-alphabet}
 	C \times \set{\mbox{parent},\mbox{self}} \times \set{\mbox{increment}, \mbox{transfer}} \times C \times \set{\mbox{parent},\mbox{self}},
 \end{align}
 where every tuple contains ``self'' at least once. The counter tree induces a graph with edges labelled by ``increment'' or ``transfer'', called its associated \emph{counter configuration graph}. The vertices of this graph, called \emph{counter configurations},  are pairs $(x,c)$ where $x$ is a node of the counter tree and $c$ is a counter. The counter configuration graph contains an edge from $(x_0,c_0)$ to $(x_1,c_1)$ labelled by  $o$ if and only if  there exists a node $x$ in the counter tree whose  label contains  a tuple
  \begin{align*}
 	(c_0,\tau_0,o,c_1,\tau_1) \qquad \mbox{with} \qquad \tau_0,\tau_1 \in \set{\mbox{parent},\mbox{self}}
 \end{align*}
 such that $x_i$ is $x$ or its parent depending on whether $\tau_i$ is ``self'' or ``parent''.
 %  The edges of this graph are described by the labels of the counter tree in the natural way, e.g.~if the label of a node $x$ is a set that contains the tuple
 % \begin{align*}
 % 	(c,\mbox{parent},\mbox{increment},d,\mbox{self}),
 % \end{align*}
 % then the counter configuration graph contains an edge which begins in  counter $c$ in the parent of $x$, has color  ``increment'', and ends in counter $d$ in $x$.

 A path in the  counter configuration graph, using possibly both kinds of edges, is called a \emph{counter path}. Its value  is defined to be  the number of ``increment'' edges. The value of a counter configuration is defined to be the supremum of values of counter paths that end in it. 
When $t$ is a counter tree, then we write $\treeval t$ for the tree with the same nodes but  with alphabet $\bar \Nat^C$, where the label of a node $x$  maps $c \in C$ to the value of $(x,c)$  in the associated counter graph.

\paragraph*{\wmsoup automata.}
We now present the automaton model  used to decide \wmsoup. The syntax of a \emph{\wmsoup  consists} of:
\begin{enumerate}
	\item A  parity  automaton;
	\item A set of counters $C$, partitioned into  \emph{bounded} and \emph{unbounded} counters;
	\item For every state $q$ of the parity automaton: 
	\begin{enumerate}
		\item a set $cut(q)$ of bounded counters, called the counters cut by $q$;
		\item a set $check(q)$ of unbounded counters, called the counters checked by $q$;
		\item a subset $counterops(q)$ of the set in~\eqref{eq:counterops-alphabet}.
	\end{enumerate}
\end{enumerate}

The automaton inputs a tree  over the input alphabet of the parity automaton in the first item.
A \emph{run}  of the automaton  is a labelling of the input tree by states, consistent with the transition relation of the parity automaton. Using the sets $counterops(q)$, we get a counter tree with counters $C$, call it  $counterops(\rho)$.  By abuse of notation, we write $\treeval \rho$ for the tree $\treeval{counterops(\rho)}$, which is a tree over $\bar \Nat^C$.
 Using the sets $cut(q)$ and $check(q)$, we can talk about the nodes in a run where a bounded counter gets cut, or an unbounded counter gets checked. A run is accepting if it has the initial state in the root, and it satisfies all three  acceptance  conditions defined below. In the conditions, we define the limsup of a function ranging over a countable set to be
 \begin{align*}
 	\limsup_{x \in X} f(x) \eqdef \limsup_{n \in \Nat} f(x_n)  \qquad \mbox{for some enumeration of $X=\set{x_1,x_2,\ldots}$},
 \end{align*}
which is well-defined because it does not depend on the enumeration.
\begin{itemize}
	\item {\it Parity.} On every  path the maximal state seen infinitely often is accepting.
	\item {\it Boundedness.} If   a bounded counter $c$  is never cut in a connected\footnote{It suffices to restrict attention to maximal connected sets of nodes where $c$ is not cut, such sets are called $c$-cut factors.
} set of nodes $X$, then
	\begin{align*}
 \limsup_{x \in X}  \treeval \rho(x,c)  < \infty
	\end{align*}
	\item {\it Unboundedness.} If an unbounded counter $c$ is checked infinitely often on a path $\pi$, then
	\begin{align*}
\limsup_{x} \treeval \rho(x,c)  = \infty
	\end{align*}
	with  $x$ ranging over those nodes in $\pi$ where $c$ is checked.
\end{itemize}
The automaton accepts an input tree if it admits an accepting run.

% \paragraph*{Alternative model.} To make the   boundedness and unboundedness acceptance conditions more symmetric with  each other, we could use the following variant of the boundedness condition:
% \begin{itemize}
% 	\item {\it Path boundedness condition.} If   a bounded counter $c$  is never cut on an infinite path $\pi$, then
% 	\begin{align*}
%  \limsup_{x \in \pi}  \treeval \rho(x,c)  < \infty
% 	\end{align*}
% \end{itemize}
% Unlike for the unboundedness condition, it is important in the path boundedness condition that the infinite path $\pi$ need not begin in the root.  One can show that replacing the boundedness condition by the path boundedness condition would not change the expressive power of the model. Despite its symmetric virtues, we do not use the path boundedness condition because it is less convenient for the proof.

\paragraph*{Equivalence to logic and emptiness.} 
Below are the two main technical results about  \wmsoup automata. The two results immediately imply that satisfiability is decidable for \wmsoup logic.

\begin{theorem}\label{thm:from-wmso-up-to-biginf-smallsup}
	For every formula of existential \wmsoup one can compute a  \wmsoup automaton that accepts the same trees, and vice versa.
\end{theorem}

\begin{theorem}\label{thm:biginf smallsup-have-decidable-emptiness}
	Emptiness is decidable for  \wmsoup automata.
\end{theorem}

The proof of Theorem~\ref{thm:from-wmso-up-to-biginf-smallsup} is in the the appendix. 
The rest of this paper is devoted to describing the proof of  Theorem~\ref{thm:biginf smallsup-have-decidable-emptiness}. The proof itself is described in Section~\ref{sec:emptiness-for-biginf-smallsup-automata}, while the next  section is about profinite trees, which are used in the proof.

\begin{remark}\label{remark}
	If in the definition of the unboundedness acceptance condition, we replace $\limsup$ by $\liminf$, we get a more powerful model. The same proof as for Theorem~\ref{thm:biginf smallsup-have-decidable-emptiness} also shows that this more powerful model has decidable emptiness.
\end{remark}

%% file: proftree.tex
\section{Profinite trees and automata on them}
\label{sec:profinite-trees}
In the emptiness algorithm for  \wmsoup automata, we  use profinite trees.  The connection between boundedness problems and profiniteness was already explored in~\cite{DBLP:conf/icalp/Torunczyk12}, in the case of words.
Profinite trees are similar to profinite words, because the recognizers are \mso formulas, the difference is that the objects are (infinite) trees.
 Consider an input alphabet $A$. Fix an enumeration of all \mso formulas over the alphabet $A$. We define the distance between two trees to be $1/n$ where $n$ is the smallest number such that the $n$-th formula is true in   one of the trees but not the other. The distance itself depends on the enumeration, but the notion of an open set or Cauchy sequence does not.
  % A Cauchy sequence of trees is a sequence of trees such that for every $\epsilon>0$, almost all trees in the sequence are at distance at most $\epsilon$ from each other.
    Cauchy sequences are considered equivalent if some (equivalently, every) shuffle of them  is also a Cauchy sequence. A \emph{profinite tree} is defined to be an equivalence class of Cauchy sequences.
To avoid confusion with  profinite trees, we use  from now on the term \emph{real tree} instead of tree. Therefore, a profinite tree is a limit of a sequence of real trees. Every real tree is also a profinite tree, as a limit of a constant sequence.

% 
% \paragraph*{Continuous functions.} If a   function from real trees to a complete metric space is continuous,  i.e.~it maps Cauchy sequences to Cauchy sequences, then it can be uniquely  extended to a continuous function from profinite trees, by defining
% \begin{align*}
% 	f(\lim t_n) \eqdef \lim f(t_n).
% \end{align*}
% It is not difficult to show that a function  from real trees  to real trees   is continuous if and only if for every set of real trees recognised by a parity automaton, its  inverse image is also recognised by a parity automaton.
% All the above definitions make sense for partial functions, by coding  the undefined value as a special element in the space of profinite trees, at distance 1 from all other trees, real or profinite.

\paragraph*{Evaluating \mso formulas on profinite trees.}
A Cauchy sequence is said to satisfy an \mso formula  if  almost all trees in the sequence satisfy it.  A Cauchy sequence satisfies either an \mso formula, or its negation. Equivalent Cauchy sequences satisfy the same \mso formulas, and therefore satisfaction of \mso formulas is meaningful for profinite trees: a profinite tree is said  to satisfy an \mso formula if this is true for some (equivalently, every) Cauchy sequence that tends to it.  Formulas of \mso  are the only ones that can be extended to profinite trees in this way; one can show that if $L$ is a set of real trees that is not \mso-definable (for instance, $L$ is defined by a formula of \wmsoup that is not in \mso), then there is a Cauchy sequence which has infinitely many elements in $L$ and infinitely many elements outside $L$. Summing up,  it makes sense to ask if a profinite tree satisfies a formula of \mso, but it does not make sense to ask if it satisfies a formula of \wmsoup.

\vspace{-0.2cm}

\paragraph*{Profinite subtrees.}
 The \emph{topological closure} of a binary relation on real trees is defined to be the pairs of profinite trees that are limits of pairs of real trees in the binary relation; with the metric in the product being the maximum of distances over coordinates.   
Define the \emph{profinite subtree} relation to be the topological closure of the subtree relation. 
A real tree might have profinite subtrees that are not real. For example, consider a real tree  $t$  such that for every $n$, some subtree  $s_n$ of $t$ has exactly one $a$, which occurs at depth $n$ on the   leftmost branch. By compactness, the sequence $s_1,s_2,\ldots$ has a convergent subsequence, whose limit is not a real tree, but is a profinite  subtree of $t$.

\begin{figure}[htbp]
	\centering		\includegraphics[scale=0.4]{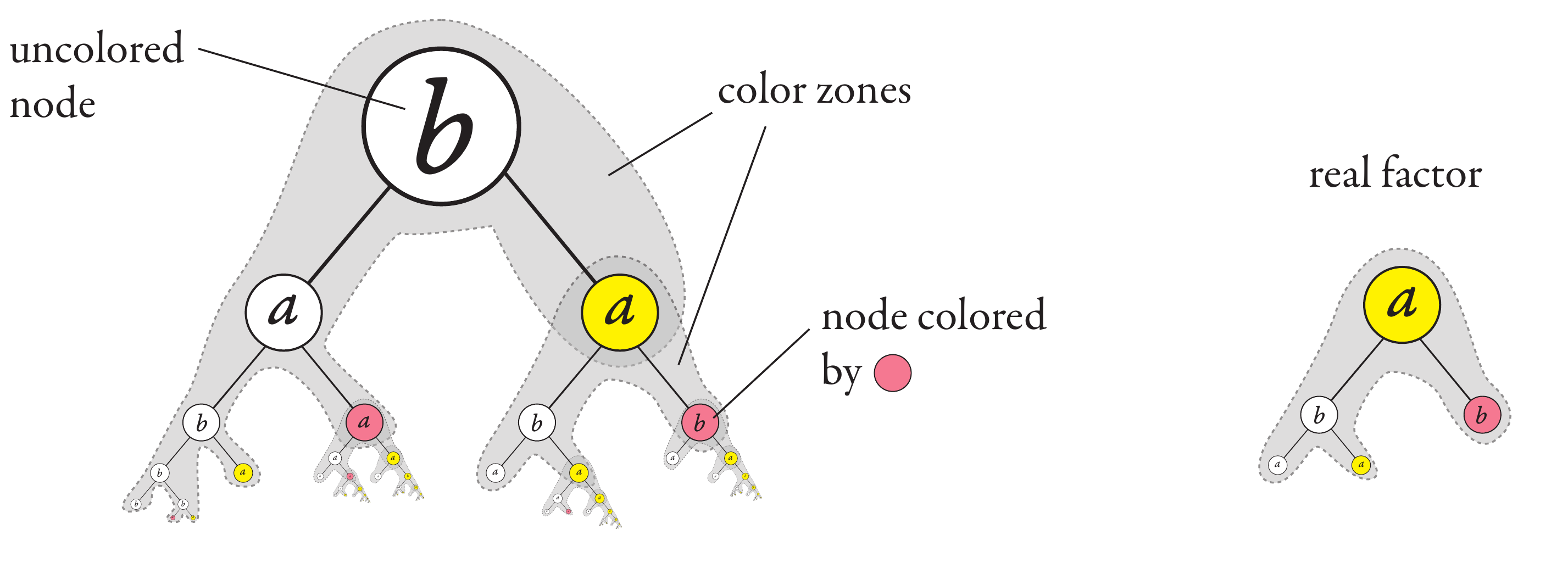}
	\caption[dupa]{A real $\set{\includegraphics[scale=0.07]{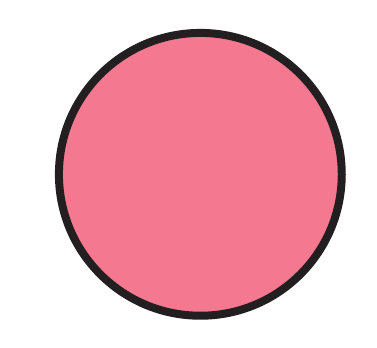},\includegraphics[scale=0.07]{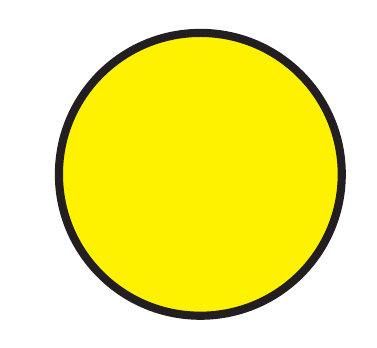}}$-colored  tree over $\set{a,b}$, together with  a real factor. Uncolored nodes are white. Note how color  zones overlap on colored nodes.}
	\label{fig:label}
\end{figure}

\vspace{-0.2cm}
\paragraph*{Partially colored trees.} Let $A$ and $Q$ be finite sets.
A \emph{partially $Q$-colored tree over $A$} is a  tree, possibly profinite, over the alphabet  $A \times (Q \cup \set \bot)$. Suppose that $\rho$ is a real partially $Q$-colored tree over $A$. If a node  has second coordinate $q\in Q$, then  we say that it is colored by $q$. When the second coordinate is $\bot$, then the node is called uncolored. A \emph{color zone} of $\rho$ is a connected set of nodes $X$ in $\rho$ such that:
\begin{itemize}
	\item the unique minimal element of $X$ is either the root of $\rho$ or is colored;
	\item maximal elements of $X$ are either leaves of $\rho$ or are colored;
	\item all other elements of $X$ are uncolored.
\end{itemize}
A real tree is called \emph{real factor}  of $\rho$ if  it is  obtained from $\rho$ by only keeping the nodes in some color zone.  These notions are illustrated in Figure~\ref{fig:label}.
The notions of defined color,  color zone and real factor are only meaningful when $\rho$ is a real tree.  When $\rho$ is not a real tree, then we can still use \mso-definable properties, such as ``the root has undefined color'' or ``only the leaves and root have defined color''.
  Define the \emph{profinite factor} relation to be the topological closure of the real factor relation.  
% ; if $\rho$ is a profinite partially $Q$-colored tree over $A$ then any profintie factor of $\rho$ is called a \emph{$\rho$-factor}.

% A \emph{factorisation} of a profinite tree $t$ is a partially $\set{0,1}$-colored tree $f$ over the input alphabet of $t$ which projects to $t$ on the $A$ coordinate.

\vspace{-0.2cm}

\paragraph*{Generalized parity automata.}
A transition in a parity automaton  can be visualized as a little tree, with one or three nodes,  all of them colored by states. We introduce a generalized model, where transitions can be arbitrary trees, possibly infinite, and possibly profinite.
A \emph{generalized parity automaton} consists of: a totally ordered set of states $Q$,  a subset of accepting states, an input alphabet, and a set of transitions, which is an arbitrary set of $Q$-colored profinite trees over the input alphabet.  An input to the automaton is a profinite tree over the input alphabet. A run over such an input is a partially $Q$-colored profinite tree over the input alphabet, call it $\rho$, which projects to the input on the coordinate corresponding to the input alphabet. By projection we mean the topological closure of the projection relation on real trees.  A run $\rho$ is accepting if all of its profinite factors are transitions, and it satisfies the \mso properties ``the root is uncolored'' and  ``on every infinite path where colored nodes appear infinitely often, the maximal color seen infinitely often is accepting''. (The  transitions where the root is uncolored  play the role of the initial state.) There might be some infinite paths which have  colors finitely often, because some transitions might have infinite paths. 
Every profinite factor of a run will necessarily satisfy the \mso property ``every node that is not the root or a leaf is uncolored'', therefore it only makes sense  to have transitions that satisfy this property. It is not difficult to show that if a run satisfies the property ``the root is uncolored'', which is the case for every accepting run, then the run has  a unique profinite factor that satisfies this property. 

A run is called \emph{regular} if it has finitely many profinite subtrees rooted in colored nodes. For a generalized parity automaton $\Aa$, define $L(\Aa)$ to be the set of profinite trees accepted by $\Aa$, and let $\lreg\Aa$ be the subset of those profinite trees which are accepted via a regular run. The following theorem shows that two sets have the same topological closure (denoted by a bar on top), i.e.~the smaller set is dense in the bigger one.
 
\begin{theorem}\label{cor:dense}
  $\overline{\lreg \Aa}= \overline{L(\Aa)}$ 	holds for every generalized parity automaton $\Aa$.
\end{theorem}

%% file: automaton-chains.tex
\subsection{Automaton chains}
Generalised parity automata are too general to be useful. For instance, every set of profinite trees is recognised by a generalised parity automaton, which has no states, and uses the recognised set as its transitions. Also, these automata do not allow a finite representation, and therefore cannot be used in algorithms. 
The emptiness algorithm for  \wmsoup automata uses a special case of generalised parity automata, called \emph{automaton chains}, which can be represented in a finite way. Roughly speaking, an automaton chain is a generalised parity automaton where the set of transitions is the set of profinite trees defined by a simpler automaton  chain, with the additional requirement that one cost function is bounded and another cost function is unbounded. The definitions of cost functions and automaton chains are given below.

\paragraph*{Cost functions.} 
A \emph{cost function} on trees is a function $\alpha$ from real trees to $\bar \Nat$, such that  the inverse image of every finite number $n \in \Nat$ is definable in \mso.  As proposed by Toru\'nczyk in~\cite{DBLP:conf/icalp/Torunczyk12}, a cost function $\alpha$ can be applied  to a profinite tree $t$ by defining $\alpha(t)$ to be a finite number $n \in \Nat$ if  $t$ satisfies the \mso property ``has value $n$ under $\alpha$'', and to be $\infty$ otherwise.
Cost functions on finite words were introduced by Colcombet in~\cite{Colcombet09} and then extended to finite trees, infinite words and infinite trees.  The specific variant of cost functions that we use is  the logic \emph{cost \wmso} that was proposed by Vanden Boom in~\cite{Boom11}. A sentence of this logic is built the same way as a sentence of \wmso over infinite trees, except that it can use an additional predicate ``$X$ is small'', which takes a set $X$ as a parameter, and can only  be used under an even number of negations. The predicate can be used for different sets, like in the following example, call it $\alpha$:
\begin{align*}
	\exists X  \ \exists Y \ X \mbox{ is small} \ \land \ Y  \ \mbox{is small}\ \land \ (\forall x \ a(x) \Rightarrow x \in X) \  \land \ (\forall y \ b(y) \Rightarrow y \in Y) \ 
\end{align*} The cost function defined by a sentence  of  cost \wmso maps a tree to   the smallest number $n$ such that the sentence becomes true after ``$X$ is small'' is replaced by $|X|<n$. If such a number does not exist, the result is $\infty$.  In the case of the example $\alpha$ above, the function maps a tree to the  number of $a$'s or to the number of  $b$'s, whichever is bigger.

\paragraph*{Automaton chains.} We now define automaton chains, by induction on a parameter called \emph{depth}.
A automaton chain of depth $0$ is any parity automaton. For $n > 0$, an automaton chain of depth $n$ is a generalised parity automaton $\Aa$ whose set of transitions is 
\begin{align*}
	\set{t : \mbox{$t$ is accepted by $\Bb$ and $\alpha(t)<\infty$ and $\beta(t)=\infty$}}
\end{align*}
 for some automaton chain $\Bb$ of smaller depth and   some cost functions $\alpha,\beta$ that are definable in cost \wmso. 
An automaton chain can be represented in a finite way and therefore used as an input for an algorithm, such as in the following lemmas.
	
	\begin{lemma}\label{lem:automaton-chain}
		Nonemptiness is  decidable for automaton chains.
		\end{lemma}
		
		\begin{lemma}\label{lem:automaton-chain-mso}
			Automaton chains are effectively closed under intersection with  \mso formulas.
		\end{lemma}

%% file: flat-emptiness.tex
\section{Emptiness of  \wmsoup  automata}
\label{sec:emptiness-for-biginf-smallsup-automata}
In this section, we describe the proof of Theorem~\ref{thm:biginf smallsup-have-decidable-emptiness}, which says that emptiness is decidable for  \wmsoup automata. We   reduce emptiness for  \wmsoup automata to emptiness of  automaton chains, which is decidable by Lemma~\ref{lem:automaton-chain}.

\paragraph*{A normal form.}
We begin by normalising  the automaton. 
A counter $c$ is called \emph{separated} in a counter tree  if the counter tree does not contain edges that involve $c$ and and some other counter. A counter $c$ is called \emph{root-directed} if every counter edge involving $c$ is directed toward the root. A \wmsoup automaton is said to be in \emph{normal form}  if:
\begin{enumerate}
	\item[(a)] for every run, in the counter graph generated by the automaton, every bounded counter is separated and root-directed.	\item[(b)] there is a total order on the states which is consistent with the order from the parity condition, and a mapping  which  maps every state $q$ to sets of counters $larcut(q)$ and $larcheck(q)$ with the following property. For every  run and every  finite path in the run that starts and ends in state $q$ and does not visit bigger states in the meantime, 
	\begin{itemize}
		\item   the  counters checked in the path are exactly $larcheck(q)$;
				\item   the counters cut in  the path are exactly $larcut(q)$.
	\end{itemize}
\end{enumerate}

\begin{lemma}\label{lem:lar}
	For every  \wmsoup automaton one can compute an equivalent one in  normal form.
\end{lemma}
In the proof, to achieve property (b), we use  the latest appearance record data structure introduced by McNaughton in~\cite{mcnaughtonlar}. 

\paragraph*{Partial runs.} Let $\Aa$ be a  \wmsoup automaton that we want to test for emptiness. 
 Thanks to Lemma~\ref{lem:lar}, we  assume without loss of generality that it is in  normal form.  In the emptiness algorithm, we describe properties of pieces of runs of $\Aa$, called partial runs, and defined below. Recall that in a parity automaton, there are two types of transitions $\delta_0$ and $\delta_2$, for leaves and non-leaves, respectively. A \emph{partial run} of a parity automaton is a labelling of the input tree by states which respect $\delta_2$ in nodes with two children, but need not respect $\delta_0$ in leaves. A \emph{partial run} of a  \wmsoup automaton is a partial run of the underlying parity automaton. A partial run is called \emph{accepting} if it satisfies the parity, boundedness and unboundedness acceptance conditions. 
An accepting run of $\Aa$ is a partial accepting run where the root has the initial state and for every leaf, its (state, label) pair is in $\delta_0$.
Note that every finite partial run is an accepting partial run.

\paragraph*{Chain automata recognising accepting runs.} 
For a state $q$ of $\Aa$, consider the following sets of  real trees over the alphabet $A \times Q$, where $A$ is the input alphabet of $\Aa$ and $Q$ is its state space:
\begin{itemize}
	\item[$\rrr_q$] accepting partial runs where  states strictly bigger than  $q$ appear only in nodes with finitely many descendants;
	\item [$\rrr_{q*}$]  the subset of  $\rrr_q$ where state $q$ is allowed only finitely often on every path. 
\end{itemize}
 Note that if $q$ is a parity-rejecting state of the automaton $\Aa$, then $R_q=R_{q*}$.
By induction on $q$ in the order on states from the assumption on $\Aa$ being in normal form, we  define automaton chains $\Rr_{q}$ and $\Rr_{q*}$ such that
	\begin{align}\label{eq:chains-do-it} 
		\overline{\rrr_{q*}} = \overline{L(\Rr_{q*})}  \qquad\mbox{and}\qquad 		\overline{\rrr_{q}} = \overline{L(\Rr_q)}.
	\end{align}
The definition of $\Rr_q$ and $\Rr_{q*}$  is given below. The proof of~\eqref{eq:chains-do-it} is in the appendix.

\paragraph*{The automaton $\Rr_{q*}$.} The automaton $\Rr_{q*}$ has a unique state, call it ``state'', which  is rejecting, meaning that it must appear finitely often on every path. 
A transition of this automaton is any profinite partially $\set{\mbox{``state''}}$-colored tree $\sigma$ over $A \times Q$ such that:
\begin{enumerate}
	\item the projection of $\sigma$ onto  the $A \times Q$ coordinate belongs  to $\overline{\rrr_p}$, where $p$ is the predecessor of $q$ in the order on states; and
	\item for every root-to-leaf path in $\sigma$ which ends in a leaf with  defined color ``state'', the maximal value of the $Q$ coordinate is $q$.
\end{enumerate}
Property 1 is recognised by an automaton chain by the induction assumption. Property 2 is \mso-definable, and therefore the conjunction of properties 1 and 2 is recognised by an automaton chain thanks to Lemma~\ref{lem:automaton-chain-mso}. It follows that $\Rr_{q*}$ is a degenerate form of an automaton chain where the cost functions $\alpha$ and $\beta$ are not used.  This degenerate form is a special case of an automaton chain, by taking $\alpha$ to be the constant $0$ and $\beta$ to be the constant $\infty$.

\paragraph*{The automaton $\Rr_{q}$.} If $q$ is a parity-rejecting state of $\Aa$, then $\Rr_q$ is equal to $\Rr_{q*}$. Otherwise, it is defined as follows.
The automaton $\Rr_{q}$  has a unique state, call it ``state'', which is accepting, meaning that it can appear infinitely often on a path.   A transition  of this automaton is  
any profinite partially $\set{\mbox{``state''}}$-colored tree $\sigma$ over $A \times Q$ such that:
	 \begin{enumerate}
	 	\item the projection of $\sigma$ onto  the $A \times Q$ coordinate belongs  to $\overline{\rrr_{q*}}$; and
	 	\item for every
		 root-to-leaf path in $\sigma$ which ends in a leaf with  defined color ``state'', the maximal value of the $Q$ coordinate is $q$.
		 
		\item $\alpha(\sigma)<\infty$ holds for the   cost function defined by
		\begin{align*}
			\alpha(\sigma) =\qquad \max_{c} \max_x \quad \treeval\sigma(x,c)
		\end{align*}
		with $c$ ranging over bounded counters not in $larcut(q)$ and $x$ ranging over nodes which do not have an ancestor where $c$ is cut.
		\item $\beta(\sigma)=\infty$ holds for the cost function defined by
		\begin{align*}
			\beta(\sigma)= 
			\begin{dcases}
								\min_c \min_x \max_y \quad \treeval\sigma(y,c) & \mbox{if the root of $\sigma$ has defined color ``state''}\\
				\infty & \mbox{otherwise}
			\end{dcases}	
		\end{align*}
		with $c$ ranging over  unbounded counters in $larcheck(q)$, $x$ ranging over leaves with   defined color ``state'', and $y$ ranging over ancestors of $x$ where $c$ is checked.
	 \end{enumerate}
As  for the automaton $\Rr_{q*}$,  the conjunction of properties 1 and 2 is recognised by an automaton chain, and therefore $\Rr_q$ is an automaton chain.

\begin{mproof}[of Theorem~\ref{thm:biginf smallsup-have-decidable-emptiness}] If $q$ is the maximal state of $\Aa$, then  $R_{q}$ is the set of all partial accepting runs. Therefore, the automaton $\Aa$ is nonempty  if and only if $\Rr_{q}$ accepts some tree which is an accepting run of the underlying parity automaton in $\Aa$. This is decidable by Lemmas~\ref{lem:automaton-chain} and~\ref{lem:automaton-chain-mso}
\end{mproof}

%% file: conclusions.tex
\section{Conclusions}
\vspace{-0.2cm}
This paper shows that satisfiability is decidable for \wmsoup on infinite trees.
We conjecture  the logic remains decidable after  adding the $\mathsf R$ quantifier from~\cite{DBLP:conf/fsttcs/BojanczykT09}.  
We also conjecture that  the methods developed here, maybe the automaton mentioned in Remark~\ref{remark}, can be used to decide satisfiability of tree languages of the form ``every path is in $L$'', with  $L$ being  $\omega$B- or $\omega$S-regular languages of infinite words, as defined in~\cite{BojanczykColcombet06}.

\vspace{-0.3cm}

%% file: appendix-overview.tex
In this part of the appendix, we prove Theorem~\ref{thm:from-wmso-up-to-biginf-smallsup}, which says that \wmsoup automata recognise exactly the tree languages definable in existential \wmsoup logic.  The whole energy of the proof goes into the logic-to-automata direction.
The proof is 
spread across Section~\ref{sec:abstract-nesting} -- \ref{sec:flat}. Here is an overview of the content of these sections.

\begin{itemize}
	\item[\ref{sec:abstract-nesting}.] In Section~\ref{sec:abstract-nesting}, we define the notion of {nesting closure} of a class of languages $\Ll$. Basing on~\cite{BojanczykTorunczyk12}, we show sufficient conditions for the nesting closure to be closed under weak set quantification and unbounding  quantification.
	\item[\ref{sec:weighted-words}.] In~\cite{Bojanczyk11}, it is shown that, over infinite infinite words, \wmsou has the same expressive power as a deterministic automaton model called {max-automata}. In the proof of Theorem~\ref{thm:from-wmso-up-to-biginf-smallsup}, we use a slightly generalised version of this result, for \emph{weighted words}, i.e.~words where every position carries, apart from its label, a vector of natural numbers.  The equivalence of \wmsou and max-automata for weighted infinite words is shown in Section~\ref{sec:weighted-words}.
	\item[\ref{sec:nested}.] In Section~\ref{sec:nested}, we introduce nested counter languages, which are the nesting closure (as defined in Section~\ref{sec:abstract-nesting}) of a class of languages called atomic counter languages. We prove that nested counter languages are exactly the languages definable in \wmsoup. For the more difficult right-to-left inclusion, closure of nested counter languages under weak set quantification and unbounding quantification follows from the results presented in Section~\ref{sec:abstract-nesting}. The difficult part is closure under path quantification, which is the main content of Section~\ref{sec:nested}. When proving closure under path quantification, we use the equivalence of \wmsou and max-automata over weighted words, as proved in Section~\ref{sec:weighted-words}.
	\item[\ref{sec:flat}.] In Section~\ref{sec:flat}, we complete the proof of Theorem~\ref{thm:from-wmso-up-to-biginf-smallsup}, by showing that every nested counter language is recognised by a \wmsoup automaton. The nesting itself is not difficult, the main difficult is showing that the atomic counter languages are recognised by \wmsoup automata.  
\end{itemize}

\newpage

%% file: appendix-abstract-nesting.tex
\section{Nesting closure}
\label{sec:abstract-nesting}
In this section, we define an abstract notion of the \emph{nesting closure} of a class of languages.  The nesting closure
is essentially the same as~\cite{BojanczykTorunczyk12}, but the definition used in this paper passes through transducers.  The idea of nesting languages is not new, for instance it is implicit in weak alternating automata, and explicit in the Comp classes known from hierarchies in the $\mu$-calculus.
\paragraph*{Lookahead transducers.} A \emph{lookahead transducer}  consists of
\begin{itemize}
	\item An input alphabet $A$ and an output alphabet $B$;
	\item Tree languages $L_1,\ldots,L_k$ over the input alphabet, called the \emph{lookahead};
	\item A set of states $Q$ with a distinguished initial state;
	\item A transition function
	\begin{align*}
		\delta : Q \times A \times  \set{0,1}^k \to B \times Q \times Q.
	\end{align*}
\end{itemize}
The transducer is run on a tree over the input alphabet. It begins in the root in the initial state. Suppose that the automaton is in a node $x$ in state $q$, and that $a$ is the label of $x$ in the input tree. Let 
\begin{align*}
	\delta(q,a,a_1,\ldots,a_k) = (b,q_0,q_1) \qquad \mbox{where }a_i = \begin{cases}
		1 & \mbox{if $t|_x \in L_i$}\\
		0 & \mbox{otherwise}
	\end{cases}
\end{align*}
In the above $t|_x$ denotes the subtree of $t$ rooted in $x$.
The automaton assigns label $b$ to the node $x$, and sends states $q_0$ and $q_1$ to the left and right children of $x$, respectively. This way, the transducer induces a function
\begin{align*}
	f : \trees A \to \trees B,
\end{align*}
which is called the function recognised by the transducer. 

Note that in principle, the transducer does not even need to have explicit access to the label of the current node, because this can be simulated by having for every label $a$ of the input alphabet a lookahead language $L_a$ which contains trees with root label $a$. Nevertheless, we include the label of the current node in the transition function so that simple transformations, e.g.~the identity transformation, can be realised without using lookahead. Lookahead  transducers without any lookahead correspond to deterministic top-down transducers. 

\paragraph*{Nesting closure.}
The \emph{nesting closure} of a class of languages, called the \emph{basis}, is defined to be the least class of languages and transducers such that:
\begin{enumerate}
	\item  the nesting closure contains  every lookahead transducer where all the  lookahead languages are in the nesting closure;
	\item if  the nesting closure contains 
	\begin{align*}
		f : \trees A \to \trees B,
	\end{align*} and    $L$ is a tree language over $B$ that is  in the basis or definable in \wmso, 
then the nesting closure contains $f^{-1}(L)$.
\end{enumerate}

Note that a language or transducer in the nesting closure  has a natural finite representation, and therefore it makes sense to talk about computing such a language or transducer. In particular, the \emph{nesting depth} of a language or transducer is defined in the natural way: the nesting depth of a transducer is one plus the maximal depth of its lookahead languages, while the nesting depth of $f^{-1}(L)$ is defined to be one plus the nesting depth of $f$.

The first two levels of nesting depth are as follows.
Nesting depth one is  lookahead transducers without lookahead, i.e.~deterministic top-down transducers. Nesting depth two is inverse images of basis languages or \wmso definable languages under deterministic top-down transducers. 

\begin{lemma}\label{lem:closure}
	If the nesting closure of a class of languages  contains 
	\begin{align*}
		f : \trees A \to \trees B \qquad g : \trees B \to \trees C \qquad L,K \subseteq \trees B
	\end{align*}
	then it also contains the composition $f \circ g$, the inverse image $f^{-1}(L)$, the complement of $L$, the union $L \cup K$ and the intersection $L \cap K$.
\end{lemma}
\begin{mproof}
	A lookahead transducer can label the root of the tree by the result of any Boolean combination of its lookahead languages. This shows the case of Boolean operations.

Note that the inverse image does not immediately follow from the definition, since the definition only allows the inverse image of a language in the basis, and not  in the nesting closure. The proof of closure under composition and inverse image is by parallel induction on the nesting depth of $g$ and $L$.
\end{mproof}

The lemma above implies  that nesting closure is indeed a closure operator, i.e.~applying the nesting closure a second time does not add any new languages or transducers.

We will now show sufficient conditions for the nesting closure to be closed under weak set quantification and under unbounding quantification. 
To discuss closure  under  quantification of a class of languages,  free variables need to be encoded  into the trees. 
We write $A \times 2$ instead of $A \times \set{0,1}$. To encode a set of nodes $X$ in a tree $t$ over alphabet $A$, we write $t \otimes X$ for the tree over alphabet $A \times 2$  obtained from $t$ by extending the label of every node by a bit that indicates if the node belongs to $X$.  Therefore, the quantifiers studied in this paper can be seen as 
 a language operations, defined by
	\begin{eqnarray*}
		\exists_{\mathrm{fin}} X L  &\eqdef& \set{t : t \otimes X \in L \mbox{ for some finite set of nodes $X$}}\\
		\mathsf U X L  &\eqdef& \set{t :t \otimes X \in L \mbox{  for  arbitrarily large finite sets $X$} }.\\	
			\exists_{\mathrm{path}} \pi L  &\eqdef& \set{t : t \otimes \pi \in L \mbox{  for some path $\pi$}}\\
	\end{eqnarray*}
In the following sections, we give sufficient conditions for the nesting closure to be closed under weak set quantification and unbounded quantification. This part of the paper uses the techniques from~\cite{BojanczykTorunczyk12}, and in the case of unbounded quantification we simply use a result from~\cite{BojanczykTorunczyk12}.

\subsection{Closure under weak quantification.}

A \emph{tree congruence} over alphabet $A$ is an equivalence relation on trees over $A$ such that the equivalence class of a tree is uniquely determined by its root label and the equivalence classes of its left and right subtrees. An equivalence relation on trees is said to \emph{saturate} a language if the language is a union of equivalence classes of the equivalence relation. 
 A class of languages is called \emph{derivative closed} if for every language $L$ there exists a tree congruence which saturates $L$, and has finitely many equivalence classes, all of which are languages in $\Ll$.  We also require  the construction to be effective, i.e.~based on a representation of $L$ one can compute representations of the equivalence classes.

\begin{lemma}\label{lem:derivative-closed}
	If a class $\Ll$ is derivative closed, then so is its nesting closure.
\end{lemma}
\begin{mproof}
Induction on the nesting depth. A language in the nesting closure is of the form $f^{-1}(L_0)$ where  $L_0 \in \Ll$ and 
	\begin{align*}
		f : \trees A \to \trees B
	\end{align*}
	is a lookahead transducer with lookahead languages $L_1,\ldots,L_k$. By induction assumption, for every $i \in \set{0,\ldots,k}$ there is a tree congruence $\sim_i$ which saturates $L_i$ and has finitely many equivalence classes in the nesting closure of $\Ll$.   Define an equivalence relation $\sim$ on trees over $A$, which considers trees equivalent if they are $\sim_i$-equivalent for every $i \in \set{1,\ldots,k}$, and their images under $f$ are $\sim_0$-equivalent. It is not difficult to see that this is a tree congruence and all of its equivalence classes are in the nesting closure of $\Ll$.
\end{mproof}

For a tree congruence $\sim$ over trees over alphabet $A$, define its characteristic transducer to be the function
\begin{align*}
	f : \trees A \to \trees {(\trees A)/_{\sim}}
\end{align*}
which labels every node in a tree $t$ by the $\sim$-equivalence class of its subtree. 
% 
% \begin{lemma}\label{lem:characteristic-transducer}
% 	If a class $\Ll$ is derivative closed, then for every $L$ in the nesting closure of $\Ll$, there exists a tree congruence $\sim$ which saturates $L$ and whose characteristic transducer is in the nesting closure of $\Ll$.
% \end{lemma}
% \begin{mproof}
% 	Follows immediately from Lemma~\ref{lem:derivative-closed}.
% \end{mproof}
The proof of the following lemma is similar to the transformation from \wmso to weak alternating automata, as shown by Muller, Saoudi and Schupp in reference~\cite{DBLP:conf/icalp/MullerSS86} from the bibliography of the appendix. A similar construction, with an abstract framework, is in~\cite{DBLP:conf/fsttcs/BojanczykT09}. 

\begin{lemma}\label{lem:abstract-closed-under-weak-quantification}
	If a class $\Ll$ is derivative closed, then languages in its nesting closure is closed under weak quantification.
\end{lemma}
\begin{mproof}
	Let $L$ be a language over alphabet $A \times \set {0,1}$ that is  in the nesting closure of $\Ll$. 	By Lemma~\ref{lem:derivative-closed},  there is a tree congruence $\sim$ which saturates  $L$  and whose characteristic transducer, call it $f$, is also in the nesting closure of $\Ll$.  Without loss of generality, we assume that $f$ also outputs the original label of each node.
It is not difficult to define a formula $\varphi$ of \wmso such that 
\begin{align*}
	f(t \otimes \emptyset) \models \varphi \qquad \mbox{iff} \qquad t \otimes X \in L \mbox{ holds for some finite $X$}.
\end{align*}
The formula $\varphi$ guesses the set $X$, and then evaluates the tree congruence $\sim$, using the values supplied by $f(t \otimes \emptyset)$ for nodes that have no elements of $X$  in their subtree.
\end{mproof}

\subsection{Closure under unbounding quantification}
Consider the set   of trees over alphabet $\set{\epsilon,\mathrm{inc},\mathrm{reset}}$, such that for every $n$, there is some finite path with at least $n$ occurrences of the letter inc, but no occurrence of the letter reset.  This language is called the \emph{basic counter  language}. Note that this language is prefix-independent, and saturated by a tree congruence with two equivalence classes: the language and its complement.

  It is not difficult to see that the nesting closure of the class that contains only the basic counter language is equal to the nested limsup automata defined in~\cite{BojanczykTorunczyk12}. Therefore, by Proposition 7 from~\cite{BojanczykTorunczyk12}, we have the following result.
\begin{lemma}\label{lem:applied-12}
	If $\Ll$ contains the basic counter language, then for every \wmso-definable language $L$,  the nesting closure of $\Ll$ contains 
	\begin{align*}
		\set{ t : \mbox{$t \otimes X \in L$ for arbitrarily large sets $X$}}.
			\end{align*}
\end{lemma}

 \begin{lemma}\label{lem:abstract-closed-under-unbounding-quantification}
 		If a class $\Ll$ is derivative closed and contains the  basic counter language,  then its nesting closure is closed under unbounding quantification.
 \end{lemma}
\begin{mproof}
We need to show that if a language over the alphabet $A \times \set{0,1}$ is in the nesting closure of $\Ll$, then so is:
\begin{align}\label{eq:chain-unbounding-closure}
	\set{ t  :  t \otimes X \in L \mbox{ for arbitrarily large finite sets $X$}}
\end{align}
By Lemma~\ref{lem:derivative-closed}, there is a tree congruence  $\sim$ that saturates $L$. Let $f$ be the characteristic transducer of $\sim$, which is in the nesting closure of $\Ll$. It follows that the function $t \mapsto t \otimes f(t \otimes \emptyset)$ is also in the nesting closure of $\Ll$.
By using the definition of a tree congruence, it is not difficult to show that for every $\sim$-equivalence class $\tau$, there is a \wmso formula such that 
\begin{align*}
	t \otimes f(t \otimes \emptyset)  \models \varphi(X) \qquad \mbox{iff} \qquad t \otimes X \in L.
\end{align*}
By Lemma~\ref{lem:applied-12}, and closure under inverse images of transducers, we get the desired result.
\end{mproof}

% 
% \paragraph*{Relationship with automata for \wmsou.} The nesting  closure of basic counter languages is essentially the nested model  used in the paper~\cite{BojanczykTorunczyk12} for the logic \wmsou, with the following  difference. From Lemmas~\ref{lem:abstract-closed-under-weak-quantification} and~\ref{lem:abstract-closed-under-unbounding-quantification} it follows that the nesting closure of basic counter languages contains the logic \wmsou; the converse inclusion is easy to show. Therefore, the nesting closure of basic counter languages is equal to \wmsou; which is the 

%% file: appendix-weighted.tex
\section{Weighted words}
\label{sec:weighted-words}
In this part of the appendix, we  introduce weighted words. We prove that the correspondence of \wmsou and max-automata, which is proved for words without weights in~\cite{Bojanczyk11}, extends to weighted words.

\paragraph*{Weighted words.} Define a \emph{weighted alphabet} to be a set $\Sigma$  partitioned into sets
\begin{align*}
	\Sigma = \weightlab \Sigma \cup \weightwei \Sigma
\end{align*} 
called the \emph{label symbols} and \emph{weight symbols}.   We use $\Sigma$ instead of $A$ to distinguish between weighted alphabets and normal alphabets.
A \emph{weighted word or tree} is one where the alphabet is the (infinite) set
\begin{align*}
	\weightlab \Sigma \times \bar \Nat^{\weightwei \Sigma}.
\end{align*}
The sets of weighted words and trees over  $\Sigma$ are denoted by 
\begin{align*}
	\weightedwords \Sigma \qquad\mbox{and}\qquad \weightedtrees \Sigma.
\end{align*}
For the words, we only use words of infinite length, for the trees we use possibly infinite trees where every node has zero or two children.

For  a weight symbol $b$,  define the $b$-weight of a position in a weighted word to be the number on coordinate $b$ in the vector labelling that position. Define the $b$-weight of a set of positions to be the sum of all $b$-weights of positions in the set. For instance, if every position has $b$-weight $1$, then the $b$-weight of a set is its size. If some position has $b$-weight $\infty$, then the $b$-weight of every set containing it will also be $\infty$.
 
 \paragraph*{Weighted \wmsou.} To express properties of  weighted  words,
we use \emph{weighted \wmsou}. The syntax is the same as for \wmsou on infinite words over the alphabet containing the label symbols,  except that for weight symbol   $b$ of the weight alphabet there are:
\begin{itemize}
	\item a predicate $b(x)$ which is true in positions with nonzero $b$-weight;
	\item a quantifier $ 	\mathsf U_b X \varphi(X)$
 % \begin{align}\label{eq:weighted-unbounded-quantifier}
 % 	\mathsf U_b X \varphi(X)
 % \end{align}
 which is true if the $b$-weights of sets satisfying $X$ are unbounded (which may be achieved by a single set with $b$-weight $\infty$).
\end{itemize} 
\paragraph*{Weighted max-automata.} For a position in a weighted word
\begin{align*} 
	w \in \weightedwords \Sigma
\end{align*}
 define its profile to be the following information: the label of the position, the set of weight symbols for which the position has nonzero weight, and the set of weight symbols for which the position has infinite weight.
A weighted max-automaton over consists of:
\begin{itemize}
	\item An input weighted alphabet $\Sigma$.
		\item A set $C$ of counters.
	\item A deterministic finite automaton whose input alphabet is profiles over $\Sigma$
	\item For every transition of the deterministic finite automaton,  a sequence of counter operations from the set
	\begin{eqnarray*}
				c := c + 1 \qquad
		c := c + b \qquad
						c := 0 \qquad
												c := \max(d,e)
	\end{eqnarray*}
	where $c,d,e$ range over counters in $C$ and $b$  ranges over weight symbols in~$\Sigma$.
	\item An acceptance condition, which is a family $\Uu$ of sets of counters.
\end{itemize}
The automaton is executed as follows on a weighted word $w$. It begins in the initial state with a counter valuation 
\begin{align*}
	v_0 : C \to \Nat
\end{align*}
that maps every counter to $0$. Suppose that after reading the first $n-1$ letters of $w$, the  state of the automaton is $q_{n-1}$ and its counter valuation  is 
\begin{align*}
	v_{n-1} : C \to \Nat.
\end{align*}
The automaton performs the transition corresponding to the profile of the $n$-th position in the input word, and it performs the counter operations for that transition, with $c := c +b$ meaning that $c$ is incremented by the $b$-weight of position $n$. The automaton accepts if the set $\Uu$ contains
\begin{align*}
	\set{ c \in C : \limsup v_n(c) = \infty},
\end{align*}
which is the set of counter that are unbounded throughout the run. The set $\Uu$ is similar to Muller acceptance condition.

Without  weights, weighted \wmsou is the same as \wmsou, and   weighted max-automata are the same as the max-automata defined in~\cite{Bojanczyk11}. Theorem 5 in~\cite{Bojanczyk11} of says that, without weights, max-automata and \wmsou have the same expressive power, and translations both ways are effective. Below we generalise this result to weighted words.

\begin{theorem}\label{thm:weighted-max-logic}
	For every formula of weighted \wmsou there is an equivalent weighted max-automaton, and vice versa.
\end{theorem}
The rest of Appendix~\ref{sec:weighted-words} is devoted to proving the above theorem. The proof is by a reduction to the case without weights.
	We only prove the more difficult logic-to-automata direction. The automata-to-logic part is shown the same way as Lemma 6 in~\cite{Bojanczyk11}, and  is not used in this paper. 
	
	Fix a weighted alphabet $\Sigma$. Let $b_1,\ldots,b_k$ be the weight symbols of $\Sigma$. Define the block encoding of a word 
	\begin{align*}
		w \in \weightedwords \Sigma
	\end{align*}
	to be the word $[w]$ obtained from $w$ by 
	by replacing position each position of the word $w$ with the word
	\begin{align*}
		a v_1 \cdots v_k    \qquad \mbox{with $v_i=$} \begin{cases}
			b_i^{n_i} & \mbox{when $n_i< \infty$}\\
			\infty & \mbox{otherwise}
		\end{cases}
	\end{align*}
	where $a$ is the label of the position, and $n_i$ is the $b_i$-weight of the position. The alphabet of $[w]$, which is a normal (unweighted word) is all symbols in $\Sigma$ (both label and weighted) plus the letter $\infty$.
By doing a syntactic transformation, it is not difficult to compute for every formula $\varphi$ of weighted \wmsou a formula $[\varphi]$ of \wmsou such that 
\begin{align*}
	w \models \varphi \qquad \mbox{iff} \qquad [w] \models [\varphi].
\end{align*}
By equivalence of \wmsou and max-automata in the case without weights,  for the formula $[\varphi]$ one can compute   a max-automaton $\Aa_\varphi$ such that for every weighted word $w$,
\begin{align*}
	[w] \models [\varphi] \qquad \mbox{iff} \qquad \Aa_\varphi \mbox{ accepts }[w].
\end{align*}

If $w$ is a weighted word, and $n$ is a nonzero natural number, then define $n \cdot w$ by multiplying all weights by $n$, in particular zero weights remain zero weights. Since weighted \wmsou is invariant under such multiplication, we have
\begin{align*}
	w \models \varphi \qquad \mbox{iff} \qquad  n \cdot w \models \varphi \qquad \mbox{ for every $n > 0$}.
\end{align*}

Let $n$ be the number of states in the automaton $\Aa_\varphi$.  By combining the above observations, we see that 
\begin{align*}
w \models \varphi \qquad \mbox{iff} \qquad \Aa_\varphi \mbox{ accepts }[n! \cdot w].
\end{align*}
The proof is concluded by the following lemma.
\begin{lemma}\label{lem:factorial}
There is a weighted max-automaton $\Bb$ such that 
	\begin{align*}
		\Bb \mbox{ accepts } w \quad \mbox{iff} \quad \Aa_\varphi \mbox{ accepts }[n! \cdot w] \qquad \mbox{for every } w \in \weightedwords \Sigma. 
	\end{align*}
\end{lemma}
\begin{mproof}

	Consider what happens when the automaton $\Aa_\varphi$ is processing the block encoding of the $i$-th letter of $n! \cdot w$. Let the encoding of this letter be 
	\begin{align}\label{eq:block-encoding-of-a-single-letter}
		a v_1 \cdots v_k    \qquad \mbox{with $v_i=$} \begin{cases}
			b_i^{n_i} & \mbox{when $n_i< \infty$}\\
			\infty & \mbox{otherwise}
		\end{cases}
	\end{align}
	Each number $n_i$ is divisible by $n!$.
	The key observation is that when a deterministic automaton has $n$ states, then for every word $w$,  the state transformation  induced by $w^{n! \cdot m}$ does not depend on $m$, as long as $m$ is nonzero.	
	By this  observation, for every fixed $\Sigma$-profile $\tau$ and  every state $q$ of the automaton $\Aa_\varphi$ there are sequences of counter operations 
	\begin{align*}
		u_1,\ldots,u_k \qquad \mbox{and} \qquad w_1,\ldots,w_k
	\end{align*}
on the counters of the automaton $\Aa_\varphi$	such that when the automaton is in state $q$ and  reads a word as in~\eqref{eq:block-encoding-of-a-single-letter} corresponding to a position with profile $\tau$, then sequence of counter operations that it produces is  exactly
	\begin{align*}
		u_1 w_1^{m_1} \cdots u_k w_k^{m_k} \qquad \mbox{with $m_i=$}\begin{cases}
			n_i & \mbox{if $n_i<\infty$}\\
			0 & \mbox{otherwise}
		\end{cases}
	\end{align*}
	This behaviour can be simulated by a weighted max-automaton.
\end{mproof}

%% file: appendix-nested-counter.tex
\section{Nested counter languages}
\label{sec:nested}
In this part of the appendix, we introduce nested counter languages and transducers, which are obtained by applying the nesting closure, as defined in Section~\ref{sec:abstract-nesting}, to a particular class of languages called {atomic counter languages}. The main result is Theorem~\ref{thm:equal-to-nested}, which says that nested counter languages are exactly the same as languages definable in \wmsoup.

\paragraph*{Atomic counter languages.} Let $C$ be a set of counters and $\Uu$ a family of subsets of $C$.  A counter $c \in C$ is called \emph{tail unbounded} on a path $\pi$ in a counter tree
\begin{align*}
	t \in \countertrees C
\end{align*}
if  the value of counter $c$ is unbounded on path $\pi$ in  every counter tree that agrees with $t$ on the labels of all but at most finitely many nodes. For a family $\Uu$ of subsets of $C$, we say that a path $\pi$ is $\Uu$-accepting in $t$ if 
 $\Uu$  contains the set of counters which are tail unbounded on the path.  An \emph{atomic counter language}  is a language of the form
\begin{align*}
	\set{ t \in \countertrees{C} : \mbox{every infinite path is $\Uu$-accepting}}
\end{align*}
for some set of counters $C$ and family of subsets $\Uu$. Note also that the basic counter language from Lemma~\ref{lem:abstract-closed-under-unbounding-quantification} is a special case of  an atomic counter language (up to a different encoding of the input). The notion of tail unbounded is defined so that every atomic counter language is prefix independent in the following sense: if $L$ is an atomic counter language, then a tree $t$ belongs to $L$ if and only if every subtree of $t$ belongs to $L$.

In this section, we prove  that the nesting closure of atomic counter languages is equivalent to the logic \wmsoup.
We use the name \emph{nested counter language} for a language in the nesting closure of atomic counter languages. Likewise we define \emph{nested counter transducers}.

\begin{theorem}\label{thm:equal-to-nested}
		Nested counter languages are exactly the languages definable in \wmsoup and translations both ways are effective.
\end{theorem}

The rest of Appendix~\ref{sec:nested} is devoted to proving the theorem. We begin with the easier inclusion, from nested counter languages to \wmsoup.

\paragraph*{Nested counter languages are \wmsoup  definable.}  The automata-to-logic direction is by a  straightforward induction on the nesting depth. Call a tree transducer
\begin{align*}
	f : \trees A \to \trees B
\end{align*}
 \wmsoup definable if for every letter $b \in B$ of the output alphabet, there is a formula $\varphi_b(x)$ over the input alphabet $A$ with one free node variable, such that for every tree $t$ over the input alphabet $A$, a node of $t$ has label $b$ in $f(t)$ if and only if formula $\varphi_b(x)$ is true in $t$. 

By induction on the nesting depth, we prove that every nested counter language and transducer is definable in \wmsoup. The induction step is straightforward, because it is not difficult to see  that languages definable in \wmsoup are preserved under  inverse images of  \wmsoup definable transducers, and every deterministic transducer with lookahead languages in \wmsoup is definable in \wmsoup.  The only interesting case is the induction base, i.e.~that atomic counter languages are definable in \wmsoup.

To show this, it suffices to show that for every set of counters $C$ and counter $c$, there is a formula of \wmsoup $\varphi_c(\pi)$, with a free path variable $\pi$,  which says that  counter $c$ is tail unbounded on path $\pi$ in a counter tree $t$ over counter $C$. Using Boolean combinations of this formula, one can say that a path is $\Uu$-accepting, and using path quantification, one can say that a tree is $\Uu$-accepting.  To define the formula $\varphi_c(\pi)$, we observe that  for every
	 counters $c,d $ one can write a formula $\varphi_{cd}(x,y)$ of \wmso, which selects a pair of nodes $(x,y)$ if and only if there is a counter path from $(x,c)$ to $(x,y)$ which does at least one increment. Using this formula, one can write a formula $\psi_c(x,X)$ of \wmso, which selects a set of nodes if and only if there is a counter path that begins in $x$, and for every node $y \in X$ the path does an increment when leaving $X$. Therefore, a counter $c$ is unbounded on a path $\pi$ if 
 	 \begin{align*}
\mathsf U X \ \exists x \ x \in \pi  \ \land \ \psi_c(x,X).
 	 \end{align*} 
	 To say that the automaton is tail unbounded, one needs to  additionally say that the above formula remains true after the counter tree is modified in any finite way, which can easily be done using quantification over finite sets.

\paragraph*{\wmsoup definable languages are nested counter languages.} As usual, the more difficult implication in Theorem~\ref{thm:equal-to-nested} is going from the logic to nested counter languages and transducers.
We need to show that nested counter languages are closed under weak set quantification, unbounding quantification, and path quantification.

By being prefix closed, for  every atomic counter languages $L$ the equivalence relation with two equivalence classes, namely $L$ and its complement, is a tree congruence. Therefore, atomic counter languages are closed under derivatives, and so Lemma~\ref{lem:abstract-closed-under-weak-quantification}, can be used to conclude that nested counter languages are closed under weak set quantification.
As we have remarked above, the basic counter language from  Lemma~\ref{lem:abstract-closed-under-unbounding-quantification} is a special case of an atomic counter language, and so Lemma~\ref{lem:abstract-closed-under-unbounding-quantification} can be used to conclude that nested counter languages are closed under unbounding quantification. The rest of this section is devoted to path quantification.

The rest of Section~\ref{sec:nested} is devoted to showing  that nested counter languages are closed under path  quantification.  In the presence of first-order quantification, it suffices to quantify over \emph{full paths}, i.e.~paths that begin in the root. This is because a non-full path $\pi$  can be described by the pair $(\sigma,x)$ where $x$ is the first node of $\pi$ and $\sigma \supseteq \pi$ is the full path obtained from $\pi$ by adding all ancestors of $x$.
 The proof strategy is as follows. Let $L$ be a nested counter language over alphabet $A \times 2$. Our goal is to show that that
\begin{align}\label{eq:path-quantification}
	\set{ t \in \trees A : t \otimes \pi \in L \mbox{ for every full path $\pi$}}
\end{align}
is also a nested counter language.  The main result is Lemma~\ref{lem:path-separation-language}, which says that the language~\eqref{eq:path-quantification} is equivalent to first running a transducer $f$ on the tree $t$, and then checking that every path in $f(t)$ satisfies a formula of weighted \wmsou, with the weights corresponding to counter values encoded in the tree $f(t)$. In Lemma~\ref{lem:weighted-on-all-paths}, we show that running a weighted \wmsou formula on all paths can be implemented by a nested counter language, thus proving closure under universal path quantification. The precise definitions of how weighted trees are computed by transducers are given below, followed by Lemmas~\ref{lem:weighted-on-all-paths} and Lemma~\ref{lem:path-separation-language}.

\paragraph*{Encoding a weighted tree.} Consider a weighted alphabet $\Sigma$. To encode a weighted tree over this alphabet, we use a tree  $s \otimes t$, where $s$ is a tree over the label alphabet of $\Sigma$ and $t$ is a counter tree over the weight alphabet of $\Sigma$. The tree $s \otimes t$ is said to be the \emph{counter tree encoding} of a weighted tree over $\Sigma$ if for every node of the weighted tree, its  label  is given by $s$ and its $b$-weight is given by the value of counter $b$.  A \emph{counter tree encoding} of a function
\begin{align*}
	f : \trees A \to \weightedtrees \Sigma
\end{align*}
is a function which, given an input $t$, produces a weighted tree encoding of the weighted tree $f(t)$. If a function $f$ admits a counter tree encoding (which is not necessarily unique) that is a nested counter  transducer, then we say that $f$ is \emph{recognised by a counter  transducer}.

Define the \emph{path word} of a path $\pi$ in a tree $t$, denoted by $t \restrict \pi$, to be the infinite word consisting of the  labels occurring on $\pi$. We will be mainly interested in the path word for full paths, i.e.~infinite paths that begin in the root.

 \begin{lemma}\label{lem:weighted-on-all-paths}
 	Let $\varphi$ be a weighted \wmsou formula over a weighted alphabet $\Sigma$, and let 
	\begin{align*}
		f : \trees A \to \weightedtrees \Sigma
	\end{align*}
	be recognised by a nested counter automaton.  The set of trees $t$ such that
	\begin{align*}
f(t) \restrict \pi \models \varphi \qquad \mbox{for every full path $\pi$}
	\end{align*}
	is a nested counter language over the alphabet $A$.
 \end{lemma}
 \begin{mproof}
	 Let $B$ be the label of alphabet of $\Sigma$, and let $C$ be the counter alpahbet. The assumption of the lemma is that there is a nested counter transducer 
	 \begin{align*}
	 	g : \trees A \to \trees B \otimes \countertrees C
	 \end{align*}
	 such that for every input tree $t$, the output $g(t)$ is a counter encoding of $f(t)$.
	  By Theorem~\ref{thm:weighted-max-logic}, the formula $\varphi$ is equivalent to a weighted max-automaton, call it~$\Aa$. Let $D$ be the counters of the automaton.  By   describing the  transitions of a weighted max-automaton  in terms of a counter tree, one can a nested counter transducer 
 \begin{align*}
 	h : \trees B \otimes \countertrees C  \to \countertrees {C \cup D}
 \end{align*}
 such that for every input $s \otimes t$ and every node $x$ of this tree and every counter $c \in C$, the value of counter $d \in D$ in node $x$ of $h(s \otimes t)$ is equal to the value of counter $d$ in the run of the automaton $\Aa$ after reading the path word that corresponds to the path in the weighted tree encoded by  $s \otimes t$ which goes from the root to node $x$. We leave the definition of the transducer $h$ to the reader, we only notice that it in order to correctly simulate the automaton $\Aa$, the transducer $h$ needs lookahead automata which compute for each node $x$ and counter $c \in C$, whether the value of counter $c$ in node $x$ is zero, nonzero but finite, or infinite. 
 
The statement of the lemma is obtained by composing $g$ with $h$, and then testing an atomic counter language on the image of this composition, which tests that the  acceptance condition of the weighted max-automaton is satisfied.
 \end{mproof}

  \begin{lemma}\label{lem:path-separation-language}
  	For every nested counter language
 	\begin{align*}
 		L \subseteq \trees {A \times 2}
 	\end{align*}
 	there exists a weighted alphabet $\Sigma$, a function
 	\begin{align*}
 		\help L : \trees A \to weightedtrees(\Sigma)
 	\end{align*}
	recognised by a nested counter transducer, 
 	and a formula $\varphi$ of weighted \wmsou over $\Sigma$ such that for every tree $t$ over $A$ and every full path $\pi$,
 \begin{align*}
 	t \otimes \pi \in L \qquad \mbox{iff} \qquad \help L(t) \restrict \pi \models \varphi.
 \end{align*}
  \end{lemma}
  
  As remarked previously, Lemmas~\ref{lem:weighted-on-all-paths}  and~\ref{lem:path-separation-language} imply that nested counter languages are closed under universal path quantification, therefore completing the proof of Theorem~\ref{thm:equal-to-nested}. 
  
  The rest of Section~\ref{sec:nested} is devoted to proving 
  Lemma~\ref{lem:path-separation-language}. The proof is by induction on the nesting depth of the language. However, when doing the induction, we pass through nested counter transducers, so we need a variant  of Lemma~\ref{lem:path-separation-language} for transducer, which is stated in Lemma~\ref{lem:path-separation-transducer}. Before stating the lemma, we  define transducers from weighted words to normal infinite words.
  
  In the same spirit  as when defining tree transducers definable in \wmsou, we  use weighted \wmsou transducers which map an (infinite) weighted word to an infinite word over a finite alphabet. 
  We say that a function
  \begin{align*}
  	f :  \weightedwords \Sigma \to A^\omega 
  \end{align*}
  is definable in weighted \wmsou, if for every letter $a$ of the output alphabet $A$, there is a formula $\varphi_a(x)$ of weighted \wmsou over the input alphabet $\Sigma$, such that for every input word $w$, the label of position $x$ in $f(w)$ is $b$ if and only if $\varphi_b(x)$ is satisfied in $w$.

  \begin{lemma}\label{lem:path-separation-transducer}
 	For every nested counter transducer 
 	\begin{align*}
 		f : \trees{A \times 2} \to \trees B
 	\end{align*}
 	there exists a nested counter transducer 
 			 recognising a function
 		 	\begin{align*}
 		 		\help f : \trees A \to weightedtrees(\Sigma)
 		 	\end{align*}
 			and a  weighted \wmsou transducer
 			\begin{align*}
 				\varphi : weightedwords(\Sigma) \to B^\omega
 			\end{align*}
 			 such that for every tree $t$ over  $A$ and  every full path $\pi$,
 			\begin{align*}
 				f(t \otimes \pi) \restrict \pi =  \varphi(\help f(t) \restrict \pi)%\alpha ( t \otimes g_1(t) \otimes \cdots \otimes g_k(t) \restrict \pi)
 			\end{align*}
 	
  \end{lemma}
 
  Lemmas~\ref{lem:path-separation-language} and~\ref{lem:path-separation-transducer}  are proved by mutually recursive induction on the nesting depth.
  The induction step for transducers, corresponding to Lemma~\ref{lem:path-separation-transducer}, is simple and given below. The induction steps for languages, corresponding to Lemma~\ref{lem:path-separation-language}, is more involved. In Section~\ref{sec:side-values} we present some results on counter trees, and the actual proof of Lemma~\ref{lem:path-separation-language} is given in Section~\ref{sec:separation-induction-step-languages}.

%% file: appendix-path-separation.tex
\begin{mproof}[of Lemma~\ref{lem:path-separation-transducer}]
	Consider a nested counter transducer 
	\begin{align*}
		f : \trees{A \times 2} \to \trees B.
	\end{align*}
	  	 Let $L_1,\ldots,L_k$ be the lookahead languages of $f$. 	 
Let $t$ and $\pi$ be as in the assumption of the lemma.
	Define $x_i$ to be the $i$-th node on $\pi$, and define 
	\begin{align*}
		a_i \in A \times 2^k
	\end{align*}
	to be the label of $t$ in $x_i$ and the bit vector indicating which lookahead languages contain the subtree of $t \otimes \pi$ rooted in $a_i$. Apply the induction assumption of Lemma~\ref{lem:path-separation-language} to the lookahead languages,  yielding  for every $i \in \set{1,\ldots,k}$ a transducer $\help {L_i}$ and a  formula $\varphi_i$. By using these transducers and formlas, one can compute  a weighted \wmsou transducer which inputs the word
	\begin{align*}
		t \otimes \help {L_1} (t) \otimes \cdots \otimes \help {L_k} (t) \restrict \pi
	\end{align*}
	and outputs $a_1 a_2 \cdots$. 
	  By simulating the control structure of the transducer $f$, one can see that there is a \wmso transducer 
	\begin{align*}
		g  : (A \times 2^k)^\omega \to B^\omega
	\end{align*}
	which inputs $a_1 a_2 \cdots$ and outputs $f(t \otimes \pi) \restrict \pi$. Composing these two transducers we get the statement of the lemma.
\end{mproof}
	 
	 \subsection{Side values}
	 \label{sec:side-values}
	 This section is devoted to proving Lemmas~\ref{lem:recognise-sidevalues} and~\ref{lem:counter-unbounded-separation}, which are used in the proof  of 	 Lemma~\ref{lem:path-separation-language}. Roughly speaking, Lemma~\ref{lem:counter-unbounded-separation}  says that   in order to determine whether a counter $c$ is unbounded on a path  in  a counter tree, one only needs to look at the labels of the counter tree  on the path, as well as an appropriate summary of what is happening outside the path. Furthermore, as shown in Lemma~\ref{lem:recognise-sidevalues}, these summaries can be produced by a nested counter transducer. 
	 
	 We begin by defining the summaries, which say for each node of a counter tree, what happens in the subtree of its sibling.
	  For a counter tree $t$ over counter $C$,  a node $x$ of $t$ and  counters $c,d \in C$ define
	  	\begin{align}\label{eq:binary-side-value}
	  		\sidevalue(t,c,d,x) \in \bar \Nat
	  	\end{align}
	  	to be the least upper bound on the  number of increments in a counter path in $t$ such that:
	  	\begin{enumerate}
	  		\item the counter path begins in counter $c$ in the parent of  $x$;
	  		\item the counter path ends in counter $d$ in the parent of $x$;
	  		\item the rest of the counter path is in the subtree of the sibling  of $x$.
	  	\end{enumerate}
		The counter path mentioned above looks like this:
		\begin{center}
			\includegraphics[scale=0.3]{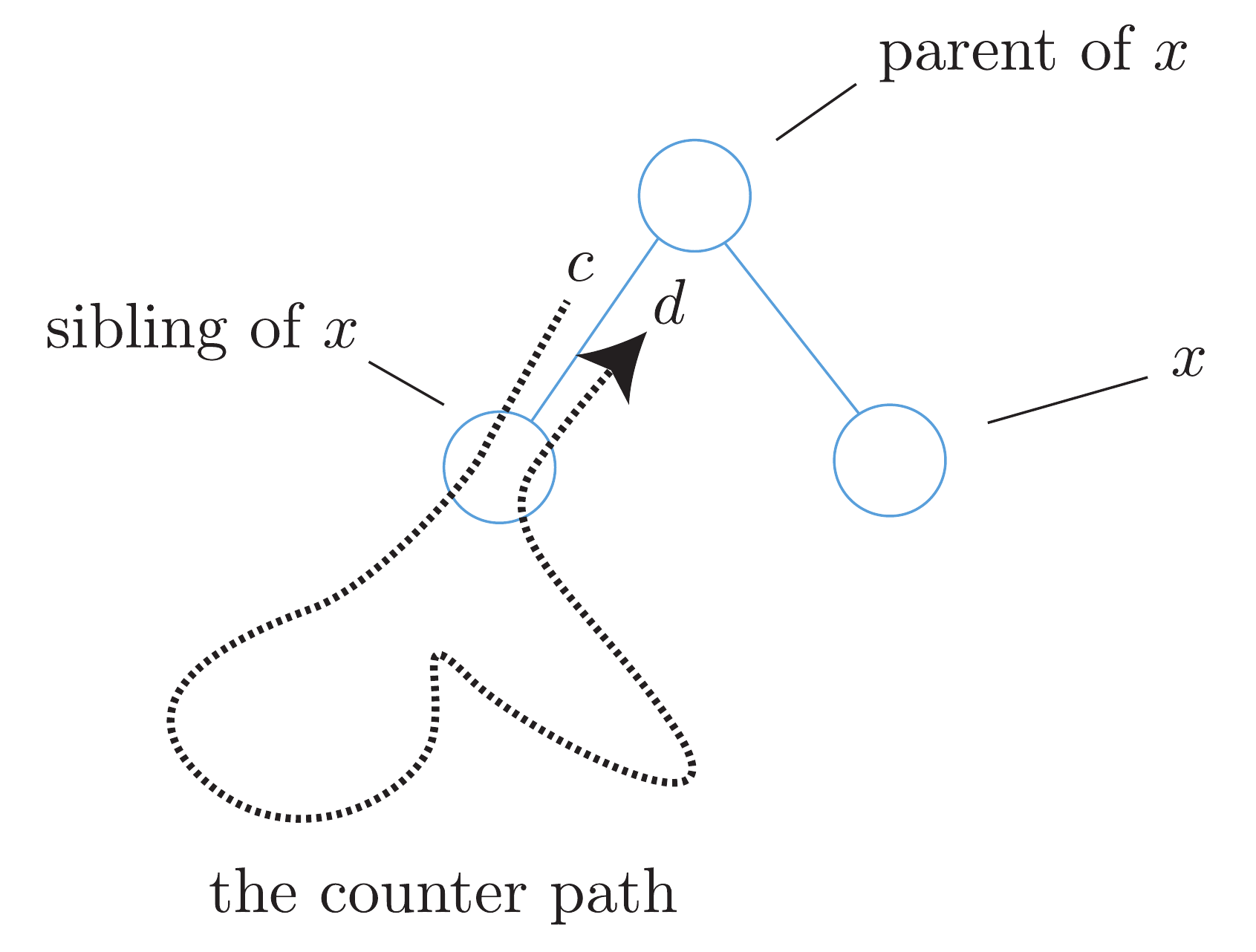}
		\end{center}
		If the parameter $c$ is not supplied, then 
		\begin{align}\label{eq:unary-side-value}
			\sidevalue(t,d,x) \in \bar \Nat
		\end{align}
		is defined as above, except that instead of condition 1, we say that  the counter path begins in the subtree of the sibling of $x$, in some counter. Such a counter path looks like this
		\begin{center}
			\includegraphics[scale=0.3]{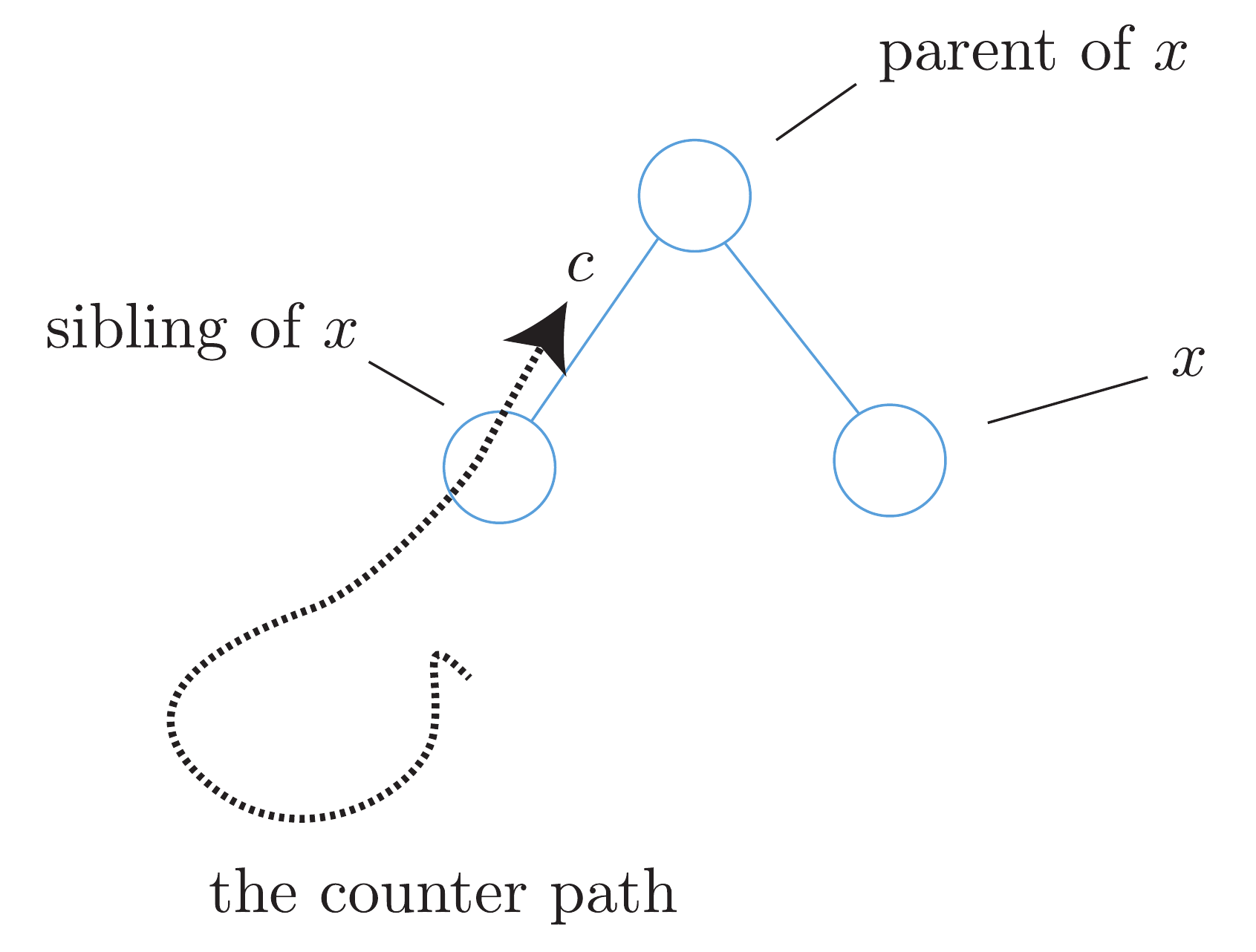}
		\end{center}
		
		\begin{lemma}\label{lem:side-value-is-side-dependent}
			The values~\eqref{eq:binary-side-value} and~\eqref{eq:unary-side-value} depend only on the subtree of $t$ rooted in the sibling of $x$. In other words, there is a function
\begin{align*}
	\svalue : \countertrees C \times (C \cup C^2) \to \bar \Nat
\end{align*}
such that  every counter tree $t$, node $x$ and counter $c$ and $d$ satisfy
\begin{align*}
	\sidevalue(t,c,d,x) = \svalue(t|_x,c,d) \qquad\mbox{and}\qquad
		\sidevalue(t,d,x) = \svalue(t|_x,d).
\end{align*}
		\end{lemma}
		\begin{mproof}
			Follows straight from the definition, with the only subtle point being  that the values do not depend on the label of the parent of $x$. This is because in a counter tree, the edges between a node and its child are encoded in the label of the child.
		\end{mproof}

		 Consider a nested counter transducer 
		  \begin{align*}
		  f : \trees A \to  \countertrees C.
		  \end{align*}
		  Let $Q$ be the states of $f$.
		  For a state $q \in Q$, define $f_q$ to be the transducer obtained from $f$ by changing the inititial state to $q$.  Let $\svalue_f(t)$ to be the weighted tree over  weight alphabet 
		 $ 	Q \times C^2 \cup Q \times C$ 
		 defined by
		  \begin{eqnarray*}
		 \svalue_f(t)(x,q,c,d) &= &\svalue(f_q(t|_x),c,d)\\
		 		 \svalue_f(t)(x,q,d) &=& \svalue(f_q(t|_x),d)
		  \end{eqnarray*}
			\begin{lemma}\label{lem:recognise-sidevalues}
	 		For every  nested counter transducer
	 	\begin{align*}
	 	 f : 	\trees A \to \countertrees C.
	 	\end{align*}
	 	the function $\svalue_f$ is recognised by a nested counter transducer, with the same nesting depth as $f$.
	 	\end{lemma}
		\begin{mproof}
			The same method as in converting a two-way automaton on trees to a one-way automaton, e.g.~when proving that tree-walking automata can be simulated by bottom-up branching automata.
		\end{mproof}

	 \begin{lemma}\label{lem:counter-unbounded-separation}
	 		For every  nested counter transducer
	 	\begin{align*}
	 	 f : 	\trees A \to \countertrees C.
	 	\end{align*}
	 	and counter $c \in C$ there exists a 		
	 	 formula $\varphi$ of weighted \wmsou   such that for every tree $t$ over $A$ and path $\pi$, counter $c$ is unbounded on path $\pi$ if and only if
	 	\begin{align*}
	  \svalue_f(t) \otimes f(t) \restrict \pi \models \varphi.
	 	\end{align*}
	 \end{lemma}
	 \begin{mproof}
		 The word  $\svalue_f(t) \restrict \pi$ gives all information about the behaviour of counter paths outside the path $\pi$.
	 \end{mproof}

	 \subsection{Proof of Lemma~\ref{lem:path-separation-language}}
	 \label{sec:separation-induction-step-languages}
We are now ready to prove Lemma~\ref{lem:path-separation-language}, which finishes the proof of Theorem~\ref{thm:equal-to-nested}. Consider a nested counter language
	 \begin{align*}
	 	L \subseteq \trees{A \times 2}
	 \end{align*} 
	  as in the assumption of Lemma~\ref{lem:path-separation-language}. We need to show the conclusion of Lemma~\ref{lem:path-separation-language}, which says  that there exists a weighted alphabet $\Sigma$, a function
 	\begin{align*}
 		\help L : \trees A \to \weightedtrees \Sigma
 	\end{align*}
	recogognised by a nested counter transducer, 
 	and a formula $\varphi$ of weighted \wmsou over $\Sigma$ such that for every tree $t$ over $A$ and every full path $\pi$,
 \begin{align*}
 	t \otimes \pi \in L \qquad \mbox{iff} \qquad \help L(t) \restrict \pi \models \varphi.
 \end{align*}
	 By the definition of a nested counter language, 
	  \begin{align*}
L = f^{-1}(K) \qquad \mbox{for some }	 f : \trees {A \times 2} 	 \to \trees B,
	  \end{align*}
	  such that $f$ is a nested counter transducer and $K$ is either definable in \wmso or an atomic counter language. Let $Q$ be the states of the transducer $f$. Define 
\begin{align*}
	\rho : \trees{ A \times 2} \to \trees Q
\end{align*}
to be the transducer which relabels each node by the state of the transducer $f$ that is used in that node.  Note that if $t$ is a tree over $A$ and $\pi$ is a path, then the label of a node $x$ in $f(t \otimes \pi)$ is uniquely determined by its labels in $t \otimes \pi$ and $\rho(t \otimes \pi)$.
The  transducer $\rho$ has the same nesting depth as $f$, and therefore by induction assumption we can apply Lemma~\ref{lem:path-separation-transducer} to it, yielding a transducer $\help \rho$.
	
We treat separately the cases when $K$ is definable in \wmso, and when $K$ is an atomic counter language.
	 
	 \subsubsection{$K$ is definable in \wmso.}  Let $k$ be the quantifier depth of the \wmso formula defining $K$. Define the $k$-type of a tree to be the set of \wmso formulas of quantifier depth that are true in the tree. It is well known that there are finitely many $k$-types, and each one is definable in \wmso. Also, having the same $k$-type is a tree congruence.
	 
	 For an infinite  path $\pi$ define the  \emph{$\pi$-child} of   a node $x \in \pi$ to be the unique child of $x$ that is  on the path $\pi$. Likewise we define the \emph{non-$\pi$-child}, which is the sibling of the $\pi$-child. Define the \emph{$k$-profile} of an infinite path $\pi$ in a tree $t$ to be the infinite word, where the $i$-th letter contains the following information about the $i$-th node of the path: the label of the node, the $k$-type of the subtree of $t$ rooted in its non-$\pi$-child, and a bit saying if  the $\pi$-child of the node is a left or right child. Using the composition method, one shows that  for every $k$-type $\tau$, there is a formula $\varphi_\tau$ of \wmso over infinite words such that $t \otimes \pi$ has $k$-type $\tau$ if and only if $\varphi_\tau$ is true in the infinite word that is the $k$-profile of $\pi$ in $t$.
	 
	 Recall that we write $f_q$ for the transducer obtained from $f$ by changing the initial state to $q$.
	 Define $\gamma$ be the function which inputs a tree $t$ over $A$ and outputs a tree $\gamma$ with the same nodes such that the label of every node $x$ is  the function 
	 \begin{align*}
	 (q,i) \in Q \times 2  \qquad \mapsto \qquad	\mbox{$k$-type of the subtree of $f_q(t\otimes \emptyset)$ rooted in the $i$-th child of $x$}.
	 \end{align*}
It  is not difficult to show that $\gamma$ is a nested counter transducer, from the assumption that $f$ is a nested counter transducer.  Define
	 \begin{align*}
	 	\help L(t) \eqdef  t \otimes \mathrm{child}(t) \otimes  \gamma(t) \otimes \help \rho(t)
	 \end{align*}
	 where  $\mathrm{child}(t)$ labels each node by the information saying whether or not the node is the root, a left child, or a right child. We claim that $\help L$ satisfies the properties in the statement of Lemma~\ref{lem:path-separation-language}. It is not difficult to see that there is weighted \wmsou transducer which inputs the word $\help L (t) \restrict \pi$ and outputs the $k$-profile of the path $\pi$ in the tree $f(t \otimes \pi)$.  Based on this $k$-profile, membership in $K$ can be determined using only weak quantification.
	 \subsubsection{$K$ is an atomic counter language.}
Consider now the case where $L$ is the preimage $f^{-1}(K)$ where  $K$ is an atomic counter language, i.e.~there is a set of counter $C$ and a family $\Uu$ of subsets of $C$ such that $K$ is the set of counter trees over $C$ where every path is $\Uu$-accepting.  The condition $f(t \otimes \pi) \in K$ can be decomposed as a conjunction of two conditions:
	 \begin{enumerate}
		 			 	\item  path $\pi$ is $\Uu$-accepting in $f(t \otimes \pi)$;
		\item every path $\sigma \neq \pi$ is $\Uu$-accepting in $f(t \otimes \pi))$.
	 \end{enumerate} 	
	 For each  condition $i \in \set{1,2}$ above,  we define a transducer $\help {L,i}$ and a formula $\varphi_i$ of weighted \wmsou such that 
	 $t$ and $\pi$ satisfy condition $i$ if and only if 
\begin{align*}
\help{L,i}(t) \restrict \pi \models \varphi_i.
\end{align*}	 
The lemma will then follow by taking $\help L$ to be the product of the  transducers $\help {L,i}$ and taking $\varphi$ to be the conjunction of the formulas $\varphi_i$.

\paragraph*{Condition 1.}
	  Define 
	 \begin{align*}
	 	\help {L,1}(t) \eqdef \help f(t) \otimes \svalue_f(t \otimes \emptyset)
	 \end{align*}
We need to show that there is formula of weighted \wmsou, which is true in 
\begin{align}\label{eq:side-word-on-input}
		 	\help {L,1}(t) \restrict \pi
\end{align}
	 if and only if path $\pi$ in the counter tree $f(t \otimes \pi)$ is $\Uu$-accepting.	 
For every counter $c \in C$, 
	  apply Lemma~\ref{lem:counter-unbounded-separation}  to $f$,
	 yielding a formula $\varphi_c$ such that for  every tree $t$ over $A$ and path $\pi$ in $t$, counter $c$ is unbounded on path $\pi$  in $f(t \otimes \pi)$ if and only if formula $\varphi_c$ is true in the word
	 	 	\begin{align}\label{eq:side-word-on-output}
	 	  \svalue_f(t \otimes \pi) \otimes f(t \otimes \pi) \restrict \pi
	 	 	\end{align}
	Therefore, it suffices to show that there is a weighted \wmsou transducer  $\alpha$ such that for every tree $t$ and path $\pi$, when given input~\eqref{eq:side-word-on-input}, it produces output~\eqref{eq:side-word-on-output}. Once such a transducer $\alpha$ has been defined, condition 1 is implemented by checking that $\Uu$ contains the set of counter $c$ such that $\varphi_c$ is true in the image of the word~\eqref{eq:side-word-on-input} under the transducer $\alpha$.
		
By induction assumption, the word $f(t \otimes \pi) \restrict \pi$  can be produced by a weighted \wmsou transducer based on the word $\help f(t) \restrict \pi$. To compute the word $\svalue_f(t \otimes \pi) \restrict \pi$, we observe that for every node $x$ in a path $\pi$, the subtree of the sibling of $x$ is the same in $t \otimes \pi$ and in $t \otimes \emptyset$. Therefore, from  Lemma~\ref{lem:side-value-is-side-dependent}, it follows that 
	 \begin{align}\label{eq:side-depends-on-side}
	 	\svalue_f(t \otimes \pi) \restrict \pi \qquad = \qquad \svalue_f(t \otimes \emptyset) \restrict \pi.
	 \end{align}	
	 	By Lemma~\ref{lem:recognise-sidevalues}, the function $t \mapsto \svalue_f(t \otimes \emptyset)$ is recognised by a nested counter transducer.
		
	\paragraph*{Condition 2.}	 
	Recall that condition 2 says that  every path $\sigma \neq \pi$ is $\Uu$-accepting  in $f(t \otimes \pi)$.  Because being $\Uu$-accepting is prefix-independent, this is equivalent to saying that  for every node $x \not \in \pi$, the language $K$ contains the subtree of $f(t \otimes \pi)$ rooted in $x$. Furthermore, since $K$ is closed under subtrees, it follows that only nodes $x$ with parent in $\pi$ need be checked. Summing up, condition 2 is equivalent to
	\begin{align}\label{eq:pre-restated-condition-two}
		f(t \otimes \pi)|_x \in K \qquad \mbox{for every $x$ whose sibling is in $\pi$}
	\end{align}
Recall that  $f_q$ is obtained from $f$ by changing the  initial state  to $q$.  Define
	\begin{align*}
		L_q \eqdef \set {t \in \trees A : f_q(t \otimes \emptyset) \in K},
	\end{align*}
	which is a nested counter language.
If  $q_x$ denotes be the state of the transducer $f$ that is used in node $x$, then~\eqref{eq:pre-restated-condition-two} is equivalent to 
		\begin{align}\label{eq:restated-condition-two}
			t|_{x} \in L_{q_x} \qquad \mbox{for every $x$ whose sibling is in $\pi$}
		\end{align}
	Let $\gamma$ be the transducer which labels every node $x$ of a tree $t$ by the set of states $q$ such that the subtree rooted in the sibling of $x$ belongs to $L_q$. 	(The root, which has no siblings, gets labelled by the empty set.) This is a nested counter transducer, since every language $L_q$ is a nested counter language.
 Define 
\begin{align*}
	\help {L,2}(t) \eqdef t \otimes \mathrm{childnumber}(t) \otimes \gamma(t) \otimes \help \rho(t).
\end{align*}
We claim that the word $\help{L,2}(t) \restrict \pi$ provides sufficient information to check condition 2. By~\eqref{eq:restated-condition-two}, condition 2 holds if and only if for every nonroot node $x \in \pi$, the label of $x$ in $\gamma(t)$ contains that state of the transducer $f$ in the sibling of $x$. This state can be deduced based on the child number of $x$,  the label of $x$ and the state of $f$ in the parent of $x$, all of these are given by  $\help {L,2}(t)$. This completes the proof of Lemma~\ref{lem:path-separation-language}, and therefore the proof of Theorem~\ref{thm:equal-to-nested}.

%% file: appendix-flat.tex
\section{\wmsoup automata}
\label{sec:flat}
In this part of the appendix, we prove Theorem~\ref{thm:from-wmso-up-to-biginf-smallsup}, which says that 	for every formula of existential \wmsoup one can compute a  \wmsoup automaton that accepts the same trees, and vice versa. The automata-to-logic is straightforward, shown the same way as the automata-to-logic direction in Theorem~\ref{thm:equal-to-nested}. For the logic-to-automata part, we only need to translate nested counter languages into \wmsoup automata, since Theorem~\ref{thm:equal-to-nested} says that the logic \wmsoup defines exactly  nested counter languages.   

We prove a slightly stronger result, because we will show that it suffices to use \wmsoup automata which satisfy condition (a) of the definition of normal form. Recall that condition (a) says that for every run, in the counter graph generated by the automaton, every bounded counter is separated and root-directed.

\begin{theorem}\label{thm:flat-capture-nested}
	Every nested counter language is recognised by a  \wmsoup automaton, which satisfies condition (a) of the definition of normal form.
\end{theorem}

For the rest of this section, all produced \wmsoup automata satisfy condition (a), so we do not mention it explicitly. In other words, from now on by ``\wmsoup automaton'' we mean ``\wmsoup automaton which satisfies condition (a)''.
In Section~\ref{sec:atomic-ones-recognised-by-wmsoup-automata}, we show that every atomic counter language is recognised by a \wmsoup automaton. In Section~\ref{sec:nested-ones-recognised-by-wmsoup-automata}, we generalise this result to nested counter languages.

\subsection{\wmsoup automata contain languages definable in \wmsou}
\label{sec:contain-wmsou}
In this  section we show  that \wmsoup automata capture all languages definable in \wmsou, i.e.~definable without the path quantifier.  To show this, we prove that \wmsoup automata generalise the automaton model from~\cite{BojanczykTorunczyk12}, which is  at least as expressive as \wmsou.

We begin by  proving  some simple closure properties of languages recognised by \wmsoup automata.  A \emph{relabelling} is an arbitrary function from an input alphabet  to an output alphabet. Such a function is lifted to trees in the natural way.

\begin{lemma}\label{lem:puzzles-closed-under-intersection}
	Languages recognised by  \wmsoup automata are closed under union,  intersection, inverse images under relabellings, images under relabellings and contain all \mso-definable languages.
\end{lemma}
\begin{mproof}
	Inverse images are immediate: an automaton can ignore part of its input.
	Union and images are by nondeterminism. For intersection,  the cartesian product works. The underlying parity automaton can be used to recognise any \mso-definable language, without using any counters. \end{mproof}
	
\paragraph*{Puzzles.}
	Define a \emph{cut-increment-reset  tree} to be a tree where every node is labelled by a subset of ``cut'', ``increment'', ``reset'', and ``$\infty$''. An increment node is one whose label contains ``increment'', likewise we define cut and reset nodes.
Define the value of a path in such a tree to be the maximal number  of increment nodes  in a subpath that does not contain reset nodes. Define the value of a node to be the supremum of values of paths that begin in the node and do not contain cut nodes. Define the  \emph{puzzle language} to be the set of cut-increment-reset  trees where a node has $\infty$ in its label if and only if its value is~$\infty$.

Note that unlike counter paths  in counter trees, which can go both ways, the paths in the definition of the puzzle language are normal paths in trees, i.e.~totally ordered connected sets of nodes. In this sense, the puzzle language uses one-way counter paths.

\begin{lemma}\label{lem:puzzle-language-recognised}
	The puzzle language is recognised by a  \wmsoup automaton. 
\end{lemma}	
\begin{mproof}
	The puzzle language is the intersection of two languages: the \emph{bounded puzzle language}, which says that every node without $\infty$ in its label has value strictly smaller than $\infty$; and the \emph{unbouned puzzle language}, which says that every node with $\infty$ in its label has value $\infty$. By closure of \wmsoup automata under intersection, it suffices to show that each of these languages is recognised by a \wmsoup automaton. 
	
	The bounded puzzle language is immediate, by using the cuts in the definition of \wmsoup automata. This lemma is actually the only place in the proof where we produce a new bounded counter, in all other places bounded counters are only inherited from already produced automata. Note that in the bounded puzzle language, there is only one bounded counter, so it is necessarily separated, and it is root-directed. Therefore, the \wmsoup automaton for the bounded puzzle language satisfies condition (a) in the definition of normal form.
	
	The rest of the proof is devoted to  the unbounded counter language.
	
	A cut-increment-reset tree $t$ can be encoded as a counter tree $\underline t$ over counters $c,d$ as follows. The nodes of $t$ and $\underline t$ are the same. If the label of a node $x$ in $t$ does not contain ``cut'' then  the counter tree  $\underline t$ contains a transfer edge from $d$ in $x$ to $d$ in its parent, and a transfer edge from $c$ in $x$ to $d$ in its parent.  If the label of $x$ 
	contains neither ``reset'' nor ``cut'', then $\underline t$ contains an  edge from counter $c$ in $x$ to counter $c$ in its parent; this edge is an increment edge if $x$ is an increment node, otherwise the edge is a transfer edge. It is easy to see that the value of a node $x$ in $t$ is equal to the value of counter $d$ in the same node of $\underline t$. 
	
	The \wmsoup automaton for the unbounded puzzle language works as follows. Given an input cut-increment-reset tree, it computes the tree $\underline t$. The counters of the automaton are $c$ and $d$, as in $\underline t$, and both are unbounded counters. The automaton also guesses a subset of positions where  counter $d$ is checked (in an accepting run, counter $d$ is unbounded on any infinite chain of positions where it is checked). The parity automaton underlying the \wmsoup automaton verifies that for every node that has $\infty$ in its label, there is some infinite path that begins in the position, contains infinitely many nodes where $d$ is checked, and does pass through a node whose label includes ``cut''.
\end{mproof}

\begin{lemma}\label{lem:contain-wmsou}
	Every language definable in \wmsou is recognised by a \wmsoup automaton.
\end{lemma}
\begin{mproof}
	In~\cite{BojanczykTorunczyk12} an automaton model called a \emph{puzzle} is introduced,  and is proved to be at least as expressive as \wmsou. Every  language recognised by a puzzle is of the form
	\begin{align*}
	\set{t : \mbox{$t \otimes t_1 \otimes \cdots \otimes t_n \in L$ holds for some  $t_1,\ldots,t_n$ in the puzzle langauge}} 
	\end{align*}
where $L$ is \mso-definable. By Lemma~\ref{lem:puzzle-language-recognised}, and the closure  properties from Lemma~\ref{lem:puzzles-closed-under-intersection},  every language recognised by a puzzle is also recognised by a  \wmsoup automaton.
\end{mproof}
\subsection{Atomic counter languages}
\label{sec:atomic-ones-recognised-by-wmsoup-automata}
This section is devoted to proving the following lemma.
		\begin{lemma}\label{lem:atomic-ones-recognised-by-wmsoup-automata}
			Atomic counter languages are recognised by \wmsoup automata.
		\end{lemma}

\label{page:restricted-value}
If $X$ is a set of nodes in  a counter tree $t$  over counters $C$, then define  the $X$-restricted value of counter $c$ in node $x$, denoted by
\begin{align*}
	\treeval t _X(x,c),
\end{align*} 
to be the supremum of values of counter paths that end in $(c,x)$, and which do not visit any ancestors of $x$ in the set $X$.
% Counter $c$  is tail unbounded on a path $\pi$ if and only if 
% \begin{align*}
% 	\limsup_{x \in \pi} \treeval t _X(x,c)  = \infty
% \end{align*}
% holds for some (equivalently, every) set $X$ through which $\pi$ passes infinitely often.
% If $X$ is a set of nodes in a tree, define an $X$-zone to be a maximal connected set of nodes $Y$ such that $Y$ is disjoint with $X$, with the possible exception of the root of $Y$.
A set of nodes $X$ in  a counter tree $t$ over counters $C$  is called a  witness for counter $c \in C$  if 
	\begin{enumerate}
		\item for every connected set $Y$ disjoint with $X$,
		\begin{align*}
			\limsup_{x \in Y}  \treeval t_{X}(x,c)  < \infty
		\end{align*}
		\item for every path $\pi$ that passes infinitely often through $X$,
		\begin{align*}
			\limsup_{x \in \pi} \treeval t _{X}(x,c) = \infty
		\end{align*}
		%  there is a finite bound on the $X$-restricted value of counter $c$ in zone $Y$.
		% \item for every 
		% \item $s$ is obtained from $t$ by deleting finitely many counter edges in every $X$-zone;
		% \item every counter path in $s$ passes through at most one node from $X$;
		% \item \label{it:record-bounded} Counter $c$ is  bounded in every $X$-zone of $s$;
		% \item \label{it:record-unbounded} Counter $c$ is   unbounded on every infinite chain $Y \subseteq X$.
	\end{enumerate}
	In Lemma~\ref{lem:c-witness-witnesses} we show that if $X$ is a witness for $c$, then counter $c$ is tail unbounded exactly on those paths that visit $X$ infinitely often. 
 Lemma~\ref{lem:c-witness-exists} shows that a witness  always exists, and Lemma~\ref{lem:c-witness-computable} shows that being a witness can be recognised by a \wmsoup automaton. These lemmas imply Lemma~\ref{lem:atomic-ones-recognised-by-wmsoup-automata}.

\begin{lemma}\label{lem:c-witness-witnesses}
	If $X$ is a witness for $c$ in a  counter tree, then counter $c$ is tail unbounded exactly on those paths that visit $X$ infinitely often.
	\end{lemma}
\begin{mproof}
	For a node $x$, define $t_x$ to be the counter tree obtained from $t$ by removing all counter edges that involve ancestors of $x$. Clearly $t$ and $t_x$ differ on finitely many nodes, and for every descendant $y$ of $x$, the value of counter $c$ in $y$ is lowest in $t_x$, as compared to other counter trees that agree with $t$ on descendants of $x$.  It follows that counter $c$ is tail unbounded on path $\pi$ in counter tree $t$ if and only if 
	\begin{align}\label{eq:a-characterisation-of-tail-unbounded}
		\limsup_{y \in \pi} \treeval{t_x}(y,c) = \infty \qquad \mbox{for all $x \in \pi$}
	\end{align}

	Suppose that $\pi$ passes through finitely many nodes of $X$. Let $x$ be the last of node $X$ that is visited by $\pi$, if no such node is visited then let $x$ be the root.  The definition of $X$ being a witness for $c$ implies that 
	\begin{align*}
		\limsup_{y \in \pi} \treeval{t_x}(y,c) < \infty
	\end{align*}
	and therefore counter $c$ is tail bounded on path $\pi$ in $t$. 
	For the converse implication, if $\pi$  passes infinitely often through $X$, then~\eqref{eq:a-characterisation-of-tail-unbounded} holds and therefore counter $c$  is tail unbounded on patph $\pi$ in $t$.
\end{mproof}

\begin{lemma}\label{lem:c-witness-exists}
	In every counter tree, every counter admits a  witness.
\end{lemma}
\begin{mproof}
	Construct a family of sets $X_0,X_1,\ldots$ by induction as follows. The set $X_0$ contains only the root. Suppose that $X_0,\ldots,X_{i-1}$ have already been defined, and let $X_{<i}$ be their union. Define $X_i$ to be the set of nodes where the $X_{<i}$-restricted value of counter $c$ is at least $i$, and which are minimal (closest to the root) for this property. Define $X$ to be the union of all sets $X_i$. 
	
	Let $x$ be a node, and let $i$ be the largest number such that $x$ has an ancestor in $X_i$.  It is not difficult to see that the  $X$-restricted value of counter $c$ in $x$ is at most $i$ when $x \not \in X$ and at least $i$ when $x \in X$. This implies that $X$ is a witness for counter $c$.
\end{mproof}

\begin{lemma}\label{lem:c-witness-computable}
	There exists a \wmsoup automaton recognising the language
	\begin{align*}
		\set{t  \otimes X : \mbox{$t \in \countertrees C$ and $X$ witness for $c$ in  $t$}}
	\end{align*}
\end{lemma}
\begin{mproof}
	The first condition in the definition of a witness for $c$ is definable in \wmsou, and therefore by Lemma~\ref{lem:contain-wmsou} it is recognised by a \wmsoup automaton. We focus on the second condition, which says that 
	\begin{align}\label{eq:second-condition-of-witness-definition}
		\limsup_{x \in \pi} \treeval t_{X}(x,c) = \infty \qquad \mbox{for every path $\pi$ that visits $X$ infinitely often.}
	\end{align}

	For counter trees $s,t$ we write $s \subseteq t$ if the counter trees have the same nodes, and every counter edge in $s$ is a counter edge in $t$. 	We claim that~\eqref{eq:second-condition-of-witness-definition} is  equivalent to saying that some  $s \subseteq t$ satisfies:
		\begin{enumerate}
			\item For every path $\pi$ that passes infinitely often through $X$, 
			\begin{align*}
				\limsup_{x \in \pi \cap X} \treeval s(x,c) =\infty
			\end{align*}
			\item If $x < y$ are in $X$, then no counter path in $s$ passes through both $x$ and $y$.
		\end{enumerate}
	Condition 1 is easily seen to be recognised by a \wmsoup automaton based on $s$; while condition 2 is \mso definable and therefore also recognised by a \wmsoup automaton. Therefore, a \wmsoup automaton can check if there is some $s \subseteq t$ which satisfies 1 and 2.
	
	It remains to show that~\eqref{eq:second-condition-of-witness-definition}   is equivalent to the existence of $s \subseteq t$ satisfying conditions 1 and 2 above. The right-to-left implication is immediate. For the left-to-right implication, let $t$ be a counter tree satisfying~\eqref{eq:second-condition-of-witness-definition}. Using induction, it is not difficult to  define counter trees
	\begin{align*}
		s_0 \subseteq s_1 \subseteq \cdots \qquad \subseteq t
	\end{align*}
	which satisfy condition 2, have finitely many counter edges, and such that for every $n$, if a path $\pi$ passes infinitely often through $X$, then 
	\begin{align*}
		\sup_{x \in X} \treeval{s_n}(x,c)\ge n.
	\end{align*}
	The tree $s$ is then taken to be the limit, i.e.~union, of the trees $s_n$. 
\end{mproof}

 \begin{mproof}[of Lemma~\ref{lem:atomic-ones-recognised-by-wmsoup-automata}]
 Let $L$ be an atomic counter language, i.e.~there exists a set of counters $C$ and a family of subsets $\Uu$ such that $L$ contains a counter tree over $C$ if and only if for every infinite path, $\Uu$ contains the  counters which are tail unbounded on the path. Let the counters be $c_1,\ldots,c_n$.  Let $K$ be the set of trees 
 \begin{align*}
 	t \otimes X_1 \otimes \cdots \otimes X_n \qquad t \in \countertrees C
 \end{align*}
 such that a) for every for every $i \in \set{1,\ldots,n}$, the set of nodes $X_i$ is a witness for $c_i$; and b) for every infinite path $\pi$, the set $\Uu$ contains the set of counters $c_i$ such that the path passes infinitely often through $X_i$. By Lemma~\ref{lem:c-witness-computable}, condition a) is recognised by a \wmsoup automaton. Since \wmsoup automata generalise parity automata, condition b) is also recognised by \wmsoup automata. Therefore, $K$ is recognised by a \wmsoup automaton. By Lemma~\ref{lem:c-witness-exists}, for every counter tree $t$ there exist sets $X_1,\ldots,X_n$ such that condition a) holds.  By Lemma~\ref{lem:c-witness-witnesses}, condition b) holds if and only if $t$ belongs to $L$. Therefore, $L$ is the image of $K$ under the relabeling which ignores the sets $X_1,\ldots,X_n$; and the lemma follows because the sets can be guessed using nondeterminism.
 \end{mproof}

		\subsection{Nested counter languages}
		\label{sec:nested-ones-recognised-by-wmsoup-automata}

For a language $L$, define its \emph{characteristic language} to  be:
\begin{align*}
								\set{t \otimes X : t|_x \not \in L \mbox{ if and only if $x \in X$}},
	\end{align*}

In the following lemma, we use a transducer without lookahead, which is the same thing as a deterministic top-down tree transducer.
\begin{lemma}\label{lem:half-characteristic}
	For every  transducer $f$ without lookahead, and every atomic counter language $L$,  the characteristic language of $f^{-1}(L)$ is recognised by  a \wmsoup automaton.
\end{lemma}

Before proving the above lemma, we show how it completes the proof of Theorem~\ref{thm:flat-capture-nested}. 

\begin{mproof}[of Theorem~\ref{thm:flat-capture-nested}]
	To complete the proof, we need to show that every nested counter language is recognised by a \wmsoup automaton. 
		We  prove a stronger result: for ever nested counter language, its characteristic language is recognised by a \wmsoup automaton. This is proved by induction on the nesting depth.
		
		Consider a nested counter language, which is of the  form $f^{-1}(L)$, where $L$ is an atomic counter language or \wmso-definable, and $f$ is in a nested counter transducer.
	 Let  the  lookahead languages of the transducer $f$ be $L_1,\ldots,L_k$.  By definition of a lookahead transducer, there is a deterministic transducer $g$ without lookahead such that for every tree $t$, 
	 \begin{align*}
	 	f(t) = g(t \otimes X_1 \otimes \cdots \otimes X_k) 
	 \end{align*}
	 where $X_i$ is the set of nodes in $t$ whose subtree belongs to $L_i$, or equivalently, $X_i$ is the unique set such that $\set{t \otimes X_i}$ is in the characteristic language of  $L_i$. The automaton recognising the characteristic language of $f^{-1}(L)$ works as follows. Given on input a tree  $t \otimes X$, it 
 guesses  sets $X_1,\ldots,X_k$ and checks that a) the trees
 \begin{align*}
 	t \otimes X_1,\ldots,t \otimes X_k
 \end{align*}
belong respectively to the  characteristic languages of $L_1,\ldots,L_k$, which can be done by the induction assumption, and b) that   the tree 
 \begin{align*}
 	t \otimes X_1 \otimes \cdots \otimes X_k \otimes X
 \end{align*} belongs to the characteristic language of $g^{-1}(L)$. When $L$ is an atomic counter language, then condition b) can be done by a \wmsoup automaton thanks to Lemma~\ref{lem:half-characteristic}. When $L$ is definable in \wmso, then condition b) is also definable in \wmso, and therefore it can be done by a \wmsoup automaton.
\end{mproof}

We are therefore left with proving Lemma~\ref{lem:half-characteristic}.  Define  the positive and negative characteristic languages of $L$ to be, respectively:
	\begin{eqnarray*}
		\set{t \otimes X : t|_x \in L \mbox{ for every $x \in X$}}\label{eq:half-chararacteristic-positive} \qquad				\set{t \otimes X : t|_x \not \in L \mbox{ for every $x \not\in X$}}\label{eq:half-chararacteristic-negative}.
		\end{eqnarray*}
		Unlike for the characteristic language, where $X$ is uniquely determined by $t$, for a given $t$ there are many sets $X$ such that $t \otimes X$ is in the positive characteristic language, although there is a unique greatest set $X$ with $t \otimes X$ in the positive characteristic language. The dual  property holds for the negative characteristic language.
The characteristic language is the intersection of the positive and negative characteristic languages. Therefore, to prove Lemma~\ref{lem:half-characteristic}, it suffices to show that both the positive and negative characteristic languages of $f^{-1}(L)$ are recognised by \wmsoup automata. We begin with the positive one.

		\begin{lemma}\label{lem:positive-all-paths}
			Let $L$ and $f$ be as in the assumption of Lemma~\ref{lem:half-characteristic}. 
			There exists a  function
			\begin{align*}
				g : \trees {A} \to \weightedtrees \Sigma
			\end{align*}
			recognised by a transducer without lookahead and  a  formula of $\psi$ of weighted \wmsou over weighted alphabet $\Sigma$ such that a tree $t \otimes X$ belongs to the positive characteristic language of $f^{-1}(L)$ if and only if 
			\begin{align*}
				g(t \otimes X)  \restrict \pi \models \psi \qquad \mbox{for every full path $\pi$}.
			\end{align*}
		\end{lemma}
		\begin{mproof}Recall the function $\svalue_f$ defined in Section~\ref{sec:side-values}.
	By Lemma~\ref{lem:recognise-sidevalues}, this function is recognised by a transducer without lookahead.  				By   Lemma~\ref{lem:counter-unbounded-separation}, for every counter $c \in C$ there is a formula $\varphi_c$ of weighted \wmsou such that for every full path $\pi$, 
			\begin{align*}
				f(t) \otimes \svalue_f(t) \restrict \pi \models  \varphi_c
			\end{align*}
			holds if and only if counter $c$ is tail unbounded on  $\pi$ in $f(t)$. Let $\Uu$ be the family of accepting sets in the definition of the language $L$. By using a Boolean combination, one gets a formula $\varphi$ of weighted \wmsou such that  for every full path $\pi$, 
			\begin{align*}
				f(t) \otimes \svalue_f(t) \restrict \pi \models  \varphi
			\end{align*}
			is true if and only if  $\pi$ is $\Uu$-accepting in $f(t)$.  Since $f(t)$ has no lookahead, the  labels  of the output $f(t)$ on the path $\pi$ can be computed, using weak quantification, based only on the labels on the input $t$ on the same path. Therefore, there is a formula   $\varphi'$ of weighted \wmsou such that for every full path~$\pi$, 
			\begin{align*}
				t \otimes \svalue_f(t) \restrict \pi \models  \varphi'
			\end{align*}
			is true if and only if  $\pi$ is $\Uu$-accepting in $f(t)$. Finally, let $\psi$  be obtained from $\varphi'$ by extending the  input alphabet with an additional bit, and requiring $\varphi'$ to occur in every suffix of the input word which begins in a position marked by the additional bit. Because 
			\begin{align*}
				\svalue_f(t)|_x = \svalue_f(t|_x) \qquad \mbox{for every node $x$},
			\end{align*}
			the formula $\psi$ satisfies the condition in the statement of the lemma, with $g$ defined by
			\begin{align*}
				t \otimes X \qquad \mapsto \qquad t \otimes \svalue_f(t) \otimes X
			\end{align*}
		\end{mproof}
		
		\begin{lemma}\label{lem:positive-characteristic}
			Let $L$ and $f$ be as in the assumption of Lemma~\ref{lem:half-characteristic}.  The positive characteristic language of $f^{-1}(L)$ is recognised by a \wmsoup automaton.
		\end{lemma}
		\begin{mproof}
			Apply Lemma~\ref{lem:positive-all-paths}, yielding a transducer $g$ and a formula $\psi$ such that the $t \otimes X$ belongs to the positive characteristic language of $f^{-1}(L)$ if and only if
			\begin{align}\label{eq:positive-all-paths}
				g(t \otimes X)  \restrict \pi \models \psi \qquad \mbox{for every full path $\pi$}.
			\end{align}
			By Lemma~\ref{lem:weighted-on-all-paths}, there exists a nested counter transducer
							\begin{align*}
								h : \trees {A \times 2} \to \countertrees D
							\end{align*}
					and an atomic counter language $K$ over counters $D$ such that~\eqref{eq:positive-all-paths} is equivalent to
							\begin{align*}
								h(t \otimes X)  \in K.
							\end{align*}  By the proof of Lemma~\ref{lem:weighted-on-all-paths}, the transducer $h$ has no lookahead because $g$ has no lookahead. The language $K$, as an atomic counter language, is recognised by a \wmsoup automaton thanks to Lemma~\ref{lem:atomic-ones-recognised-by-wmsoup-automata}. 
							The result follows by closure of   \wmsoup automata under inverse images of deterministic transducers without lookahead.\end{mproof}
		
		It remains to show that the negative characteristic language
		\begin{align*}
			\set {t \otimes X : f(t|_x) \not \in L \mbox{ for every $x \not \in X$}}
		\end{align*}
		 is recognised by a  \wmsoup automaton. Using nondeterminism, we reduce this language to a positive characteristic language (for some different $f$ and $L$).
	Let $C$ be the counters and let $\Uu$ be the family of counters in the atomic counter language $L$. If $f(t|_x)$ is not in $L$, this means that there must be some path, call it $\pi_x$, which begins in $x$ and is $\Uu$-rejecting in the tree $f(t|_x)$.
	
	This motivates the following definition.   An \emph{$(f,L)$-reject path}  in a tree $t$ over alphabet $A$ is an infinite path $\pi$, whose source node $x$ is  not necessarily the root, such that $\pi$ is $\Uu$-rejecting in $f(t|_x)$.
 A tree $t \otimes X$ belongs to the negative characteristic language of $f^{-1}(L)$ if and only if for every $x \not \in X$ there is a $(f,L)$-reject path that begins in $x$.
	
We say that an infinite path $\pi$  \emph{merges} with a node $x$ that is an ancestor of the source node of $\pi$ if   in the two runs of the transducer $f$ obtained by starting the transducer in node $x$ and in the source node of $\pi$, the same states are reached in some node of  $\pi$.  

\begin{lemma}\label{lem:merge-reject-paths} Let $L$ and $f$ be as in the assumption of Lemma~\ref{lem:half-characteristic}. For every node $x$ in a  tree $t$ over alphabet $A$,   $f(t|_x) \not \in L$ if and only if some $(f,L)$-reject path $\pi$ merges with $x$.
	\end{lemma}
	\begin{mproof} For the left-to-right implication in the lemma, we can simply take a path that begins in $x$, and is $\Uu$-rejecting in $f(t|_x)$. For the right-to-left implication, we use   prefix independence of $\Uu$-rejecting paths.
	\end{mproof}
	
	If $X,Y$ are sets of nodes, define an $(X,Y)$-path to be a path  contained in $X$ such that $Y$ contains the source node of the path and no other nodes of the path. The path can be finite or infinite.
	\begin{lemma}\label{lem:reject-path-witnesses}
		Let $L$ and $f$ be as in the assumption of Lemma~\ref{lem:half-characteristic}.
		For every input tree $t$ and state $q$ of the transducer $f$ there exist sets $X_q,Y_q$ such that:
		\begin{enumerate}
			\item every infinite $(X_q,Y_q)$-path is an $(f,L)$-reject path for every $q$;
			\item for every  $x$ with $f(t|_x) \not \in L$, some infinite $(X_q,Y_q)$-path merges with $x$ for some $q$;
		\end{enumerate}
		\end{lemma}
	\begin{mproof}
		Let $x_1,x_2,\ldots$ be the enumeration of the nodes of $t$ in breadth-first search order.   By induction on  $n \in \set{1,2,\ldots}$ we define sets $X_{qn},Y_{qn}$ for states $q$ of $f$
		such that the conditions in the statement of the lemma hold, with the second condition restricted only to nodes $x$ in $\set{x_1,\ldots,x_n}$.  The construction is left to the reader; the sets in the statement of the lemma are obtained by taking the limit.
	\end{mproof}

\begin{lemma}\label{lem:code-positive-in-negative}
	Let $L$ and $f$ be as in the assumption of Lemma~\ref{lem:half-characteristic}.
	The following language is recognised by a \wmsoup automaton:
	\begin{align*}
		\set{t \otimes X \otimes Y : \mbox{every infinite $(X,Y)$-path is an $(f,L)$-reject path}}
	\end{align*}
\end{lemma}
\begin{mproof}
Let $c$ be a fresh counter not in $C$.	Define a transducer 
	\begin{align*}
		f' : \trees {A \times 2 \times 2} \to \countertrees {C \cup \set c}
	\end{align*}
such that for every tree $t$ over alphabet $A$ and sets of nodes $X,Y$, the output $f_1(t \otimes X \otimes Y)$ is defined as follows. The  counter edges concerning $C$ are defined as in $f(t)$.  If the path from the root to a node $x$ is an $(X,Y)$-path, then counter $c$ is transferred with an increment from $x$ to its parent. The operations on counter $c$  are defined so that  counter $c$ is tail unbounded on a path $\pi$ in 
	\begin{align}\label{eq:reject-path-transducer}
		f'((t \otimes X \otimes Y)|_x)
	\end{align}
	if and only if $\pi$ is an infinite $(X,Y)$-path.  It follows that $t \otimes X$ belongs to the language in the statement of the lemma if and only if for every node $x$ in $t$, the subtree~\eqref{eq:reject-path-transducer} belongs to the atomic counter language over counters $C \cup \set c$ with the acceptance condition $\Uu'$ defined by
	\begin{align*}
		D \in \Uu' \qquad\mbox{iff} \qquad  c \in D \Rightarrow D \cap C \not \in \Uu.
	\end{align*}
This property can be recognised by a \wmsoup automaton thanks to Lemma~\ref{lem:positive-characteristic}.

\end{mproof}

The lemmas above imply that the negative characteristic language of $f^{-1}(L)$ is recognised by a \wmsoup automaton. Given an input $t \otimes X$ the automaton works as follows. The automaton guesses sets $X_q,Y_q$ as in Lemma~\ref{lem:reject-path-witnesses}. It checks that for every node $x \not \in X$,  some infinite $(X_q,Y_q)$-path merges with $x$, which can be done by a \wmsoup automaton because this  is an \mso-definable property. Finally, it checks condition 1 in Lemma~\ref{lem:reject-path-witnesses}, which can be done by a \wmsoup automaton thanks to  Lemma~\ref{lem:code-positive-in-negative}. 

This completes the proof of Lemma~\ref{lem:half-characteristic}, and therefore also the proof Theorem~\ref{thm:flat-capture-nested}.

%% file: appendix-overview-emptiness.tex
In this part of the appendix, we complete the proof of  Theorem~\ref{thm:biginf smallsup-have-decidable-emptiness}, which says that emptiness is deciable for \wmsoup automata.
The plan of this part is given below. 

\begin{itemize}
	\item[\ref{sec:appendix-profinite}.] In Section~\ref{sec:appendix-profinite} we prove results about generalised parity automata, in particular  we show Theorem~\ref{cor:dense} which says  that  the regular accepting runs are dense in all accepting runs.
	\item[\ref{sec:appendix-chains}.] In Section~\ref{sec:appendix-chains} we show that the automaton chains  have decidable emptiness. The proof is by reduction a theorem of Vanden Boom in~\cite{Boom11}, which establishes decidability for the  domination problem for cost functions on infinite trees that are definable in cost \wmso. 
	\item[\ref{sec:normal-form-appendix}] In Section~\ref{sec:normal-form-appendix} we prove Lemma~\ref{lem:lar} which says that every automaton can be transformed into normal form.  Property (a) in the definition of normal form  holds for any automaton produced as a result of Theorem~\ref{thm:flat-capture-nested}. Property (b) is achieved using the \textsc{lar} construction of McNaughton.
	\item[\ref{sec:profinite-characterisation-appendix}.] In Section~\ref{sec:profinite-characterisation-appendix}, we prove that the reduction from emptiness of \wmsoup automata  to emptiness for  automaton chains, as presented in Section~\ref{sec:emptiness-for-biginf-smallsup-automata}, is correct. 
\end{itemize}

\newpage

%% file: appendix-profinite.tex
\section{Profinite trees}
\label{sec:appendix-profinite}
In this section of the appendix, we prove results about generalised parity automata. Section~\ref{sec:automata-with-closed-transitions} shows that automata with closed transitions recognise closed languages. Section~\ref{sec:proof-cor-dense}, shows that  regular accepting runs are dense in all accepting runs.  Section~\ref{sec:approximations} shows how a profinite tree can be converted into a similar tree, with the similarity growing on each path.

\subsection{Automata with closed transitions}
\label{sec:automata-with-closed-transitions}
In this section we prove Lemma~\ref{lem:closed-transitions}, which says that if the transitions of  a generalised parity automaton are a closed set, then so is the recognised language. % We begin with the following simple observation, which is given without proof.
% \begin{lemma}\label{lem:having-factor-clopen}
% 	Let $A$ and $Q$ be a finite sets. 
% 	For every \mso formula $\varphi$ over partially $Q$-clored trees over $A$, there is an \mso formula $\psi$ over partially $Q$-colored trees over $A$ such that 
% 	\begin{align*}
% 		\mbox{$\rho$ satisfies $\psi$} \qquad \mbox{iff} \qquad \mbox{some profinite $\rho$-factor satisfies $\varphi$}.
% 	\end{align*}
% \end{lemma}

\begin{lemma}\label{lem:closed-transitions}
	If the transitions of  a generalised parity automaton are a closed set, then so is the recognised language.
\end{lemma}
\begin{mproof}
	Let $(t_n)$ be a sequence of profinite trees accepted by the automaton, whose limit is a tree $t$. We need to show that the limit is also accepted.  For each tree $t_n$, let $\rho_n$ be an accepting run, which is again a profinite tree. By compactness and extracting a subsequence, we can assume without loss of generality that the sequence $(\rho_n)_n$ tends to a run, call it $\rho$. Therefore, the  lemma boils down to showing that the limit of a sequence of accepting runs is also accepting.  The parity condition is \mso-definable, it is closed, and therefore preserved in the limit. It remains to show that every  profinite factor of the limit $\rho$  is a transition of the automaton. 
	
Let then $\sigma$	 be a profinite factor of $\rho$.
Let $\epsilon > 0$.	Let $X_\epsilon$ be the set of trees which have a profinite factor at distance at most $\epsilon$ from $\sigma$.  For every $\epsilon>0$, the   set of trees at distance at most $\epsilon$ from $\rho$ is \mso-definable, and therefore $X_\epsilon$  is also \mso-definable. 
 % by Lemma~\ref{lem:having-factor-clopen}.
Since the set of profinite factors of a given tree is closed, it follows that 
\begin{align*}
	\bigcap_{\epsilon>0} X_\epsilon
\end{align*}
is the set of trees that have profinite factor $\sigma$.
Since  $X_\epsilon$ is \mso-definable, it is open,  and since it contains $\rho$, it must  also contain almost all runs $\rho_n$, in particular  at least one run $\rho_n$. It follows that for every $\epsilon>0$, some run $\rho_n$ has a profinite factor that is at distance at most  $\epsilon$ from $\sigma$. Since all profinite factors of $\rho_n$ are transitions, it follows that $\sigma$ can be approximated arbtrarily closely by transitions. By the assumption that the set of transitions is closed, it follows that $\sigma$ itself is a transition.\end{mproof}

\subsection{Proof of Theorem~\ref{cor:dense}}
\label{sec:proof-cor-dense}
In this section, we prove Theorem~\ref{cor:dense}, which says that   
\begin{align*}
	 \overline{L(\Aa)} =	\overline{\lreg \Aa}
\end{align*} 	holds for every generalised parity automaton $\Aa$. The nontrivial inclusion is 
\begin{align*}
		 \overline{L(\Aa)} \subseteq	\overline{\lreg \Aa}.
\end{align*}
By properties of closure, it is sufficient to show
\begin{align}\label{eq:inclusion-in-regular-runs}
		 {L(\Aa)} \subseteq	\overline{\lreg \Aa}.
\end{align}
To prove the above inclusion, we will use the following lemma.
\begin{lemma}\label{lem:included-in-closure}
	Let $X,Y$ be sets of profinite trees. Then $X \subseteq \overline Y$ if and only if every \mso formula true in some tree from  $X$ is  also true  in some tree from  $Y$.
\end{lemma}
\begin{mproof}\label{pf:}
	By unraveling the definitions, the inclusion $X \subseteq \overline Y$ means that for every profinite tree $t \in X$  and every $\epsilon > 0$, there is a tree in $Y$ which is $\epsilon$-close to $t$.
	 
	   	For the right-to-left implication, we observe that the $\epsilon$-ball around an arbitrary profinite tree  $t$ is \mso-definable. Therefore, if $t$ is in $X$, then the  $\epsilon$-ball around it intersects $X$, and therefore it intersects $Y$. Therefore, every open set containing $t$ intersects $Y$, which means that $t$ belongs to $\overline Y$.

	For the left-to-right implication, suppose that an \mso formula is true in some  profinite tree $t \in X$. For sufficiently small $\epsilon$, the $\epsilon$-ball around $t$ contains only trees that satisfy the \mso formula. This $\epsilon$-ball must contain an element of $Y$.
\end{mproof}

By the above lemma, to prove~\eqref{eq:inclusion-in-regular-runs},   it suffices to show that for every \mso formula true in some tree accepted by $\Aa$, the same \mso formula is true in some tree accepted by $\Aa$ via a regular run. The reason for this is that, as we will show in Lemma~\ref{lem:profinite-regularity-rabin-lemma}, if a generalised parity automaton has some accepting run, then it has a regular accepting run. Before proving Lemma~\ref{lem:profinite-regularity-rabin-lemma}, we introduce some terminology.

Let $\rho$ be a regular partial $Q$-coloring of a tree $t$. We say that $\rho$ has  \emph{degree of regularity at most $k$} if  $\rho$ has at most $k$   distinct profinite subtrees that have a colored root. A partially colored tree is regular if and only if it has finite degree of regularity.
\begin{lemma}\label{lem:degree-of-regularity}
	For every $k \in \Nat$, 	the set of regular partially $Q$-colored trees with degree of regularity $\le k$ is closed.
\end{lemma}
\begin{mproof}
 A partially $Q$-colored tree does not have degree of regularity at most $k$ if and only if one can find a  set $\Gamma$ of $k+1$  \mso formulas
	which are mutually contradictory,  and each one of them is true in some subtree of $\rho$ that is rooted in a colored node. For every fixed choice of $\Gamma$, this property is \mso-definable and therefore clopen, and therefore the intersection ranging over all choices of $\Gamma$ is closed.
\end{mproof}

Let $\Aa$ be a generalised parity automaton. As for standard parity automata, nonemptiness  for generalised parity automata can be described in terms of a parity game, called  the \emph{acceptance game for $\Aa$}, which is played by players Automaton and Pathfinder. Positions of player Automaton  are states of the automaton plus a special initial position. Positions of player Pathfinder are transitions of the automaton. In the initial position, player Automaton chooses a transition where the root is uncolored; in a position corresponding to a state, player Automaton chooses a transition where the root colored by that state. In a position corresponding to a transition, player Pathfinder chooses a state that appears in some leaf that is colored by a state.  The parity condition is on the sequence of states that appears in the play, with the order on states and notion of accepting state  inherited from the automaton.

\newcommand{\rootname}{\bot}
\newcommand{\roothat}[1]{#1_{\rootname}}
\begin{lemma}\label{lem:profinite-regularity-rabin-lemma}
	For a generalised parity automaton, the following  are equivalent:
	\begin{enumerate}
		\item the automaton has an accepting run;
		\item player Automaton wins the acceptance game;
		\item the automaton has a regular accepting run.
	\end{enumerate}
\end{lemma}
\begin{mproof}
By using classical proofs, one can easily see that  the lemma is true for  real runs; i.e.~when the runs of generalised parity automata are restricted to real trees. The point of the proof is to show that the equivalence extends to profinite trees.

		Let $Q$ be the states of the automaton.
		For a transition  in the automaton, define its \emph{profile} to be the pair $(q,P)$ where: $q$ is the color of the root or $\rootname$ if the root is uncolored, and $P$ is the set of states that appear as colors of leaves.   Define  the \emph{profile of a run} $\rho$ of the automaton to be the set of profiles the  transitions used by the run. 
For  a set $X$ of transition profiles,		consider the following parity game,  call it \emph{the profile game of  $X$}, which is also played by Automaton and Pathfinder. 
		The initial position is  $\rootname$. In a  position  $q \in Q \cup \set \rootname$, player Automaton chooses a set $P \subseteq Q$ with $(q,P) \in X$, and the game continues from position $P$.   In a position  $P \subseteq Q$, player Pathfinder chooses an element  $p \in P$, and the game continues from position $p$.  The winning condition for player Automaton is that the maximal state  seen infinitely often is accepting. 
		
		We now prove the equivalence of the conditions in the lemma. Clearly 3 implies 1, so we only to show the remaining implications.

\paragraph*{1 implies 2.} 		It is not difficult to see that player Automaton wins the acceptance game of the automaton if and only if player Automaton wins the profile game of $X$, where $X$ is the set of all profiles of transitions used in the automaton. Therefore, to prove that 1 implies 2, it suffices to show  that if   $\rho$ is an  accepting run, then  player Automaton  wins the profile game of the  profile of $\rho$.		An accepting run with profile $X$ satisfies the following two \mso-definable properties: a) it has profile $X$; and b) it satisfies the parity condition, i.e.~on every infinite path with infinitely many colored nodes, the maximal color appearing infinitely often is accepting. 
		The conjunction of these properties is \mso-definable. If an \mso formula is true in some profinite tree, then it is true in some real tree. Therefore  some real tree satisfies a) and b). For real trees, a) and b) imply that Eve wins the game associated to $X$.

\paragraph*{2 implies 3.} Let the transitions of the automaton be $\Delta$. A strategy of player Automaton in the acceptance game is a function
\begin{align*}
	\sigma : \bot \cdot  (\Delta \cdot Q)^* \to \Delta.
\end{align*}
If the range of the strategy contains only real trees, then the strategy induces  a real  run, call it the \emph{unfolding of $\sigma$},  which is accepting if and only if the strategy is winning.  When the strategy is furthermore memoryless, then its unfolding is regular, and has degree of regularity bounded by the number of states.
Since the acceptance game is a parity game, if player Automaton wins it, then player Automaton wins it via a memoryless strategy, which is  a function
\begin{align*}
	\sigma :  Q \cup \set \bot \to \Delta.
\end{align*}
Because every profinite tree can be approximated arbitrarily closely by real trees, we can choose for  every $\epsilon >0$ a function 
\begin{align}
	\sigma_\epsilon :  Q \cup \set \bot \to \mbox{real trees}
\end{align}
such that for every argument $q$, $\sigma_\epsilon(q)$ is at distance at most $\epsilon$ from $\sigma(q)$ and  has the same profile. Note that $\sigma_\epsilon$ is not necessarily a strategy in the acceptance game, since its values might not be transitions. Since the unfolding of $\sigma$  satisfies the parity condition, and the function $\sigma_\epsilon$ uses the same profiles, it follows that the unfolding of $\sigma_\epsilon$, call it $\rho_\epsilon$,  also satisfies the parity condition.  The unfolding  $\rho_\epsilon$  has degree of regularity bounded by the number of states.
		By compactness, we can assume without loss of generality that there is a limit of $\rho_\epsilon$ as $\epsilon$ tends to zero.
		By Lemma~\ref{lem:degree-of-regularity}, the limit has degree of  regularity bounded by the number of states and  is therefore regular.  The parity condition is satisfied in the limit, because it is \mso-definable and therefore preserved in the limit. Finally, every   factor of the limit is of the form   $\sigma(q)$, and therefore the limit is an accepting run of $\Aa$. For the last statement, we use the fact that every $\rho_\epsilon$ has degree of regularity bounded by the number of states.	\end{mproof}

\begin{mproof}[of Theorem~\ref{cor:dense}]
We only need to show the inclusion 
\begin{align*}
	L(\Aa) \subseteq \overline{\lreg \Aa}.
\end{align*}
By Lemma~\ref{lem:included-in-closure}, it suffices to show that for every \mso formula true in some tree accepted by $\Aa$, the same \mso formula is true in some tree accepted by $\Aa$ via a regular run.  In other words, we need to show that if $\Bb$ is a (non-generalised) parity automaton, then if it $\Aa$ and $\Bb$ accept some common tree, then they accept a tree where the accepting run of $\Aa$ is regular. This is proved by applying Lemma~\ref{lem:profinite-regularity-rabin-lemma} to a product automaton. 
	
\end{mproof}
	% \begin{lemma}\label{lem:}
	% 	Suppose that 
	% 	\begin{align*}
	% 		\lim_n t_n = t
	% 	\end{align*}
	% 	is a converegent sequence of profinite trees. If $s$ has the same nodes as $t$, then there is a convergent sequence of profinite trees with
	% 	\begin{align*}
	% 		\lim_n (t_n \otimes s_n) = t \otimes s.
	% 	\end{align*}
	% \end{lemma}
	% \begin{mproof}
	% 	Suppose that the distance between $t_n$ and $t$ is $\epsilon$. We can find some $s_n$ such that the distance between $t_n \otimes s_n$ and $t \otimes s$ is at most $\epsilon'$, with $\epsilon \mapsto \epsilon'$ tending to zero.
	% \end{mproof}
% 
% By the above lemma, we  can find a sequence of factor identifiers $X_1,X_2,\ldots$ such that
% \begin{align*}
% 	(t_n \otimes \rho_n)  | X_n  \qquad \to \qquad (t \otimes \rho )|X.
% \end{align*}
% In other words, the profinite tree on the right is a limit of transitions, and therefore it is a transition itself by the assumption that transitions form a closed set.

\subsection{Approximating a profinite factorised tree}
\label{sec:approximations}
Define a \emph{factorised tree over alphabet $A$} to be a tree over alphabet $A$ together with a partial coloring that uses one color. In other words, this is a (possibly profinite) tree over the alphabet $A \times \set{c, \bot}$, where $c$ is the name of the unique color. We  denote factorised trees by $\lambda$. A \emph{factorisation} of a profinite tree $t$ over $A$ is a factorised tree that projects to $t$.
In this section, we prove Lemma~\ref{lem:unfold-simulation}, which says that a profinite factorised tree  $\lambda$ can be approximated by a real factorised tree $\real \lambda$  so that $\lambda$ and $\real \lambda$ are close, and $\real \lambda$-factors resemble $\lambda$-factors with the resemblance growing closer as the distance from the root increases.

Before proving Lemma~\ref{lem:unfold-simulation}, we need some auxiliary results. The following lemma shows  that if $f$ is a relabelling, and $f(t)$ is a limit of trees from a set $R$, then $t$ is a limit of trees with images in $R$.

\begin{lemma}\label{lem:general-topology}
	Let $A,B$ be alphabets and let  $f : A \to B$. Then 
	\begin{align*}
		f^{-1}(\overline R) \subseteq \overline{f^{-1}(R)} \qquad \mbox{for every set $R$ of profinite trees over $B$}.
	\end{align*}
\end{lemma}
\begin{mproof}
	Let $s$ be an element of the left side in the inclusion of the lemma.
	For $n \in \Nat$, let $S_n$ be the profinite trees over alphabet $A$ that are at distance $1/n$ from $s$. This set is \mso-definable and therefore clopen. The image $f(S_n)$ is also \mso-definable because \mso-definable sets are closed under images of relabelings (by using existential quantification). Since $f(s)$ belongs to both $f(S_n)$ and $\overline R$,  it follows that $f(S_n)$ and $\bar R$ have nonempty intersection. Since  $f(S_n)$ is open, it follows that $f(S_n)$ and $R$ have nonempty intersection, let $s_n$ be an element of this intersection.  Since the elements $s_n$ have smaller and smaller distances to $s$, it follows that the limit of $s_n$ must be $s$. Also every $s_n$ is in $f^{-1}(R)$, and $s$ belongs to the right side of the inclusion in the statement of the lemma.
\end{mproof}

Define a  \emph{pre-parity automaton} like a parity (or generalised parity) automaton, except that it does not have the transitions. If $\Aa$ is a pre-parity automaton and $\Delta$ is a set of transitions, then let $\Aa[\Delta]$ denote the generalised parity automaton which is like $\Aa$ and has transitions $\Delta$. Sometimes, we will denote generalised parity automata by $\Aa[\Delta]$, if we want to make explicit the transitions.

In a real  paritally colored tree $\rho$, define the \emph{color depth} of a node to be the number of colored ancestors of that node. A factor of $\rho$  at color depth $n$  is one obtained from a color zone whose root has color depth $n$. A subtree at color depth $n$ is a  subtree rooted in a node at color depth $n$. We emphasize that these notions are defined for real trees.

\begin{lemma}\label{lem:unfolder}
	Let $\Aa$ be a pre-parity automaton,   $\Delta$  a set of real trees, and $\Gamma \subseteq \bar \Delta$. Let $\epsilon > 0 $ be a real number. If $\Aa[\Gamma]$ has an accepting run, then $\Aa[\Delta]$ has a real accepting run $\rho$ such that for every $n \in \Nat$:
	\begin{enumerate}
		\item there are finitely many subtrees of $\rho$ with colored roots at color depth $n$;
		\item every factor of $\rho$ at color depth $n$ is at distance $\frac \epsilon n$ from some transition in~$\Gamma$.
	\end{enumerate}
\end{lemma}
\begin{mproof}
Let the input alphabet of the automaton be $A$ and the states be $Q$.  By Lemma~\ref{lem:profinite-regularity-rabin-lemma}, if $\Aa[\Gamma]$ is nonempty, then player Automaton wins the acceptance game. As in the proof of  Lemma~\ref{lem:profinite-regularity-rabin-lemma}, a  winning strategy is a function
\begin{align*}
	\sigma : Q \cup \set \bot \to \Gamma.
\end{align*}
By the assumption that $\Gamma \subseteq \overline \Delta$, for every $n \in \Nat$ there is a function
\begin{align*}
	\sigma_n :  Q \cup \set \bot \to \Delta
\end{align*}
such that for every argument $q$,  $\sigma_n(q)$ is at distance at most $\frac \epsilon n$ from $\sigma(q)$ and  has  the same profile. In the acceptance game for $\Aa[\Delta]$ consider a strategy for player Automaton where function $\sigma_n$ is used  in the $n$-th round. This is a winning strategy by the assumption that $\sigma$ was a winning strategy and that the profiles match for $\sigma$ and $\sigma_n$. Therefore the run of $\Aa[\Delta]$ constructed from unfolding this strategy is an accepting run. Item 1 from the statement of the lemma holds because the strategy depends only on the current state and the number of rounds played so far. Item 2 holds because $\sigma_n(q)$ is at distance $\frac \epsilon n$ from $\sigma(q)$.
\end{mproof}

 Consider a generalised parity automaton $\Aa[\Delta]$ which inputs factorised  trees over some alphabet $A$.  This automaton is called \emph{factor consistent} if for every run $\rho$ over an input $\lambda$, every $\rho$-factor projects to a $\lambda$-factor when the states of $\Aa$ are forgotten. In other words, the automaton uses states in the places where $\lambda$ has a defined colour.

\begin{lemma}\label{lem:factor-consistent}
	Every \mso-definable set of profinite factorised  trees is recognised by a generalised parity automaton which is  factor consistent, and such that the transitions are \mso-definable
\end{lemma}
\begin{mproof}
	Let $\Aa$ be a parity automaton that recognises the \mso-definable property in the assumption of the lemma. We construct an automaton which, on a given input $\lambda$, labels $\lambda$ by a run of   $\Aa$, but then removes the states from nodes that are not colored by $\lambda$.
\end{mproof}

We are now ready to state the main result of Section~\ref{sec:approximations}.

\begin{lemma}\label{lem:unfold-simulation}
Let $\lambda$ be a profinite factorised tree such that  every $\lambda$-factor is in $\bar R$ for some set of real treess $R$, and let $\epsilon > 0$ be a  real number. There exists a real factorised tree $\real \lambda$ 
	such that
	\begin{enumerate}
				\item \label{item:unfold-has-factors-in-r} all $\real \lambda$-factors are in $R$;
		\item \label{item:unfold-is-simulated} every $\real \lambda$-factor at color depth $n$ is at distance at most $\frac \epsilon n$ from some $\lambda$-factor;
		\item \label{item:unfold-is-narrow} for every $n \in \Nat$, there are finitely many $\real \lambda$-factors at color depth $n$.
				\item \label{item:unfold-is-close} $\lambda$ and $\real \lambda$ are at distance at most $\epsilon_0$;
	\end{enumerate}
\end{lemma}
\begin{mproof}
	Consider the $\epsilon$-ball around $\lambda$, which is \mso-definable.  Apply Lemma~\ref{lem:factor-consistent} to this ball, yielding a factor consistent generalised parity  automaton, call it $\Aa[\Sigma]$. Let $\real \Sigma$ be the real trees in $\Sigma$. Like for every \mso-definable set, $\Sigma = \overline{\real \Sigma}$.  Define $\Delta$ to be the real trees in $\Sigma$ and which project to $R$ once the transitions of the automaton $\Aa$ are ignored.  By assumption that all $\lambda$-factors are in $\bar R$, it follows that $\Aa[\bar \Delta]$ accepts $\lambda$, and is therefore nonempty.   	
	Let $\Gamma \subseteq \bar \Delta$ be those transitions which project to a $\lambda$-factor. The automaton $\Aa[\Gamma]$ still accepts $\lambda$.

	 Apply Lemma~\ref{lem:unfolder} to $\Aa$, $\Gamma$ and $\Delta$, yielding a real run $\rho$ of $\Aa[\Delta]$ over some input $\real \lambda$. The automaton $\Aa[\Delta]$ is factor consistent, since it is obtained from the factor consistent automaton $\Aa[\Sigma]$ by removing transitions. Therefore, the $\real \lambda$-factors are exactly the projections of the $\rho$-factors.
	  Since every element of  $\Delta$ projects to $R$, we get item~\ref{item:unfold-has-factors-in-r} in the statement of the lemma.  Items~\ref{item:unfold-is-simulated} and~\ref{item:unfold-is-narrow} are the two properties in the statement of Lemma~\ref{lem:unfolder}. 
The automaton $\Aa[\Delta]$ recognises a subset of the $\epsilon$-ball around $\lambda$, which yields  item~\ref{item:unfold-is-close}.
\end{mproof}

%% file: appendix-automaton-chains.tex
\section{Nonemptiness for automaton chains}
\label{sec:appendix-chains}
In this section we prove 
Lemma~\ref{lem:automaton-chain}, which says that  nonemptiness is  decidable for automaton chains. Lemma~\ref{lem:automaton-chain} follows immediately from Lemmas~\ref{lem:reduction-to-domination} and~\ref{lem:reduction-to-reduction-to-domination} given below.

\begin{lemma}\label{lem:reduction-to-domination}
	Given cost functions $\alpha,\beta$ of cost \wmso and a formula $\varphi$ of \mso, one can decide if there is a profinite tree  $t$ that satisfies
	\begin{align*}
		 t \models \varphi \qquad\mbox{and}\qquad \alpha(t)<\infty \qquad\mbox{and}\qquad \beta(t)=\infty.
	\end{align*}
	\end{lemma}
\begin{mproof}
	We say that a cost function $\alpha$ is dominated by a cost function $\beta$ over a set $L$ of trees if for every set of trees $K \subseteq L$, if $\beta$ is bounded over $K$ then so is $\alpha$. As shown by Vanden Boom in~\cite{Boom11}, the domination problem is decidable, assuming that $\alpha$ and $\beta$ are defined in cost \wmso, and $L$ is \mso definable. 
	It is not difficult to see that  the property in the statment of the lemma is equivalent to $\beta$ not being dominated by $\alpha$ over $\varphi$. 
\end{mproof}

\begin{lemma}\label{lem:reduction-to-reduction-to-domination}
	For every automaton chain $\Aa$ with input alphabet $A$, one can compute a relabeling $f : B \to A$, cost formulas $\alpha$, $\beta$ over cost \wmso over alpahbet $B$, and a formula $\varphi$ of \mso over alphabet $B$, such that the language recognised by $\Aa$ is the image under $f$ of the profinite trees $t$ satisfy
	\begin{align*}
		 t \models \varphi \qquad\mbox{and}\qquad \alpha(t)<\infty \qquad\mbox{and}\qquad \beta(t)=\infty
	\end{align*}
\end{lemma}
\begin{mproof}[sketch]
	The proof is by induction on the depth of the automaton chain. The language before the projection describes runs of the automaton chain.   The key observation is that for every cost function $\alpha$ definable in cost \wmso, one can compute    cost functions $\alpha_{\sup}$ and $\alpha_{\inf}$ of cost \wmso such that for every real   tree $t$ with a real  partial coloring $\rho$, 
	\begin{align}\label{eq:inf-sup-projections}
		\alpha_{\sup}(t \otimes \rho) = \sup_{\sigma}  \alpha(\sigma) \qquad 		\alpha_{\inf}(t \otimes \rho) = \inf_{\sigma}  \alpha(\sigma)
	\end{align}
	with $\sigma$ ranging over $\rho$-factors of $t$.  The formula defining  $\alpha_{\sup}$ says that for every $\rho$-colored node $x$ (or the $x$ being the root), the value of  the factor with root $x$ is small. The formula for  $\alpha_{\inf}$ is obtained using duality of cost functions definable in cost \wmso.	
It is not difficult to see that the equalities~\eqref{eq:inf-sup-projections} extend to profinite trees with profinite partial colorings.
\end{mproof}

%% file: appendix-lar-form.tex
\section{Normal form of \wmsoup automata}
\label{sec:normal-form-appendix}
In this part of the appendix, we prove Lemma~\ref{lem:lar}, which says that every automaton can be converted into the  following normal form:
\begin{enumerate}
	\item[(a)] for every run, in the counter graph generated by the automaton, every bounded counter is separated and root-directed.	\item[(b)]   for every state $q$ there are sets of counters $larcut(q)$ and $larcheck(q)$ with the following property. For every  run and every  finite path in the run that starts and ends in state $q$ and does not visit bigger states in the meantime, 
	\begin{itemize}
		\item   the  counters checked in the path are exactly $larcheck(q)$;
				\item   the counters cut in  the path are exactly $larcut(q)$;
	\end{itemize}
\end{enumerate}
Recall that in Theorem~\ref{thm:flat-capture-nested}, a nested counter language is converted into a \wmsoup automaton that satisfies condition (a).  Therefore, for the purposes of deciding satisfiability of \wmsoup logic,  the part of Lemma~\ref{lem:lar} which assures condition (a) is not actually needed, because in the satisfiability algorithm for \wmsoup logic we only use automata produced via Theorem~\ref{thm:flat-capture-nested}. Nevertheless, the part of Lemma~\ref{lem:lar} concerning (a) is still true: one can take an arbitrary \wmsoup automaton, convert it to existential \wmsoup logic, and then use Theorems~\ref{thm:equal-to-nested} and~\ref{thm:flat-capture-nested} to convert it back into a  \wmsoup automaton that satisfies condition~(a).

The main contribution of Lemma~\ref{lem:lar} concerns condition (b). This is achieved by  using the following lemma, with $A$ being the states of the  \wmsoup automaton. 

\begin{lemma}\label{lem:}
	Let $A$ be an alphabet. There is a deterministic word automaton $\Aa$ with a totally ordered state space $Q$ and a function 
	\begin{align*}
		lar : Q \to P(A)
	\end{align*}
	 such that for every word $w \in A^*$   and positions $i < j$, if state $q$ appears in positions $i<j$, and no state bigger than $q$ appears in positions $\set{i,\ldots,j}$, then the set of letters that appears in positions $\set{i,\ldots,j}$ is exactly $lar(q)$.
\end{lemma}
\begin{mproof} This automaton is the latest appearance record of McNaughton.
	A state of the automaton is a pair $(w,v)$ such that $wv \in A^*$ does not contain any letter twice. In particular, the number of states is the finite number
	\begin{align*}
		\sum_{n \le |A|} (n+1) \cdot n! 
	\end{align*}
  One invariant that will be satisfied by the automaton is that:
	\begin{quote}
		(*) If the state of the automaton after reading $u$ is $(w,v)$, then $wv$ is obtained from  $u$ by only keeping the last appearance of each letter.
	\end{quote}
	The division of $wv$ into $(w,v)$ is such that the comma indicates the place where the last letter of $wv$ was in the previous state. More specifically, the initial state is $(\epsilon,\epsilon)$ and the transition function is defined by
	\begin{align*}
		\delta((w,v),a) = \begin{cases}
		(x,ya) & \mbox{if $wv=xay$ for some $x,y$}\\
			(\epsilon,wva) & \mbox{otherwise}
		\end{cases}
	\end{align*}

Consider any total order $\le$ on the states of this automaton, such that if $v$ is shorter than $v'$, then $(w,v) < (w',v')$. Finally define  $lar(w,v)$ to be the set of letters in $v$.
To prove the statement of the lemma, consider a run
\begin{align*}
(\epsilon,\epsilon)=	(w_0,v_0) \stackrel {a_1} \longrightarrow 	(w_1,v_1) \stackrel {a_2} \longrightarrow  \cdots 	 \stackrel {a_n} \longrightarrow 	(w_n,v_n)
\end{align*}
such that for some $i < j$,
\begin{align*}
	(w_i,v_i)=(w_j,v_j) \qquad \mbox{and} \qquad (w_k,v_k) \le (w_i,v_i) \mbox{ for $k \in \set{i,\ldots,j}$}.
\end{align*}
We need to show that the set  words  $a_{i+1} \cdots a_j$ and $v_i$
have the same letters.
The transitions of the automaton are defined in such a way that the combined length $(w,v)$ grows or stays the same during a run, and therefore 
\begin{align*}
	|w_iv_i| = \cdots = |w_jv_j|.
\end{align*}
Since no bigger states appear between positions $i$ and $j$, it follows that all the words $v_i,\ldots,v_j$ are at most as long as $v_i$, and therefore all the words $w_i,\ldots,w_j$ are at least as long as $w_i$. It follows that no letters from $w_i$ can appear in  $a_{i+1} \cdots a_j$. It remains to show that all letters from $v_i$ appear in $a_{i+1} \cdots a_j$. 
In the state $(w_{j-1},v_{j-1})$, the letter $a_j=a_i$ is before all other letters that appear in  $v_i$; therefore these other letters must have been seen after the last appearance of $a_i$, which was at position $i$ or later.
\end{mproof}

%% file: appendix-profinite-characterisation.tex
\section{A profinite characterisation of partial accepting runs}
\label{sec:profinite-characterisation-appendix}

This part of the appendix is devoted to showing~\eqref{eq:chains-do-it} from Section~\ref{sec:emptiness-for-biginf-smallsup-automata}.  
Recall that in Section~\ref{sec:emptiness-for-biginf-smallsup-automata}, based on a \wmsoup automaton $\Aa$ in normal form, we define 
sets  $\rrr_{q*}, \rrr_q$ of real trees and generalised parity automata $\Rr_{q*}$ and $\Rr_{q}$. The main result in
Section~\ref{sec:emptiness-for-biginf-smallsup-automata} is then
		\begin{align}
			\overline{\rrr_{q*}} = \overline{L(\Rr_{q*})}  \qquad\mbox{and}\qquad 		\overline{\rrr_{q}} = \overline{L(\Rr_q)}\tag{\ref{eq:chains-do-it}}.
		\end{align}
As shown at the end of  Section~\ref{sec:emptiness-for-biginf-smallsup-automata},  using~\eqref{eq:chains-do-it} one can reduce emptiness of $\Aa$ to emptiness of the automata $\Rr_q$, which are automaton chains, and therefore have decidable emptiness by Lemma~\ref{lem:automaton-chain}. Therefore, to complete the emptiness algorithm for \wmsoup automata, it suffices to show~\eqref{eq:chains-do-it}, which we do in this part of the appendix. 

For the reader's convenience, we recall the definitions of $\rrr_{q*}, \rrr_q, \Rr_{q*}$ and $\Rr_{q}$. In all cases, the   input alphabet is the one used to describe runs of $\Aa$, i.e.~it is the  product $A \times Q$ where $A$ is  the input alphabet of $\Aa$  and $Q$ is  the state space of $\Aa$.
	
	The sets $\rrr_q$ and $\rrr_{q*}$ are sets of real trees.  The set $\rrr_q$ consists of  the accepting partial runs  where  states strictly bigger than  $q$ appear only in nodes with finitely many descendants. The  set $\rrr_{q*}$ is   the subset of  $\rrr_q$ where state $q$ is allowed only finitely often on every path. 
	
	The automata  $\Rr_{q*}$ and $\Rr_{q}$ are generalised parity automata which recognise properties of profinite trees. Both automata have one state only, call it ``state'', this state is rejecting in $\Rr_{q*}$ and accepting in $\Rr_{q}$.  
	A transition of $\Rr_{q*}$ is any profinite partially $\set{\mbox{``state''}}$-colored tree $\sigma$ over $A \times Q$ such that:
	\begin{enumerate}
		\item the projection of $\sigma$ onto  the $A \times Q$ coordinate belongs  to $\overline{\rrr_p}$, where $p$ is the predecessor of $q$ in the order on states; and
		\item for every root-to-leaf path in $\sigma$ which ends in a leaf with  defined color ``state'', the maximal value of the $Q$ coordinate is $q$.
	\end{enumerate}
A transition  of this automaton is  
any profinite partially $\set{\mbox{``state''}}$-colored tree $\sigma$ over $A \times Q$ such that:
	 \begin{enumerate}
	 	\item the projection of $\sigma$ onto  the $A \times Q$ coordinate belongs  to $\overline{\rrr_{q*}}$; and
	 	\item for every
		 root-to-leaf path in $\sigma$ which ends in a leaf with  defined color ``state'', the maximal value of the $Q$ coordinate is $q$.
		 
		\item[3,4.] $\alpha(\sigma)<\infty$  and  $\beta(\sigma)=\infty$ holds for  cost functions $\alpha,\beta$  defined  in Section~\ref{sec:emptiness-for-biginf-smallsup-automata};	 \end{enumerate}

The left equality of~\eqref{eq:chains-do-it} is  shown   in Section~\ref{sec:proof-of-half-level}, the right equality is shown in   Section~\ref{sec:full-levels}.

\subsection{$\overline{\rrr_{q*}} = \overline{L(\Rr_{q*})} $}
\label{sec:proof-of-half-level}
Section~\ref{sec:proof-of-half-level} is devoted to proving that 
\begin{align}
	\label{eq:inclusion-half-level}
	\overline{\rrr_{q*}} = \overline{L(\Rr_{q*})}
\end{align}
holds for every state $q$.

\subsubsection{Left-to-right inclusion.} We begin with the left-to-right inclusion in~\eqref{eq:inclusion-half-level}. We use the name \emph{zone} for a connected set of nodes in a real tree, which is also closed under siblings.  By connectedness, a zone has a unique minimal node (with respect to the descendant relation), this node is called the \emph{root} of the zone. If $X$ is a zone in a real tree $t$, then $t|X$ denotes the tree obtained from $t$ by keeping only the nodes from $X$.
We say a zone $Y$ is a \emph{well-founded extension}  of a zone $X \supseteq Y$ if $X$ and $Y$ have the same root  and there are no infinite paths in $Y-X$.
\begin{lemma}\label{lem:proto-enlarge}
Every zone  $X$ in  a partial accepting run $\rho$ of $\Aa$ admits a well-founded extension to a zone $Y$ such that $\rho|Y$ is a partial accepting run.
\end{lemma}
\begin{mproof}
		Let $c$ be an unbounded counter. We say that a run is $c$-accepting if   the unboundedness condition holds only for $c$, i.e.~on every infinite path that checks $c$ infinitely often, the limsup is infinite for the value of counter $c$  in nodes where $c$ is checked. Note that if $Y$ is a well-founded extension of  $X$, and $\rho|X$ is $c$-accepting, then also $\rho|Y$ is $c$-accepting, because a) every infinite path that is contained in $Y$ is already contained in $X$; and b) the values of counters can only grow when adding nodes to a counter tree.
	\begin{claim}
For every unbounded counter $c$, every zone  $X$  admits a well-founded  extension to a zone $Y$ such that $\rho|Y$ is $c$-accepting.
	\end{claim}
	\begin{mproof}
		Define\footnote{Similar notation, but with $X$ being a lower index,  was used in Section~\ref{page:restricted-value} for a different concept. In Section~\ref{page:restricted-value}, the counter paths were prohibited to pass through an ancestor of $x$ in the set $X$.} $\treeval \rho^X(x,c)$ to be the value of counter $c$ in node $x$, but only counting those counter paths that are entirely contained in $X$.  		This is the same as the value of counter $c$ in the node corresponding to $x$  in the counter tree corresponding to  $\rho|X$.
		It is not difficult to see that every $X$ admits a well-founded extension  $Y$ such that 
		\begin{align*}
			\treeval \rho^Y(x,c)
			\begin{cases}
				\mbox{is equal to }\treeval \rho (x,c) &\mbox{ when $\treeval \rho(x,c) < \infty$}\\
			 \mbox{is at least } |x|&  \mbox{ when $\treeval \rho(x,c)=\infty$}\\
			\end{cases} \qquad \mbox{for every $x \in X$.}
		\end{align*}
		This is the zone required by the claim.
	\end{mproof}
	
	Using the claim, we prove the lemma. Let $c_1,\ldots,c_n$ be the unbounded counters. Applying the claim $n$ times, we  find zones 
	\begin{align*}
		X=X_0,X_1,\ldots,X_n
	\end{align*} such that every $X_i \in \set{X_1,\ldots,X_n}$  is a well-founded extension of   $X_{i-1}$ and $\rho|X_i$ is $c_i$-accepting.
	The well-founded  extension relation is transitive, so $X_n$ is a well-founded extension of $X$. As we have observed, being $c$-accepting is closed under well-founded extensions, and therefore $X_n$ is $c_i$-accepting for every unbounded counter. Finally, since $\rho$ is a partial accepting run, and the boundedness and parity  acceptance conditions are preserved under removing nodes, the run $\rho|X_n$ satisfies  the boundedness and parity  acceptance conditions. 
\end{mproof}

Define $\rrr_{<q}$ to be the union of $\rrr_p$ ranging over states $p < q$.
\begin{lemma}\label{lem:enlarge-accepting-factor}
	Let $\rho \in \rrr_q$ and let $x$ be a node in $\rho$. There is a zone $X$ with root $x$ such that $\rho|X \in \rrr_{<q}$ and for every maximal node  $y \in X$, either $y$ is a leaf of $\rho$, or $q$ is the maximal state that appears on the path from $x$ to $y$.
\end{lemma}
\begin{mproof}
	Let $Z$ be the set of (not necessarily proper) descendants of $x$ that have a state bigger or equal to $q$, and are closest to $x$ for that property. 
    Apply Lemma~\ref{lem:proto-enlarge} to the zone 
\begin{align}\label{eq:first-q-on-every-path}
\set{ z : x \le y \le z \mbox{ for some $z \in Z$}}
\end{align}	
yielding a well-founded  extension $Y$ such that $\rho|Y$ is a partial accepting run. In the special case when  $x$ has a state bigger or equal to $q$, the set in~\eqref{eq:first-q-on-every-path} is equal to $Y$ and is also equal to $\set{x}$.
For every maximal node  $y \in X$, the path from $x$ to $y$ contains a node from~\eqref{eq:first-q-on-every-path}, and therefore a state bigger than equal to $q$. Define $X$ by adding to $Y$ all nodes that have an ancestor in $Y$ with a state strictly bigger than $q$.   The set $X$ is a zone and a well-founded extension of  $Y$, since the assumption  $\rho\in \rrr_q$ implies that every state strictly bigger than $q$ has finitely many descendants. In the proof of Lemma~\ref{lem:proto-enlarge}, we showed that well-founded extension preserves partial accepting runs, and therefore $\rho|X$ is a partial accepting run. Since $X$ was obtained from~\eqref{eq:first-q-on-every-path} by two consecutive well-founded extensions, and in~\eqref{eq:first-q-on-every-path} nodes with state $q$ have zero descendants, it follows that nodes with state $q$ have finitely many descendants in $X$, and therefore $\rho|X$ belongs to $\rrr_{<q}$. Finally, every maximal node  $y \in X$ which is not a leaf has an ancestor in $Z$, and therefore on the path from $x$ to $y$ there is a state bigger or equal to  $q$. Furthermore, if the state is strictly bigger than $q$, then  $y$ is a leaf of $\rho$, which finishes the proof of the lemma.
	\end{mproof}

% 
% \begin{lemma}\label{lem:finitely-enlarge}
% 	Let $\rho$ be an accepting run of $\Aa$, and let $X \subseteq Y$ be  zones in $\rho$  that have the same root and such that
% 	every path in $Y-X$ is finite. Then
% 	\begin{align*}
% 		  \rho|X \in \rrr_{<q} \qquad \mbox{implies}\qquad \rho|Y \in \rrr_{<q}.
% 	\end{align*}
% \end{lemma}
% \begin{mproof}
% 	The boundedness acceptance condition is satisfied by both runs, because they are factors of an accepting run. The unboundedness condition is accepted, because no new infinite paths are introduced by $Y$. Likewise for the condition imposed by $\rrr_{<q}$, which says that every node with state strictly bigger than $q$ has finitely many descendants.
% \end{mproof}

\begin{lemma}\label{lem:half-simple-half}
	Every run in $\rrr_{q*}$ is accepted by $\Rr_{q*}$.
\end{lemma}
\begin{mproof}
	Let $\rho \in \rrr_{q*}$. For nodes $x,y$ in $\rho$, define  $x \preceq y$ to hold if  $x$ is a descendant of $y$ (note that  $\preceq$-bigger nodes are closer to the root) and there is at least one occurrence of state $q$ on the path from $y$ to $x$. The assumption that $\rho$ belongs to $\rrr_{q*}$ implies that $q$ appears finitely often on every path, and therefore  $\preceq$ is  a well-founded  order. By induction on this order, we show that for every node $x$ in $\rho$, there is a run of $\Rr_{q*}$ on the subtree $\rho|_x$.
	
	Let $\rho$ be a run in $\rrr_{q*}$ and let $x$ be a node in $\rho$.
	Apply Lemma~\ref{lem:enlarge-accepting-factor} to $x$ and  $\rho$, yielding a zone $X$ with root $x$ such that $\rho|X \in \rrr_{<q}$. For every maximal node  $y \in X$ that is not a leaf of $\rho$, state $q$ appears on the path from $x$ to $y$, and therefore $y \prec x$. By  the induction assumption, the subtree $\rho|_x$ is accepted by $\Rr_{q*}$.  Combining the run $\rho|X$ with the accepting of $\Rr_{q*}$ on those subtrees, we get an accepting run of $\Rr_{q*}$ on $\rho$.
\end{mproof}

The above lemma says that
\begin{align*}
	\rrr_{q*} \subseteq	 L(\Rr_{q*}).
\end{align*}
Since closure is a monotone operator, the  left-to-right inclusion in~\eqref{eq:inclusion-half-level} follows.

\subsubsection{Right-to-left inclusion.}
By Lemma~\ref{cor:dense}, the set of trees which admit a regular  accepting run of $\Rr_{q*}$ is dense in all trees accepted by $\Rr_{q*}$. Therefore,  to prove the right-to-left inclusion
\begin{align*}
	\overline{\rrr_{q*}} \supseteq \overline{L(\Rr_{q*})}
\end{align*} in~\eqref{eq:inclusion-half-level}, 
it suffices to show  
\begin{align*}
		\overline{\rrr_{q*}} \supseteq \lreg{\Rr_{q*}})
\end{align*}
which is stated in the  following lemma.

% \begin{lemma}\label{lem:}
% If an open set of profinite factorised trees  contains all real regular factorised trees, then it also contains all profintie regular factorised trees.
% \end{lemma}
% \begin{mproof}
% 	Let $X$ be the open set in the statement of the lemma. Let $\lambda$ be a profinite regular factorised tree, and let 
% \end{mproof}

\begin{lemma}\label{lem:half-difficult-half}
	$\overline{\rrr_{q*}}$ contains every input accepted by $\Rr_{q*}$ via a regular run.
\end{lemma}
\begin{mproof}
	Let  $\rho$ be a profinite tree which admits a regular run, call it $\lambda$, of the  generalised parity automaton $\Rr_{q*}$. By definition of the automaton $\Rr_{q*}$, we know that every $\lambda$-factor is in $\overline{R_{<q}}$.
	
	Let $k \in \Nat$ be such that $\lambda$ has degree of regularity at most $k$. Let  $\lambda_1,\lambda_2,\ldots$ be a sequence of real factorised trees that tends to $\lambda$. By  Lemma~\ref{lem:degree-of-regularity}, we can assume without loss of generality that every $\lambda_n$ has degree of regularity at most $k$.
	Having a well-founded factorisation is \mso-definable, and therefore we can assume that every coloring $\lambda_n$ is well-founded.
	 Every real tree $\lambda_n$ satisfies the property  ``on every path, there are at most $k$ nodes colored by $\lambda_n$''. Since this property is \mso-definable, it is preserved in the limit, and therefore in $\lambda$ there are at most $k$ nodes on every path which are colored by $\lambda$.

	Choose some real number $\epsilon>0$. Apply Lemma~\ref{lem:unfold-simulation} to $\lambda$, yielding a real factorised tree  $\real \lambda$. Note that $\real \lambda$ is a factorisation of some real partial run of $\Aa$, call this real partial run $\real \rho$.	
	  If $\epsilon$ is  small enough with respect to $k$,  item~\ref{item:unfold-is-close} of Lemma~\ref{lem:unfold-simulation} which says that $\lambda$ and $\real \lambda$ are at distance at most $\epsilon$, implies that    the maximal $\real \lambda$-depth is also $k$. Item~\ref{item:unfold-is-narrow} says  that at every $\real \lambda$-depth there are finitely many $\real \lambda$-factors, and since there are finitely many $\real \lambda$-depths, it follows that  altogether there are finitely many different $\real \lambda$-factors.   Item~\ref{item:unfold-has-factors-in-r} implies that every  $\real \lambda$-factor is an accepting partial run.   This implies that  every $\real \lambda$-factor satisfies the boundedness acceptance condition. A real partial  run of $\Aa$ with finitely many factors, each of which satisfies the boundedness acceptance condition, must itself satisfy the boundedness acceptance condition. % (The previous sentence is true under the assumption that the counter trees generated by the automaton are such that there are no counter paths that connect bounded and unbounded counters; this can be assumed without loss of generality.) 
	Therefore, $\real \rho$ satisfies the boundedness acceptance condition. 
The parity and unboundedness acceptance conditions are also satisfied in $\real \rho$, because for  every infinite path in $\real \rho$, all but finitely many nodes of the path are in some $\real \lambda$-factor, where the parity and  unboundedness acceptance condition are satisfied. 
\end{mproof}

% 
% \begin{lemma}\label{lem:}
% 	Let $\rho$ be a profinite partial run and let  $f$ be a regular factorisation such that:
% 	\begin{itemize}
% 		\item every $f$-factor is a limit of real partial accepting runs;
% 		% \item Let $c$ be a bounded counter. If an infinite path in $\rho$ passes through infinitely many $f$-nodes, cuts $c$ finitely often and resets $c$ finitely often, then it increments $c$ finitely often.
% 		\item for every bounded  counter $c$, the maximal value of   $c$ in  the root $c$-cut factor of every  $f$-factor is  $<\infty$.
% 	\end{itemize}
%  Then there is some $\epsilon > 0$  such that for every real partial run $\real \rho$  with a real factorisation $f'$, if 
%  \begin{itemize}
% 	 \item every $f'$-factor of $\real \rho$ is an accepting partial run; and
%  	\item the distance between $\rho \otimes f$ and $\real \rho \otimes f'$ is at most $\epsilon$
%  \end{itemize}
% 	then $\real \rho$  satisfies the boundedness acceptance condition.
% \end{lemma}

\subsection{$\overline{\rrr_q} = \overline{L(\Rr_q)}$}
\label{sec:full-levels}
The rest of Section~\ref{sec:profinite-characterisation-appendix} is devoted to showing 
\begin{align*}
		\overline{\rrr_{q}} = \overline{L(\Rr_q)}.
\end{align*}
When $q$ is a parity-rejecting state of $\Aa$, then the equality above is the same one as in the previous section.  Therefore, we consider the case when $q$ is parity-accepting.

\subsubsection{Left-to-right inclusion.}
We begin by proving  the left-to-right inclusion. As in Section~\ref{sec:proof-of-half-level}, it suffices to show that every run in $\rrr_q$ is accepted by $\Rr_q$.

For a bounded counter $c$,	define a $c$-cut zone in a real run of $\Aa$ to be a maximal connected set of nodes where counter $c$ is not cut. Define the $c$-cut zone of a node $x$ to be the $c$-cut zone that contains $x$. The boundedness acceptance condition says that for every $c$-cut zone, the  value of $c$ has finite limsup.

Recall that a real factorisation of a real run $\rho$ is a partition of its nodes into colored and uncolored nodes. % Therefore, a real factorisation of a real run is the same thing  as a subset of nodes $X$.
% It is in this sense that we talk about  $X$-factors in a real run $\rho$, when $X$ is a set of nodes in $\rho$. In other words, an $X$-factor of $\rho$ is a tree of the form $\rho | Y$, where $Y$ is obtained by beginning in either the root of $\rho$ or in a node from $X$, and then  recursively adding  all children, until leaves or nodes from $X$ are reached.
\begin{lemma}\label{lem:enlarge-factorisatio}
Every     $\rho \in \rrr_q$ admits a real factorisation $\gamma$ such that
\begin{enumerate}
	\item \label{item:gamma-factors-in-rq} every $\gamma$-factor is in $\rrr_{q*}$;
	\item for every finite path with source and target colored by $\gamma$:
	\begin{enumerate}
		\item \label{item:maximal-state-is-q-gamma}the maximal visited state is $q$;
		\item \label{item:value-bounded-by-source-path-gamma}on nodes of the path, all bounded counters outside $larcut(q)$ are bounded by a finite number $n_x$ that only depends on the source node $x$ of the path.
	\end{enumerate}
\end{enumerate}
\end{lemma}
\begin{mproof}
For every node $x$ in $\rho$, apply Lemma~\ref{lem:enlarge-accepting-factor}, yielding a zone $Y_x$, and let $X_x$   be the maximal nodes of $Y_x$ that are not leaves of $\rho$. Lemma~\ref{lem:enlarge-accepting-factor} says that $\rho|X_x$ belongs to $R_{<q}$ and that on every path from $x$ to a node in $X_x$, the maximal visited state is $q$.  
Let $X$  be the smallest set that contains $X_{x}$ for $x$ being the root,  and such  that if $x \in X$, then $X_x \subseteq X$.  Note that $X$ does not contain the root. 
Define $\gamma$  so that it colors a node $x$ if and only if $x \in X$ and on some path that begins in $x$, there are infinitely many nodes from $X$. By construction, $\gamma$ satisfies items~\ref{item:gamma-factors-in-rq} and~\ref{item:maximal-state-is-q-gamma} in the lemma. Below we prove item~\ref{item:value-bounded-by-source-path-gamma}.

Let $\pi$ be a path as in item~\ref{item:value-bounded-by-source-path-gamma}.		For every bounded counter $c$, let $n_c$ be the maximal value of counter $c$ in the $c$-cut zone of the source node of $\pi$. This number is finite by assumption that $\rho$ is accepting.  Let $n$ be the maximal value of 		 of $n_c$,  ranging over bounded counters $c \not \in larcut(q)$.   To prove item~\ref{item:value-bounded-by-source-path-gamma}, it suffices to show that no bounded counter $c \not \in larcut(q)$ is cut on the path $\pi$. This follows from the following claim.

\begin{claim}
	Every path $\pi$ as in item~\ref{item:value-bounded-by-source-path-gamma} is contained in a finite path that begins and ends in state $q$, and visits no bigger nodes. In particular, no bounded counter $c \not \in larcut(q)$ is cut on $\pi$.
\end{claim}	
\begin{mproof}
	Let the source and target nodes of $\pi$ be $x$ and $y$. By assumption that $x$ is colored by $\gamma$,  it belongs to $X_{x'}$ for some $x'$, possibly $x'$ being the root. Therefore, there is some path from $x'$ to  $x$ such that the maximal visited state is $q$. By definition of the nodes colored by $\gamma$, the set  $X_y$ is nonempty and therefore there is some path that begins in $y$ and such that the maximal visited state is~$q$.
\end{mproof}
	\end{mproof}

In Lemma~\ref{lem:enlarge-factorisatio}, we defined a real factorisation $\gamma$ of a   real run. In the next lemma, we produce  a profinite factorisation of a real run. The lemma refers to the cost functions $\alpha$ and $\beta$ that were defined  in the definition of the automaton $\Rr_q$, which we recall here:
	\begin{itemize}
		\item 	the cost function $\alpha$ is defined by \begin{align*}
		\alpha(\sigma) =\qquad \max_{c} \max_x \quad \treeval \sigma(x,c)
	\end{align*}
		with $c$ ranging over bounded counters not in $larcut(q)$ and $x$ ranging over nodes which do not have an ancestor where $c$ is cut.
\item		the cost function $\beta$ is defined by
			\begin{align*}
				\beta(\sigma)= 
				\begin{dcases}
									\min_c \min_x \max_y \quad \treeval \sigma(y,c) & \mbox{if the root of $\sigma$ has defined color ``state''}\\
					\infty & \mbox{otherwise}
				\end{dcases}	
			\end{align*}
			with $c$ ranging over  unbounded counters in $larcheck(q)$, $x$ ranging over leaves with   defined color ``state'', and $y$ ranging over ancestors of $x$ where $c$ is checked.

	\end{itemize}

\begin{lemma}\label{lem:enlarge-factorisatio-profinite} 
	Let $\rho$ and $\gamma$ be as in Lemma~\ref{lem:enlarge-factorisatio}. For every $\gamma$-colored node $x$, there is a factorisation $\lambda$ of the subtree $\rho|_x$ such that:
	\begin{enumerate}
		\item every $\lambda$-factor of $\rho$ is in $\overline{\rrr_{q*}}$;
			\item the value of $\alpha$ is smaller than $\infty$ on every $\lambda$-factor of $\rho|_x$.
		\item[3.] the value of $\beta$ is $\infty$  on every $\lambda$-factor of $\rho|_x$.
	\end{enumerate}
\end{lemma}
\begin{mproof}
	Let $\rho$, $\gamma$ and $x$ be as in the statement of the lemma.
	Let $n_x \in \Nat$ be the bound from item~\ref{item:value-bounded-by-source-path-gamma} in Lemma~\ref{lem:enlarge-factorisatio}.
	
	\begin{claim}
		For every $n \in \Nat$, there is a factorisation $\lambda_n$ of   $\rho|_x$ such that:
		\begin{enumerate}
			\item every $\lambda_n$-factor of $\rho|_x$ is accepted by $\Rr_{q*}$;
				\item the value of $\alpha$ is at most  $n_x$ on every $\lambda_n$-factor of $\rho|_x$.
			\item[3$n$.] the value of $\beta$ is at least  $n$ on every $\lambda_n$-factor of $\rho|_x$.
		\end{enumerate}
	\end{claim}
	\begin{mproof}
		We define sets of nodes 
		\begin{align*}
			Y_0,Y_1,\ldots \subseteq  \set{y : \mbox{$y$ is a $\gamma$-colored descendant of $x$}}.
		\end{align*}
 The set $Y_0$ contains only $x$.
		 Suppose that $Y_{n-1}$ has already been defined. Define $Z_n$ to be those $\gamma$-colored nodes $z$, such that for some  ancestor $y \in Y_{n-1}$,  all unbounded counters $c \in larcheck(q)$ are checked between $y$ and $z$ and have value at least $n$. Note that every path which begins in $x$ and visits $\gamma$-colored nodes infinitely often must eventually see a node from $Z_n$. This is because for such a path, the maximal state seen infinitely often is $q$, and therefore all counters in $larcheck(q)$  are checked infinitely often, and therefore they must have unbounded values on the path.   Define $Y_n$ to be the minimal nodes of $Z_n$.    Define $\lambda_n$ to be the Boolean coloring which is defined in the nodes
		 \begin{align*}
		 Y_n \cup Y_{n+1} \cup \cdots
		 \end{align*}
		 From our observation on $Z_n$, it follows that if a path is entirely contained in an $\lambda_n$-factor, then it sees finitely many $\gamma$-colored nodes. This implies that every $\lambda_n$-factor belongs to $R_{q*}$, and is therefore accepted by $\Rr_{q*}$. Item 2 of the claim follows from item 3 of Lemma~\ref{lem:enlarge-factorisatio}.  Item 3 follows by definition of $\lambda_n$.
	\end{mproof}
	A run of $\Rr_q$ is a factorisation that satisfies properties 1 and 2 in the claim, and satisfies property 3$n$ for every $n$.
	Let $\lambda_n$ be the run of $\Rr_q$ in the statement of the claim.  Note that if $m>n$, then  $\lambda_m$ satisfies property 3$n$ as well. 
	By compactness, some subsequence of $\lambda_1,\lambda_2,\ldots$ has a limit, call it $\lambda$.
	Property 1 is \mso-definable, and therefore closed, and therefore satisfied in the limit $\lambda$. Property 2 is closed by Lemma~\ref{cor:dense}, and therefore satisfied in the limit $\lambda$.   For every $n$, property 3$n$ is \mso-definable, and therefore closed. Since property 3$n$ is satisfied by almost all elements of the sequence $\lambda_1,\lambda_2,\ldots$ it must  be satisifed in the limit $\lambda$. Therefore $\lambda$ is an accepting run of $\Rr_q$.	
\end{mproof}

\begin{lemma}\label{lem:combine-suspicious}
	Let $\rho \in \rrr_q$ and let $Y$ be a set of nodes in $\rho$ such that:
	\begin{itemize}
		\item $\Rr_q$ accepts every subtree of $\rho$ rooted in a node from $Y$;
		\item $\Rr_{q*}$ accepts the tree obtained from $\rho$ by removing nodes from $Y$ together with their subtrees.
	\end{itemize}
	Then $\Rr_q$ accepts $\rho$.
\end{lemma}
\begin{mproof}
	By combining the accepting runs.
\end{mproof}

We now show every  $\rho \in \rrr_q$ is accepted by $\Rr_q$, which completes the proof of 
\begin{align*}
	\overline{\rrr_q} \subseteq \overline{L(\Rr_q)}
\end{align*}
	Apply Lemma~\ref{lem:enlarge-factorisatio} yielding a factorisation $\gamma$. The set  $Y$ of minimal $\gamma$-colored  nodes  satisfies items 1 and 2 of Lemma~\ref{lem:combine-suspicious}, with item 2 following from Lemma~\ref{lem:enlarge-factorisatio-profinite}.

\subsubsection{Right-to-left inclusion.}

% \begin{lemma}\label{lem:}
% 	Suppose that $\rho$ is a partial run and $\lambda$ is a regular partial coloring
% \end{lemma}

We are left with showing
\begin{align*}
	\overline{\rrr_{q}} \supseteq \overline{L(\Rr_q)}.
\end{align*}
As in Section~\ref{sec:proof-of-half-level}
it suffices to show the following lemma.
\begin{lemma}\label{lem:full-difficult-half}
	$\overline{\rrr_{q}}$ contains every input accepted by $\Rr_{q}$ via a regular run.
\end{lemma}
 The rest of Section~\ref{sec:full-levels} is devoted to showing this lemma.
	 Let then $\rho$ be a tree accepted by the automaton $\Rr_q$, via a regular accepting run $\lambda$.   
	 We need to show that $\rho$ can be approximated by trees from $\rrr_q$ with arbitrary precision  $\epsilon > 0$.  Choose some $\epsilon$ that is small enough (smaller than the  $\epsilon$ from Lemma~\ref{lem:outer-inner}).  Apply Lemma~\ref{lem:unfold-simulation} to  $\lambda$ and $\epsilon$ with the set $R$ being $\rrr_{q*}$, yielding a real factorisation $\real \lambda$ of a real partial run $\real \rho$ of $\Aa$.  By item~\ref{item:unfold-is-close}  the lemma, the distance between $\lambda$ and $\real \lambda$ is at most $\epsilon$. Since $\rho$ is a projection of $\lambda$ and $\real \rho$ is a projection of $\real \lambda$, and projections are non-expansive,  
	  it follows that the distance between $\rho$ and $\real \rho$ is at most $\epsilon$. It remains to prove that $\real \rho$  is an accepting run. We first deal with the unboundedness condition, and then the boundedness condition.

\begin{lemma}\label{lem:}
	The run $\real \rho$ satisfies the unboundedness and parity conditions.
\end{lemma}
\begin{mproof}
	We only do the more interesting case of the unboundedness condition.
	
	Consider an infinite path $\pi$ in $\real \rho$.  If the path passes through finitely many nodes colored by  $\real \lambda$, then all but a finite part of the path is included in some $\real \lambda$-factor. By item~\ref{item:unfold-has-factors-in-r} of Lemma~\ref{lem:unfold-simulation} this factor is in $\rrr_{q*}$, and  therefore the unboundedness and parity acceptance conditions are satisfied by the path. 
	The more interesting case is when the path visits $\real \lambda$-colored nodes infinitely often. Toward a contradiction, suppose that some unbounded  counter $c$ is checked infinitely often by the path, but its values in checked nodes are  bounded by some $k \in \Nat$. 

	Decompose $\pi$ into finite paths where the source and target nodes are colored by $\real \lambda$, call them 
	\begin{align*}
		\pi_1,\pi_2,\ldots
	\end{align*}
	For every $n$, let $\real \sigma_n$ be the $\real \lambda$-factor that contains the path $\pi_n$.
	By item~\ref{item:unfold-is-simulated} of Lemma~\ref{lem:unfold-simulation},  there is some $\lambda$-factor, call it $\sigma_n$, which is at distance at most $\frac \epsilon n$ from $\real \sigma_n$. For every  unbounded counter $c \in larcheck(q)$, and every $m \in \Nat$, in every $\lambda$-factor with root colored by $\lambda$, every path from the root to a $\lambda$-colored node   satisfies the following \mso-definable properties:
	\begin{enumerate}
		\item the maximal state that appears on the path is $q$;
		\item  the maximal checked value of $c$ on the path is at least $m$.
	\end{enumerate}
	It follows that for every $c$ and $m$,  these properties are satisified by almost all the $\real \lambda$-paths  $\pi_n$. This implies the unboundedness. 
\end{mproof}

\paragraph*{Boundedness condition.}
We are now left with the boundedness condition, which is where we use the assumption that $\lambda$ is regular. 

Let $c$ be a bounded counter. A node $x$ in a real run is said to be a
\begin{itemize}
	\item $c$-increment if  counter $c$ in node $x$ is transferred with an increment to counter $c$ in the parent of $x$;
	% \item transfer counter $c$ if  counter $c$ in node $x$ is transferred without an increment to counter $c$ in the parent of $x$;	
	\item $c$-cut if the state in $x$ cuts counter $c$;
	 \item $c$-reset if counter $c$ in node $x$ is not transferred to 
\end{itemize}Define the $c$-value of a path $\pi$ to be the least upper bound on the  number of $c$-increment nodes that can be found between nodes $x < y \in \pi$ such that there are no $c$-resets between $x$ and $y$, and there are no $c$-cuts between the source of $\pi$ and $y$. Because the automaton is in normal form, the value of counter $c$ in a node is the maximal $c$-value of paths that leave the node.

For a finite path $\pi$ in a factorisation, define its \emph{factor decomposition} to be the unique decomposition
\begin{align*}
	\pi = \pi_0 \cdots \pi_n
\end{align*}
such  that the path $\pi_0$ contains no colored nodes (and is therefore possibly empty), and in the remaining paths the first node is colored and no other nodes are colored.% 
% Define the inner $c$-value of a path to be the biggest $c$-value among the paths $\pi_0,\ldots,\pi_n$, and define the \emph{outer $c$-value} of $\pi$ to be the number of paths $\pi_i$  such that $\pi_i$ has $c$-value at least one, and all nodes  in the paths $\pi_0,\ldots,\pi_{i-1}$ are $c$-incrementing or $c$-transferring. The $c$-value of a path is bounded from above by the product of its outer and inner $c$-values.

\begin{lemma}\label{lem:outer-inner}
	There is some $\epsilon > 0$ such that if a real partial run $\real \rho$ admits  a factorisation that is at distance at most $\epsilon$ from $\lambda$, then
	\begin{enumerate}
		\item \label{item:cut-three} Every bounded counter in $larcut(q)$ is cut in every path in $\real \rho$  that contains at least three colored nodes.
		\item \label{item:uncolored-uncut} There is some $k$ such that for every bounded counter $c \not \in larcut(q)$, the  $c$-value is at most $k$ for  paths  in $\real \rho$ that begin in a colored node and contain no other  colored nodes.
		\item \label{item:small-outer} There is some $k$ such that for  every   path $\pi$ in $\real \rho$ with color decomposition  $\pi=\pi_0 \cdots \pi_n$, there are at most $k$  paths among  $\pi_0,\ldots,\pi_n$ which have nonzero $c$-value for some bounded counter $c \not \in larcut(q)$.
	\end{enumerate}
\end{lemma}
\begin{mproof}
Note that each of the properties is open in the topological sense.
The first property is open because it is  \mso definable, while the remaining two properties are open because they are unions of  \mso definable properties, ranging over possible values of $k$.

Below we show that each of the properties from the statement of the lemma is true in $\rho$ and~$\lambda$.  Therefore, by openness, if $\real \rho$ and $\real \lambda$ are sufficiently close to $\rho$ and $\lambda$, then they will also satisfy the three properties.
	\begin{enumerate}
		\item  By definition of the automaton $\Rr_q$, every path in $\rho$ that goes from a colored node to a colored node must visit state $q$ at least once, and no bigger states.  Therefore, a path that visits three colored nodes must contain a path which goes from state $q$ to state $q$ and does not visit bigger states. Such a path  must  cut all counters in $larcut(q)$.  
		\item  By definition of the automaton $\Rr_q$,  the cost function $\alpha$ has finite value on every $\lambda$-factor of $\rho$. It follows that in every $\lambda$-factor of $\rho$, call it $\sigma$, there is some finite number $k_\sigma$ such that all paths that begin in the root of $\sigma$ have $c$-value bounded by $k_\sigma$ on every bounded counter $c \not \in larcut(q)$. Since $\lambda$ is regular, there are finitely many $\lambda$-factors, and therefore finitely many possible values of $k_\lambda$, so we can take $k$ to be the maximal value of $k_\sigma$.
		The run $\rho$ has finitely many $\lambda$-factors.  Therefore, property~\ref{item:uncolored-uncut} holds for $\rho$ and $\lambda$, giving some bound $k$. For a fixed value of $k$, property~\ref{item:uncolored-uncut} is  \mso definable. Therefore, if $\epsilon$ is small enough with respect to $k$, then property~\ref{item:uncolored-uncut} also holds for $\real \rho$ and $\real \lambda$.
		\item Let $k$ be such that $\lambda$ is $k$-regular.   Using a pumping argument, one shows that every real run with a $k$-regular  real coloring satisfies the following property:
		\begin{quote}
			 (*) if item~\ref{item:small-outer} fails for $k+1$, then some path in the run  increments counter $c$ infinitely often without ever resetting or cutting it.
		\end{quote}
		Since the implication (*) is  \mso definable, it must also be true in $\rho$ and $\lambda$.  Since the conclusion of the implication (*) is false in $\rho$, it follows that item~\ref{item:small-outer} must hold in $\rho$ and $\lambda$.  
	\end{enumerate}
\end{mproof}

\begin{lemma}\label{lem:boundedness-conclusion}
Let $\epsilon$, $\real \rho$ and  $\real \lambda$ be as in Lemma~\ref{lem:outer-inner}, and suppose also that:
	\begin{enumerate}
		\item  every  factor of $\real \rho$ satisfies the boundedness acceptance condition.
		\item for every $n \in \Nat$, there are  finitely many subtrees of $\real \rho$ rooted in colored nodes at color depth $n$.
	\end{enumerate}
	Then $\real \rho$ satisfies the boundedness acceptance condition.
\end{lemma}
\begin{mproof}
We need to show that for every node $x$ in $\real \rho$ and every bounded counter $c$,  the $c$-value is bounded for paths that begin in $x$.   We can restrict attention to paths that do not cut counter $c$. Let $\pi$ be a path that begins in $x$ and does not cut counter $c$. Let the color decomposition of $\pi$ be $\pi_0 \cdots \pi_n$. 	The  path $\pi_0$ is entirely contained in a single  factor, and therefore its $c$-value is bounded by a function of $x$ by the assumption that every  factor satisfies the boundedness acceptance condition. For the paths $\pi_1,\ldots,\pi_n$,  consider two cases, depending on whether $c$ belongs to $larcut(q)$.
\begin{itemize}
	\item  Let $c \in larcut(q)$.   By item~\ref{item:cut-three} of Lemma~\ref{lem:outer-inner}, the path $\pi$ can pass through at most three colored nodes and therefore there are at most three paths  $\pi_1,\ldots,\pi_n$. By assumption 2, there are finitely many subtrees of $\real \rho$ that are rooted in colored nodes visited by such paths. Therefore, by assumption 1, each of the paths $\pi_1,\ldots,\pi_n$ has bounded $c$-value.
\item Let $c \not \in larcut(q)$.   By item~\ref{item:small-outer}  of Lemma~\ref{lem:outer-inner}, at most $k$ paths among $\pi_0,\ldots,\pi_n$  have nonzero $c$-value. By item~\ref{item:uncolored-uncut}  of Lemma~\ref{lem:outer-inner}, each of the paths in $\pi_0,\ldots,\pi_n$ has bounded $c$-value.
\end{itemize}\	
\end{mproof}

The real run $\real \rho$ with factorisation $\real \lambda$, as defined at the beginning of the proof of Lemmma~\ref{lem:full-difficult-half}, satisfy the assumptions of Lemma~\ref{lem:boundedness-conclusion}. Therefore, the run $\real \rho$ satisfies the boundedness acceptance condition.